\documentclass[fleqn,usenatbib]{mnras}

\usepackage{newtxtext,newtxmath}

\usepackage[T1]{fontenc}
\usepackage{ae,aecompl}

\usepackage{graphicx}	
\usepackage{amsmath}	
\usepackage{amssymb}	
\graphicspath{{./}{figures/}}
\usepackage{color,subfigure,lineno} 
\usepackage{amsmath}
\let\oldAA\AA
\renewcommand{\AA}{\text{\normalfont\oldAA}}
\usepackage{rotating}
\usepackage{color}

\newcommand{\hbeta}{H{$\beta$}}
\def\CIV{C\,{\sc iv}}
\def\MgII{Mg\,{\sc ii}}
\def\FeII{Fe\,{\sc ii}}
\newcommand{\oiii}{[O {\sc iii}]}

\title[Extreme Variability Quasars]{Spectral Variability of a Sample of Extreme Variability Quasars and Implications for the \MgII\ Broad-line Region}
\author[Qian Yang et al.]{Qian Yang,$^{1}$\thanks{E-mail: qiany@illinois.edu}
Yue Shen,$^{1,2}$
Yu-Ching Chen,$^{1,2}$
Xin Liu,$^{1}$
James Annis,$^{3}$
\newauthor
Santiago Avila,$^{4}$
Emmanuel Bertin,$^{5,6}$
David Brooks,$^{7}$
Elizabeth Buckley-Geer,$^{3}$
\newauthor
Aurelio Carnero Rosell,$^{8,9}$
Matias Carrasco Kind,$^{2,1}$
Jorge Carretero,$^{10}$
\newauthor
Luiz da Costa,$^{8,11}$
Shantanu Desai,$^{12}$
H. Thomas Diehl,$^{3}$
Peter Doel,$^{7}$
Josh Frieman,$^{3,13}$
\newauthor
Juan Garcia-Bellido,$^{4}$
Enrique Gaztanaga,$^{14,15}$
David Gerdes,$^{16,17}$
Daniel Gruen,$^{18,19}$
\newauthor
Robert Gruendl,$^{2,1}$
Julia Gschwend,$^{8,11}$
Gaston Gutierrez,$^{3}$
Devon L. Hollowood,$^{20}$
\newauthor
Klaus Honscheid,$^{21,22}$
Ben Hoyle,$^{23,24}$
David James,$^{25}$
Elisabeth Krause,$^{26}$
\newauthor
Kyler Kuehn,$^{27}$
Christopher Lidman,$^{28}$
Marcos Lima,$^{8,29}$
Marcio Maia,$^{8,11}$
\newauthor
Jennifer Marshall,$^{30}$
Paul Martini,$^{31,22}$
Felipe Menanteau,$^{2,1}$
Ramon Miquel,$^{32,10}$
\newauthor
Andr\'es Plazas Malag\'on,$^{33}$
Eusebio Sanchez,$^{9}$
Vic Scarpine,$^{3}$
Rafe Schindler,$^{18}$
\newauthor
Michael Schubnell,$^{17}$
Santiago Serrano,$^{15,14}$
Ignacio Sevilla,$^{9}$
Mathew Smith,$^{34}$
\newauthor
Marcelle Soares-Santos,$^{35}$
Flavia Sobreira,$^{8,36}$
Eric Suchyta,$^{37}$
Molly Swanson,$^{2}$
\newauthor
Gregory Tarle,$^{17}$
Vinu Vikram,$^{38}$
Alistair Walker$^{39}$
\\
$^{1}$Department of Astronomy, University of Illinois at Urbana-Champaign, Urbana, IL 61801, USA \\
$^{2}$National Center for Supercomputing Applications, University of Illinois at Urbana-Champaign, Urbana, IL 61801, USA \\
$^{3}$Fermi National Accelerator Laboratory, P. O. Box 500, Batavia, IL 60510, USA\\
$^{4}$ Instituto de Fisica Teorica UAM/CSIC, Universidad Autonoma de Madrid, 28049 Madrid, Spain \\
$^{5}$ CNRS, UMR 7095, Institut d'Astrophysique de Paris, F-75014, Paris, France \\
$^{6}$ Sorbonne Universit\'es, UPMC Univ Paris 06, UMR 7095, Institut d'Astrophysique de Paris, F-75014, Paris, France \\
$^{7}$ Department of Physics \& Astronomy, University College London, Gower Street, London, WC1E 6BT, UK \\
$^{8}$ Laborat\'orio Interinstitucional de e-Astronomia - LIneA, Rua Gal. Jos\'e Cristino 77, Rio de Janeiro, RJ - 20921-400, Brazil \\
$^{9}$ Centro de Investigaciones Energ\'eticas, Medioambientales y Tecnol\'ogicas (CIEMAT), Madrid, Spain \\
$^{10}$ Institut de F\'{\i}sica d'Altes Energies (IFAE), The Barcelona Institute of Science and Technology, Campus UAB, 08193 Bellaterra \\(Barcelona) Spain \\
$^{11}$ Observat\'orio Nacional, Rua Gal. Jos\'e Cristino 77, Rio de Janeiro, RJ - 20921-400, Brazil \\
$^{12}$ Department of Physics, IIT Hyderabad, Kandi, Telangana 502285, India \\
$^{13}$ Kavli Institute for Cosmological Physics, University of Chicago, Chicago, IL 60637, USA \\
$^{14}$ Institut d'Estudis Espacials de Catalunya (IEEC), 08193 Barcelona, Spain \\
$^{15}$ Institute of Space Sciences (ICE, CSIC),  Campus UAB, Carrer de Can Magrans, s/n,  08193 Barcelona, Spain \\
$^{16}$ Department of Astronomy, University of Michigan, Ann Arbor, MI 48109, USA \\
$^{17}$ Department of Physics, University of Michigan, Ann Arbor, MI 48109, USA \\
$^{18}$ SLAC National Accelerator Laboratory, Menlo Park, CA 94025, USA \\
$^{19}$ Kavli Institute for Particle Astrophysics \& Cosmology, P. O. Box 2450, Stanford University, Stanford, CA 94305, USA \\
$^{20}$ Santa Cruz Institute for Particle Physics, Santa Cruz, CA 95064, USA \\
$^{21}$ Department of Physics, The Ohio State University, Columbus, OH 43210, USA \\
$^{22}$ Center for Cosmology and Astro-Particle Physics, The Ohio State University, Columbus, OH 43210, USA \\
$^{23}$ Max Planck Institute for Extraterrestrial Physics, Giessenbachstrasse, 85748 Garching, Germany \\
$^{24}$ Universit\"ats-Sternwarte, Fakult\"at f\"ur Physik, Ludwig-Maximilians Universit\"at M\"unchen, Scheinerstr. 1, 81679 M\"unchen, Germany \\
$^{25}$ Harvard-Smithsonian Center for Astrophysics, Cambridge, MA 02138, USA \\
$^{26}$ Department of Astronomy/Steward Observatory, 933 North Cherry Avenue, Tucson, AZ 85721-0065, USA \\
$^{27}$ Australian Astronomical Observatory, North Ryde, NSW 2113, Australia \\
$^{28}$ The Research School of Astronomy and Astrophysics, Australian National University, ACT 2601, Australia \\
$^{29}$ Departamento de F\'isica Matem\'atica, Instituto de F\'isica, Universidade de S\~ao Paulo, CP 66318, S\~ao Paulo, SP, 05314-970, Brazil \\
$^{30}$ George P. and Cynthia Woods Mitchell Institute for Fundamental Physics and Astronomy, and Department of Physics and Astronomy, \\Texas A\&M University, College Station, TX 77843,  USA \\
$^{31}$ Department of Astronomy, The Ohio State University, Columbus, OH 43210, USA \\
$^{32}$ Instituci\'o Catalana de Recerca i Estudis Avan\c{c}ats, E-08010 Barcelona, Spain \\
$^{33}$ Department of Astrophysical Sciences, Princeton University, Peyton Hall, Princeton, NJ 08544, US \\
$^{34}$ School of Physics and Astronomy, University of Southampton,  Southampton, SO17 1BJ, UK \\
$^{35}$ Department of Physics, Brandeis University, Waltham, MA 02453, USA \\
$^{36}$ Instituto de F\'isica Gleb Wataghin, Universidade Estadual de Campinas, 13083-859, Campinas, SP, Brazil \\
$^{37}$ Computer Science and Mathematics Division, Oak Ridge National Laboratory, Oak Ridge, TN 37831 \\
$^{38}$ Argonne National Laboratory, 9700 South Cass Avenue, Lemont, IL 60439, USA \\
$^{39}$ Cerro Tololo Inter-American Observatory, National Optical Astronomy Observatory, Casilla 603, La Serena, Chile \\
}

\date{Accepted XXX. Received YYY; in original form ZZZ}

\pubyear{2019}

\usepackage{eso-pic}

\AddToShipoutPictureBG*{%
 \AtPageUpperLeft{%
  \hspace{0.75\paperwidth}%
  \raisebox{-3.5\baselineskip}{%
     \makebox[0pt][l]{\textnormal{DES 2018-0434}}
}}}%

\AddToShipoutPictureBG*{%
 \AtPageUpperLeft{%
  \hspace{0.75\paperwidth}%
  \raisebox{-4.5\baselineskip}{%
     \makebox[0pt][l]{\textnormal{Fermilab PUB-37631V}}
}}}%
\begin{document}
\label{firstpage}
\pagerange{\pageref{firstpage}--\pageref{lastpage}}
\maketitle

\begin{abstract}
We present new Gemini/GMOS optical spectroscopy of 16 extreme variability quasars (EVQs) that dimmed by more than 1.5 mag in the $g$ band between the Sloan Digital Sky Survey (SDSS) and the Dark Energy Survey (DES) epochs (separated by a few years in the quasar rest frame). The quasar sample covers a redshift range of $0.5 < z < 2.1$. Nearly half of these EVQs brightened significantly (by more than 0.5 mag in the $g$ band) in a few years after reaching their previous faintest state, and some EVQs showed rapid (non-blazar) variations of greater than 1-2 mag on timescales of only months.  Leveraging on the large dynamic range in continuum variability between the earlier SDSS and the new GMOS spectra, we explore the associated variations in the broad \MgII\,$\lambda2798$ line, whose variability properties have not been well studied before. The broad \MgII\ flux varies in the same direction as the continuum flux, albeit with a smaller amplitude, which indicates at least some portion of \MgII\ is reverberating to continuum changes. However, the width (FWHM) of \MgII\ does not vary accordingly as continuum changes for most objects in the sample, in contrast to the case of the broad Balmer lines. Using the width of broad \MgII\ to estimate the black hole mass therefore introduces a luminosity-dependent bias. 
\end{abstract}



\section{Introduction} \label{sec:introduction}

The canonical unification model of AGN dictates that Type 2 objects (with only narrow emission lines) are drawn from the same underlying population as Type 1 objects (with both broad and narrow emission lines), but the AGN continuum and broad-line emission is obscured by a dust torus \citep{Antonucci1993, Urry1995}. However, this static classification scheme is challenged by an increasingly large body of discoveries of quasars that apparently change spectral types on multi-year timescales {\citep[e.g.,][]{Denney2014, LaMassa2015, MacLeod2016, McElroy2016, Ruan2016, Runnoe2016, Gezari2017, Yang2018, Stern2018, MacLeod2019}}, mostly from recent multi-epoch imaging and spectroscopic surveys of quasars. The broad Balmer emission lines, including H$\alpha$, \hbeta, and H$\gamma$, and broad helium lines were observed to have dramatically changed between the dim and bright epochs, even completely disappearing or emerging, following large-amplitude variability in the continuum. This population of changing-look quasars (CLQs) challenges the unified model of AGN, and is difficult to understand in the standard accretion disk theory given the observed short timescale of the changes. They may have profound implications for accretion physics \citep{Lawrence2018}.

The continuum radiation of quasars is observed to vary typically by 0.2 mag on timescales of months to years \citep[e.g.,][]{VandenBerk2004, Wilhite2005, Sesar2007, Schmidt2010, MacLeod2012, Morganson2014}. \citet{Rumbaugh2018} found that approximately 10\% of quasars can vary by $>1$ mag, dubbed extreme variability quasars (EVQs), over an observed baseline of $\sim 15$ years. Because the broad emission lines are presumably powered by the ionizing continuum, they will reverberate to the continuum variability on the BLR light-travel timescales of days to weeks, which is much shorter compared to the BLR dynamical timescale of the order of a few years. Thus the study of the EVQ population is not only useful to understand accretion physics in the context of extreme variability, but also important to characterize the response of the broad lines to the extreme continuum variability, which in turn will shed light on the kinematics and structure of the BLR.

While there have been numerous variability studies on the broad \hbeta\ line with, e.g., reverberation mapping {(RM)} data, similar studies on the broad \MgII\ line have been sparse \citep[e.g.,][]{Trevese2007, Woo2008, Hryniewicz2014, Cackett2015, Sun2015, Zhu2017}. However, the broad \MgII\ line is an important line, as it is used for RM \citep{Clavel1991, Reichert1994, Metzroth2006, Shen2016} and single-epoch black hole mass estimation of quasars at redshift $z>1$ \citep[e.g.,][]{McLure2002, McLure2004, Shen2008, Shen2011, Wang2009}, and thus it is important to understand the variability properties of broad \MgII\ as the continuum varies.

Assuming that the BLR is virialized, the BH mass is determined by the BLR size and the virial velocity \citep[e.g.,][]{Wandel1999}. Practically, the width of broad emission line is used as an indicator of the virial velocity assuming that the line is Doppler broadened. For individual object, the inner gas orbits faster in the virial theorem. Under these assumptions, the broad line width should increase (or decrease) as the BLR contracts (or expands) in respond to a decrease (or increase) in the continuum (i.e., the ``breathing'' of the BLR). An anti-correlation between the broad line width and the BLR size (proportional to the continuum luminosity) is expected, and has been seen in broad \hbeta\ \citep[e.g.,][]{Park2012, Shen2013}. However, the situation is much less clear for \MgII\ and \CIV\ \citep{Shen2013}. The full width at half maximum (FWHM) of broad \MgII\ is strongly correlated with that of broad \hbeta\ \citep[e.g.,][]{Shen2008, Shen2011, WangShu2019}, but the range of dispersion seen in the width of \MgII\ is smaller than that for \hbeta\ for the same objects \citep[][]{Shen2011}, indicating that \MgII\ is less variable \citep[e.g.,][]{Sun2015} and possibly has slightly different kinematic structure than broad \hbeta. In addition, there is a population of galaxy spectra with broad \MgII\ but no broad \hbeta\ \citep[e.g.,][]{Roig2014}. All these observations suggest that broad \MgII\ is somewhat different, and a systematic study of \MgII\ is required to understand its phenomenology and physics.

Motivated by the questions about how EVQs vary and how their broad emission lines vary accordingly, we select a sample of EVQs that vary by more than 1.5 mag in the $g$ band, using data from DES and SDSS. We obtain new Gemini/GMOS optical spectroscopy of 16 EVQs, and compare with earlier SDSS spectra. Our EVQ sample covers a redshift range of $0.5 < z < 2.1$ with \MgII\ coverage in both SDSS and GMOS spectra for all our targets, providing a unique opportunity to study \MgII\ variability leveraged on the extreme continuum variability. Compared with previous works on CLQs, most of our targets are at higher redshifts and probe higher quasar luminosities. 

In this work we present our new GMOS spectroscopy and study the spectral variability of these 16 EVQs between the SDSS epoch and the more recent GMOS epoch. The paper is organized as follows. Section \ref{sec:data} describes the imaging data from multiple sources and SDSS spectroscopic data. Section \ref{sec:method} outlines the target selection, spectroscopic observations using Gemini/GMOS, and spectral fitting procedures. We present our results in Section \ref{sec:results} and discuss the implications in Section \ref{sec:discussion}. We summarize in Section \ref{sec:summary}. 
We use a $\Lambda$CDM cosmology with parameters $\Omega_{\Lambda}=0.7$, $\Omega_{\rm m}=0.3$, and $H_0=70$ km s$^{-1}$ Mpc$^{-1}$.

\section{Data} \label{sec:data}
\subsection{Imaging Data}
The Dark Energy Survey (DES) observed 5000 deg$^2$ of the sky in five filters ($grizY_{\rm DES}$), using a wide-field camera (DECam) on the 4-m Blanco Telescope \citep{Flaugher2015}. The 10$\sigma$ single-epoch PSF magnitude limit in the five $grizY$ bands are 23.57, 23.34, 22.78, 22.10, and 20.69, respectively \citep{Abbott2018}. We used five years DES data starting from 2013 \citep{Abbott2018, Diehl2018}. For the preliminary year 5 data, we applied the zero-point calibration with an accuracy of $\sim 0.01 - 0.02$ mag. 

SDSS mapped the sky in five filters ($ugriz_{\rm SDSS}$) using a 2.5 m telescope \citep{Gunn2006} at the Apache Point Observatory \citep{Abazajian2009}, covering 11,663 deg$^2$ of the sky. SDSS also repeatedly imaged a $120^\circ \times 2.5^\circ$ stripe along the celestial equator centered at zero declination in the Southern Galactic Cap (``Stripe 82") from 1998 to 2007. The frequency was increased from 2005 to 2007 for the supernova survey \citep{Frieman2008}. Over the 10 year duration of the program, there are, on average, more than 60 epochs in the Stripe 82 region. 

The Pan-STARRS1 \citep[PS1;][]{Chambers2016} survey mapped three-quarters of the sky in five broadband filters ($grizy_{\rm PS1}$), using a 1.8-m telescope with a 1.4 Gigapixel camera. PS1 data were obtained from 2011 to 2014, filling the gap between the SDSS and DES. We used multi-epoch photometry from the detection catalog in the PS1 Data Release 2 (DR2). For quasars, the magnitude offset between the SDSS and PS1 (DES) $g$ bands is negligible, between $-0.053$($-$0.065) and 0.005(0.008) mag at redshift $z<2$ \citep{Yang2018}. 

\subsection{SDSS Spectroscopic Data}
We used the SDSS spectra as the earlier epoch comparison spectra. The SDSS spectroscopy from the SDSS-I/II covers a wavelength range from 3800 to 9200 ${\rm \AA}$ at an average resolution of $R = \lambda/\Delta \lambda \sim 2000$ \citep{Abazajian2009}, {ranging from $R\sim$ 1500 at 3800 ${\rm \AA}$ to $R\sim$ 2500 at 9000 ${\rm \AA}$}. 

\begin{figure*}
 \centering
 \hspace{0cm}
  \subfigure{
   \hspace{-1.2cm}
  \includegraphics[width=3.8in]{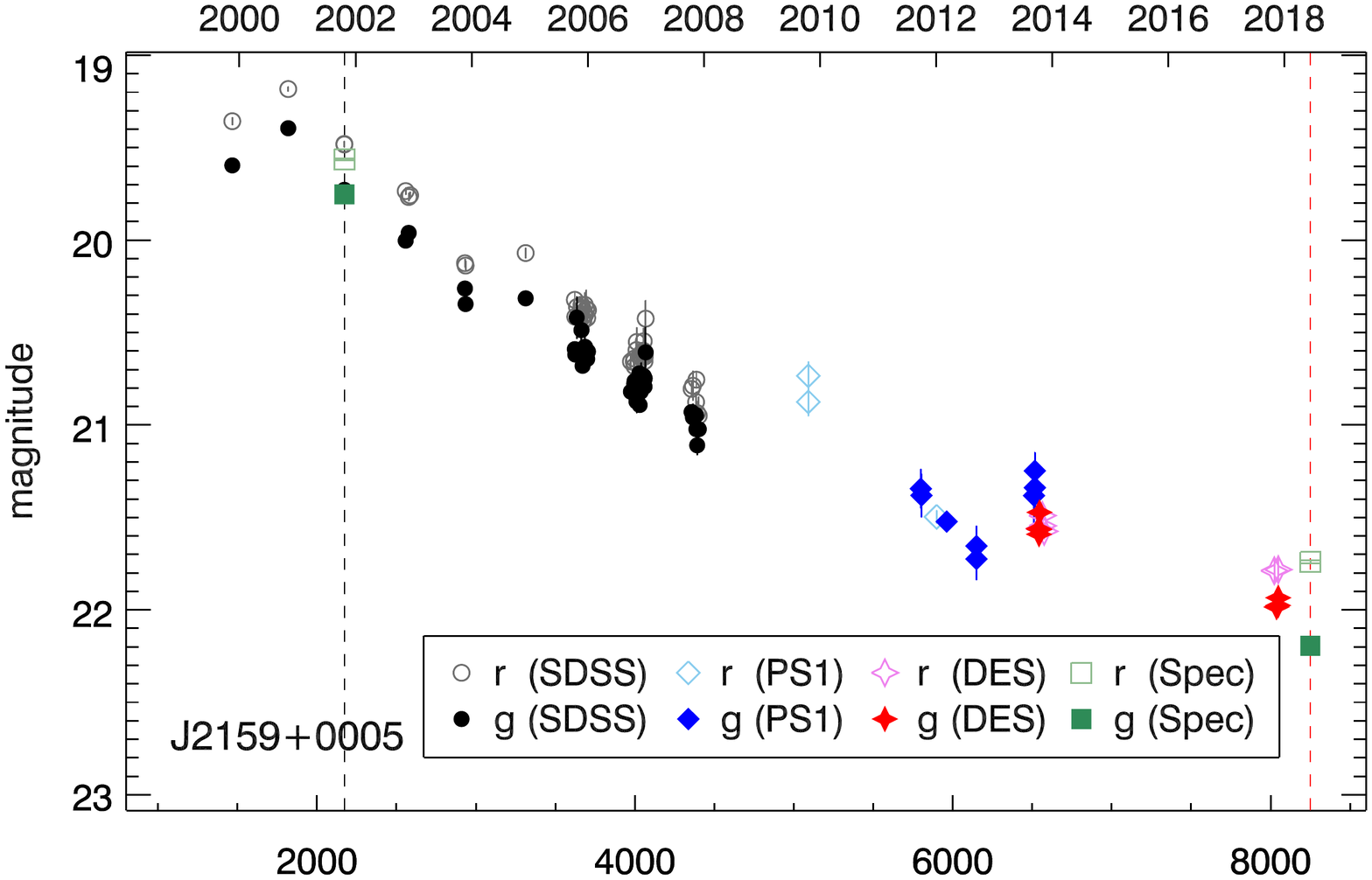}}
 \hspace{-1.4cm}
 \subfigure{
  \includegraphics[width=3.8in]{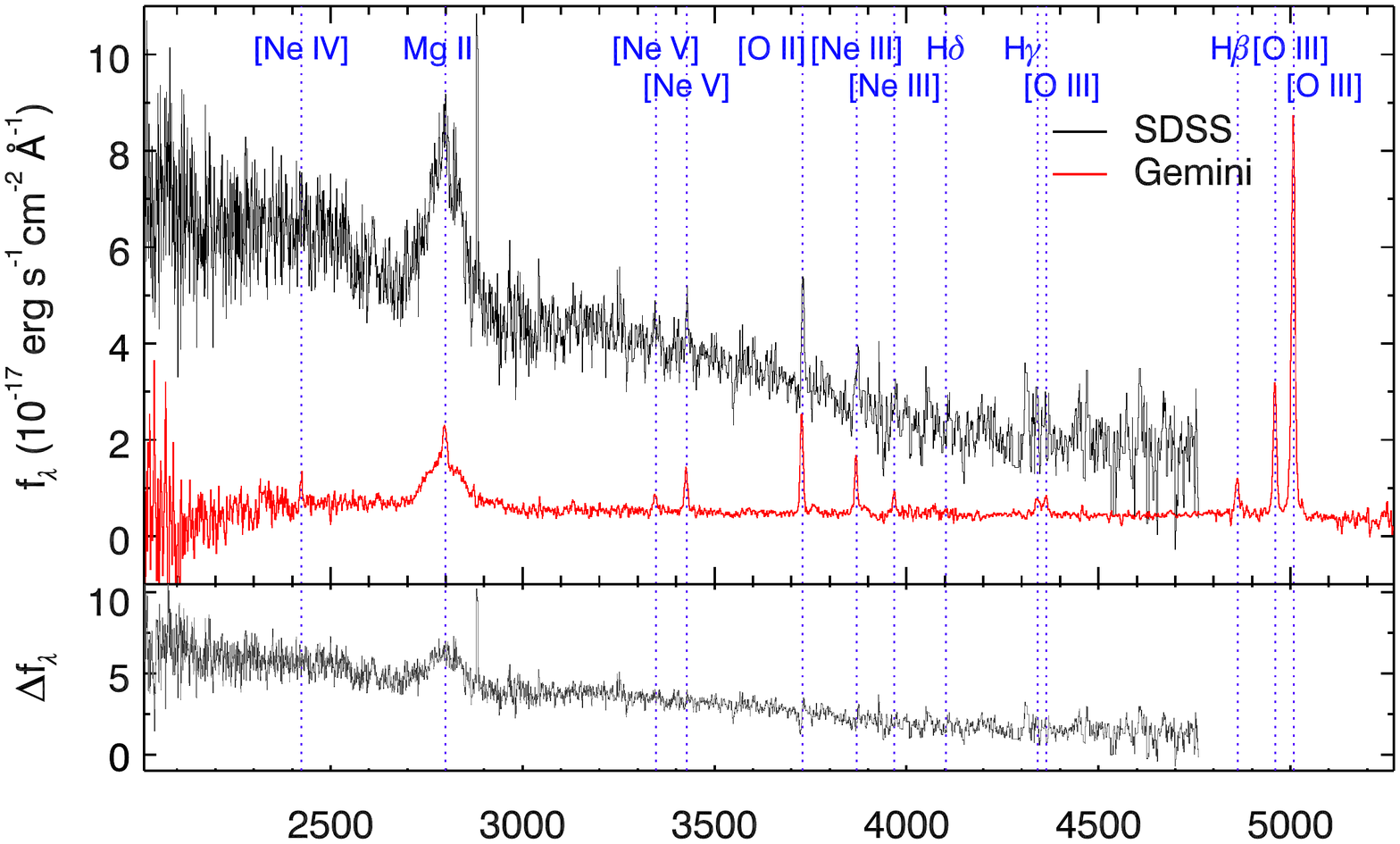}}
  
 \vspace{-2.6cm}
 
   \centering
  \hspace{0cm}
  \subfigure{
   \hspace{-1.2cm}
  \includegraphics[width=3.8in]{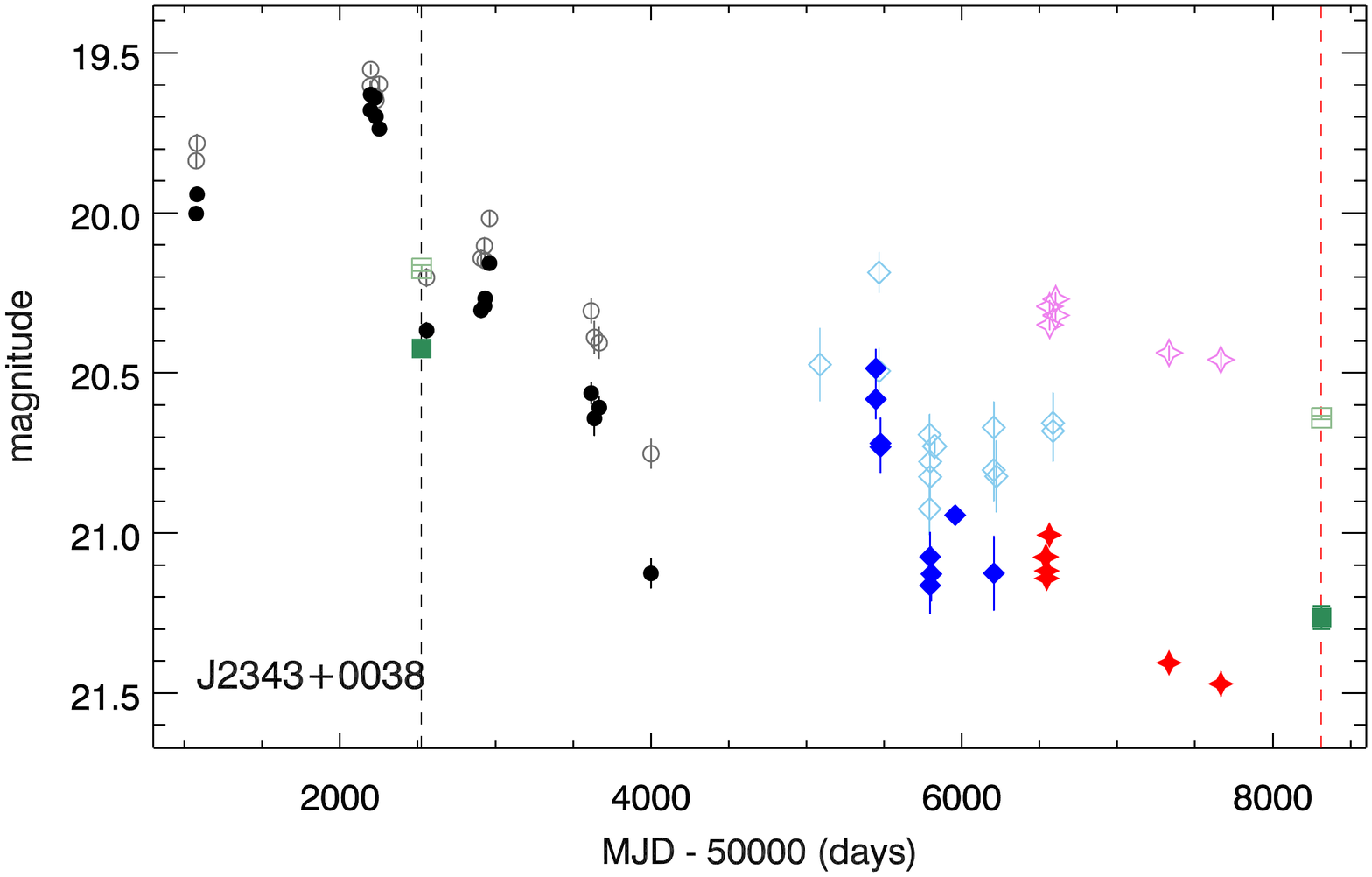}}
 \hspace{-1.4cm}
 \subfigure{
  \includegraphics[width=3.8in]{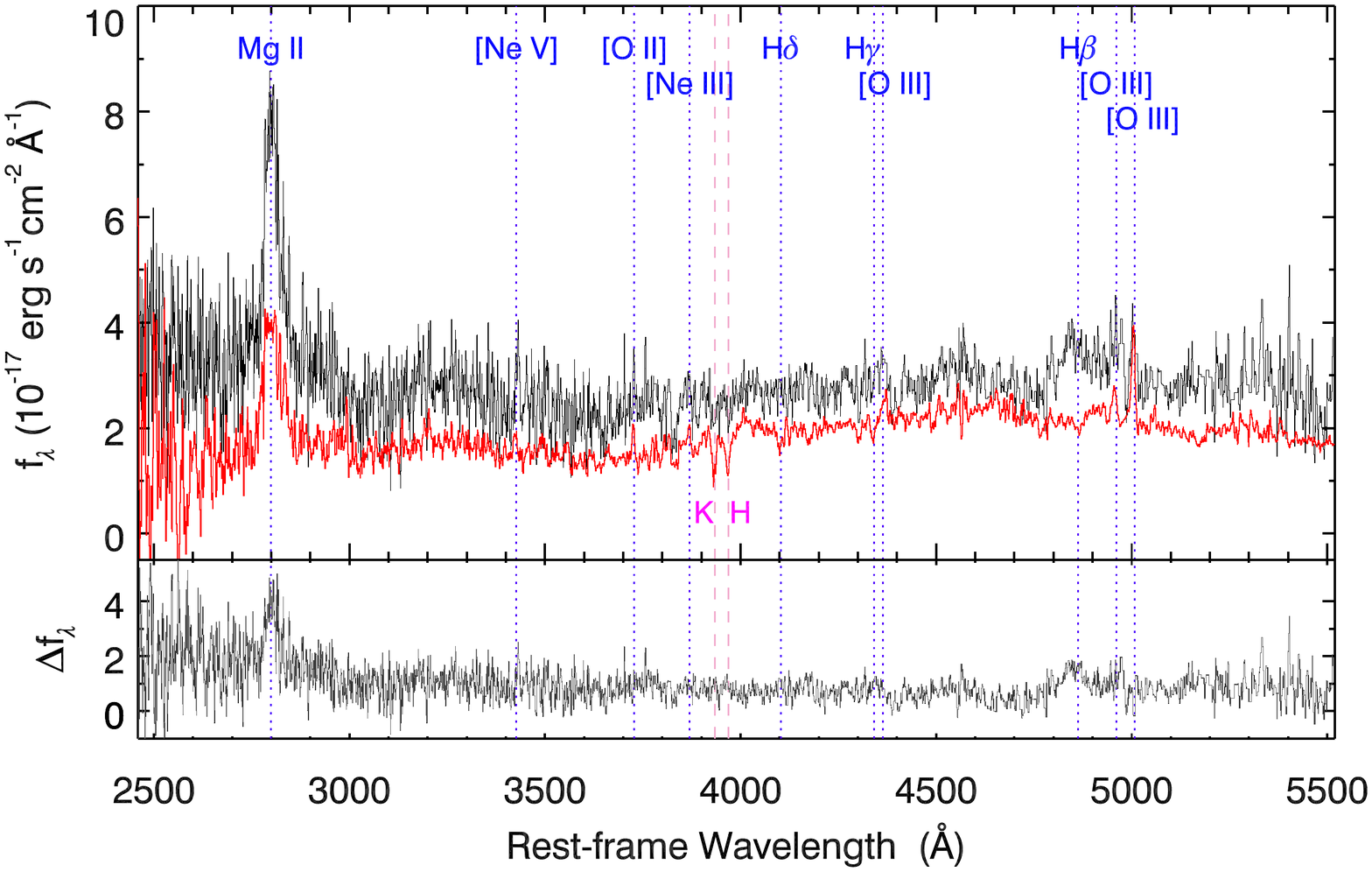}}\\
 \vspace{-0.5cm}
 
\caption{Two EVQs that have changed their spectroscopic appearance, J2159$+$0005 (upper panels) and J2343$+$0038 (bottom panels). Left: the light curves in the $g$ (filled markers) and $r$ (open markers) bands. The photometric data are from the SDSS $g/r$ (black/gray circles), PS1 $g/r$ (blue/light-blue diamonds), and DES $g/r$ (red/violet stars) bands. The vertical dashed lines mark the epochs of the SDSS (black) and recent GMOS (red) spectra. The spectrophotometric magnitude is computed from convolving the spectrum with SDSS $g$ and $r$-band filter curves (green/light-green squres).
Right: the spectrum at the earlier epoch (SDSS, black) and the most recent epoch (GMOS, red). {The spectra were smoothed with a 3-pixel boxcar.} The bottom panel shows the difference, $\Delta f_{\lambda}$, between the bright and faint epochs.
The upper-left panel shows that J2159$+$0005 faded consistently continuously since $\sim 2001$. In the dim state, there is no detectable broad \hbeta\ emission line, but broad \MgII\ is still visible. Although the SDSS spectrum does not cover \hbeta, this quasar is possibly a CLQ. If confirmed, it would be the most distant one known to date at $z=0.936$. J2343$+$0038 is a CLQ as the broad \hbeta\ emission, that is visible in the SDSS spectrum, disappears when the quasar is fainter.
\label{fig:evq_example}}
\end{figure*}

\section{Methods} \label{sec:method}
\subsection{Target Selection}
Given the large overlap between DES and SDSS within the Stripe 82 region, we carried out a systematic search for EVQs from 9,258 quasars in the Stripe 82 region \citep{MacLeod2012}, with improved calibration \citep{Ivezic2007, Sesar2007} and detailed spectral measurements from their SDSS spectra \citep{Shen2011}. Combining the SDSS Stripe 82 and DES light curves, we selected quasars with $g$-band variability larger than 1.5 mag. For the GMOS followup, we only used DES data up to Y3 (2016) at the time of the proposal. However, in this work we use all the available DES data up to Y5 to construct photometric light curves. In selecting EVQs from the combined SDSS and DES photometric light curves, we rejected points that deviate from the running median of a $\pm$100-day window by more than 0.5 mag, and noisy points with magnitude uncertainties larger than 0.15 mag. This search resulted in 146 quasars. We imposed additional criteria: (1) the quasars were restricted to be those that were in the faint state during the DES epochs; (2) the latest DES epoch (as in Y3 data at the time of the Gemini proposal) is brighter than 22 mag in $i$-band for reasonably good Gemini spectroscopy; (3) the object did not show frequent large-amplitude ($>0.5$ mag) and rapid (shorter than a month) optical variability, which may be due to blazar activity; (4) it is not a broad absorption line quasar according to the SDSS spectrum, thus excluding objects in which the variability is caused by outflows; (5) we rejected objects with poor or problematic SDSS spectra after visual inspection. We finally selected 27 EVQs, all of which have relatively smooth light curves. 

\subsection{Spectroscopic Observations}
We observed 16 of the 27 EVQs during the Gemini {2018A} run from May to September, using the Gemini GMOS-South spectrograph (summarized in Table \ref{tab:photometry}). The targets were observed with the R150 grating and a 0.5$\arcsec$ slit. We choose this configuration to balance the needs for wavelength coverage, spectral resolution and sensitivity. To mitigate the effect of CCD gaps on coverage, we split the exposures (ExpTime in Table \ref{tab:photometry}) into pairs with central wavelengths of 5800 ${\rm \AA}$ and 6000 ${\rm \AA}$, respectively. The coadded spectra cover a wavelength range of 4000 -- 10200 ${\rm \AA}$ with a spectral resolution of $R\sim630$. The impact of the spectral resolution difference between the GMOS and SDSS spectra is negligible, leading to less than 0.7\% difference in line width measurements when ${\rm FWHM} > 4000\ {\rm km\,s^{-1}}$ as in the case for the quasars we observed.

The spectra were reduced using standard IRAF\footnote{
IRAF is the Image Reduction and Analysis Facility, written and supported by the National Optical Astronomy Observatories (NOAO) operated by the Association of Universities for Research in Astronomy (AURA) under cooperative agreement with the National Science Foundation.} routines \citep{Tody1986, Tody1993}. We used two standard stars (LTT9239 or LTT7987) for our flux calibration. For the 12 objects observed before July 14, the standard stars were observed during the same night of our science observations. For the 4 objects observed on July 14 and September 5, including J2343$+$0038, J2350$+$0025, J0140$-$0035, and J0140$+$0052, we use the standard star LTT9239 observed on July 12. These four objects, as well as the standard star, were observed on photometric nights.

\hspace*{-0.5cm}
\begin{figure*}
	\includegraphics[width=6.9in]{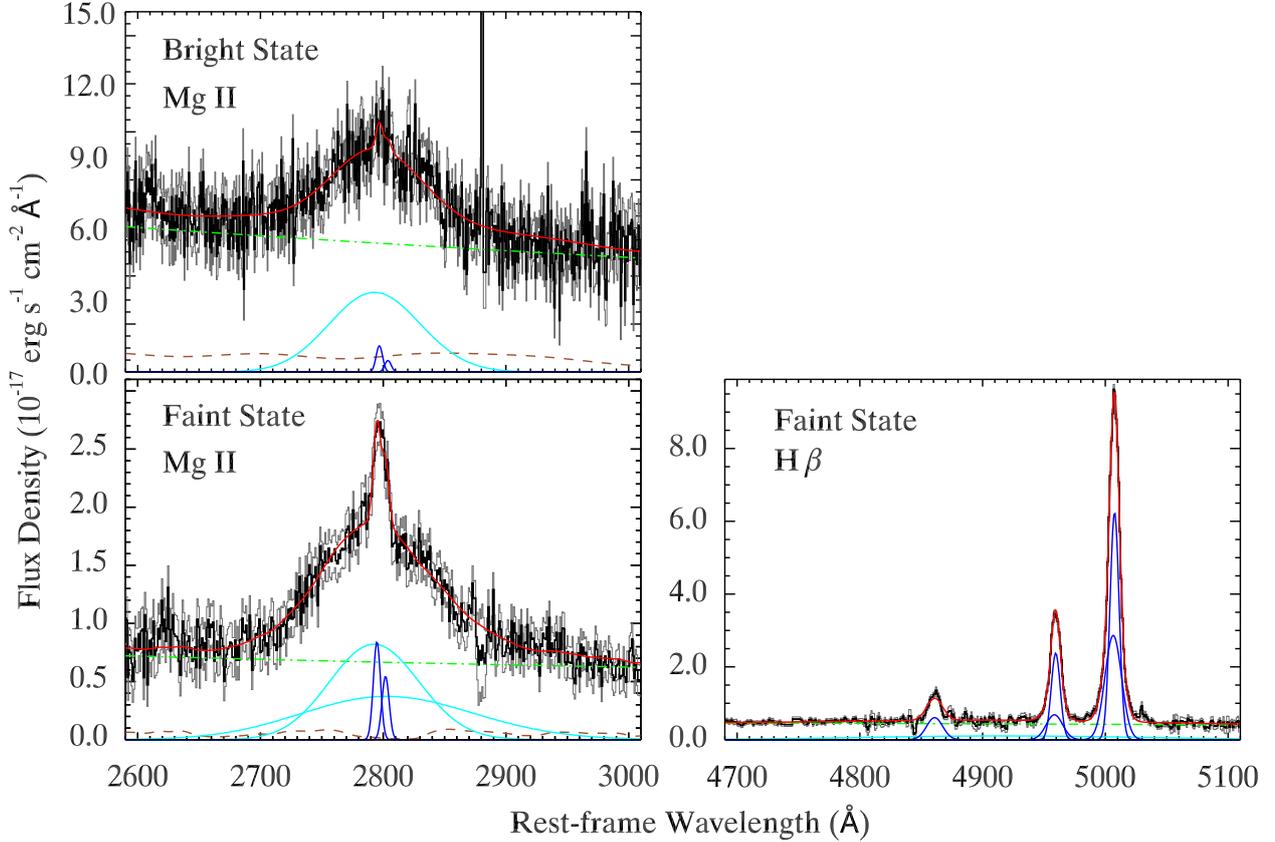}
    \caption{Spectral fitting of EVQ J2159$+$0005. In the two left panels, the region around \MgII\ is shown for the spectra obtained with SDSS (upper) and Gemini (lower). The right panel centres on \hbeta\ and \oiii${\rm \lambda \lambda}$4959,5007${\rm \AA}$. The black histogram is the spectrum; the red line is the sum of continuum (green dot-dashed line), broad (cyan) and narrow (blue) emission lines, and \FeII\ emission (brown dashed line). When the continuum faded, there is no broad \hbeta\ detected above 1$\sigma$, meanwhile the broad \MgII\ emission is still apparent. The narrow \MgII\ emission is visible on top of the broad component.} \label{fig:fitting}
\end{figure*}

\subsection{Spectral Fitting}
We fit the spectra using the spectral fitting code from \citet{Shen2018}, which models the quasar continuum, broad \FeII\ emission, and emission line components. The spectra were fit in the rest-frame of the quasar after correcting for Galactic reddening using the dust map in \citet{Schlegel1998} and {the extinction curve from \citet{Cardelli1989}}. The continuum includes a power-law continuum and a positive 3rd-order polynomial accounting for dust reddening internal to the quasar. We used empirical UV \FeII\ emission templates from the literature \citep{Vestergaard2001, Tsuzuki2006, Salviander2007} covering from 1000 ${\rm \AA}$ to 3500 ${\rm \AA}$, and an optical \FeII\ template (3686-7484 ${\rm \AA}$) from \citet{Boroson1992}. We fitted the \FeII\ emission with three free parameters: normalization, broadening, and wavelength shift. We first fit the continuum and \FeII\ emission together, choosing a few continuum windows. For instance, to avoid the influence of \MgII\ emission, we masked the region between 2675 ${\rm \AA}$ and 2925 ${\rm \AA}$. Then, we subtracted the continuum and \FeII\ emission components, and fitted the other emission lines. Following \citet{Shen2018}, we used three (one) Gaussian components for the broad (narrow) \hbeta\ line, and two Gaussian components for broad \MgII. In addition, we used two Gaussian components for the narrow \MgII\,$\lambda\lambda$2796, 2803 ${\rm \AA}$ doublet \citep{Wang2009}. To quantify the measurement uncertainties, we used a Monte Carlo approach by adding a random Gaussian deviate to the flux at each pixel, with the Gaussian $\sigma$ equal to the spectral error at that pixel. We measured the continuum/line properties and their uncertainties using the median value and the semi-amplitude of the range enclosing the 16th and 84th percentiles of the distribution from 50 trials. Our fitting results show that the narrow emission line luminosity is consistent between the SDSS and GMOS spectra, as the narrow line flux is expected to remain constant on multi-year timescales for our quasars. For example, the narrow [O II] $\lambda$3728 luminosity for J2159$+$0005 is $10^{42.34 \pm 0.05}$ and $10^{42.30 \pm 0.01}$ erg s$^{-1}$ from the SDSS and GMOS spectra, indicating that our flux calibration is reasonably accurate.

 \begin{figure*}
 \subfigure{
  \includegraphics[width=3.3in]{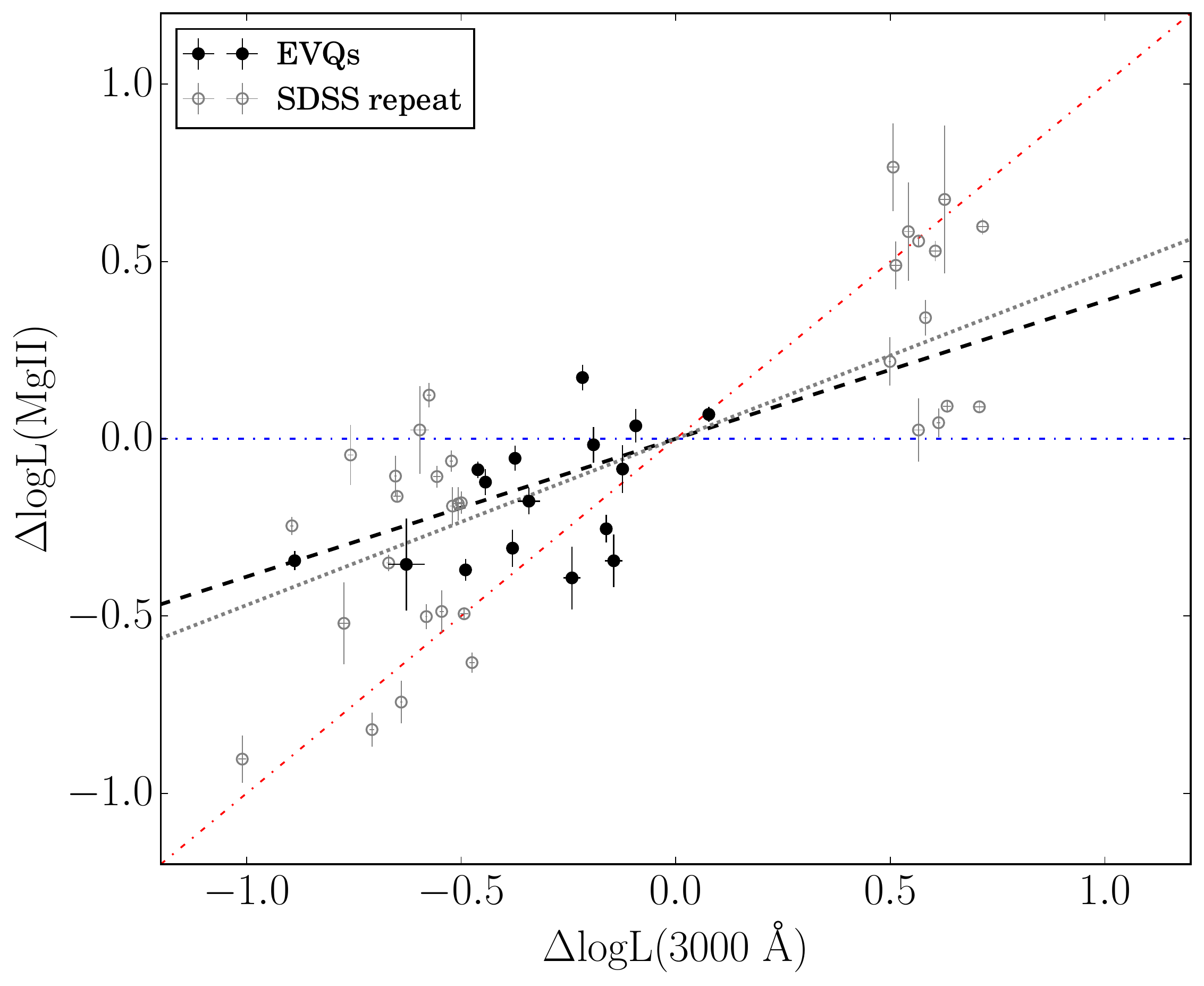}}
 \hspace{0cm}
 \subfigure{
  \includegraphics[width=3.3in]{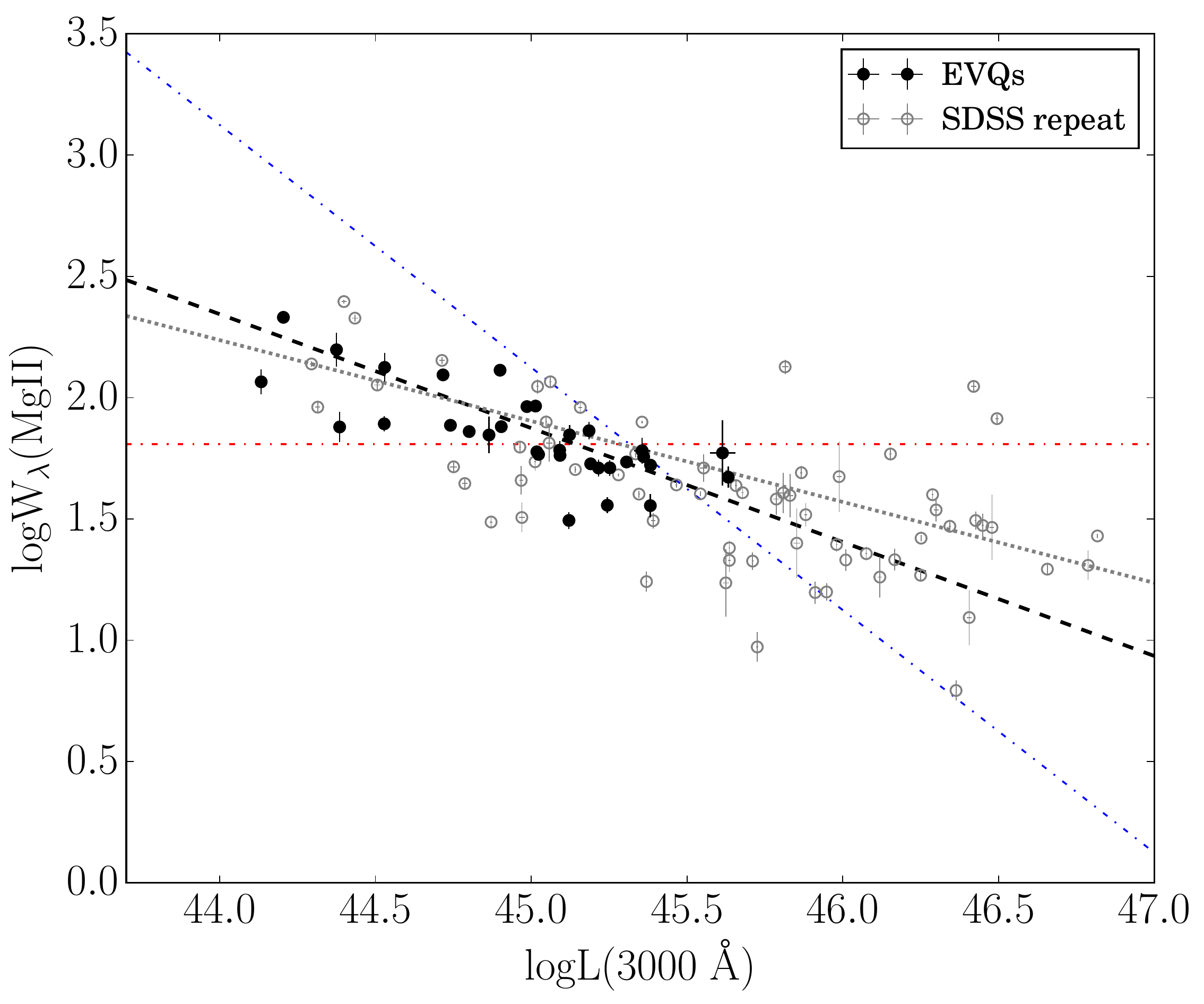}}
  \caption{Left: the broad \MgII\ luminosity variability as a function of the continuum luminosity variability between two epochs of spectroscopy. Both the line and continuum variability measures the changes from the earlier epoch to the later epoch. The filled black dots are the 16 EVQs selected from SDSS$+$DES imaging and followed-up with GMOS spectroscopy, and the open gray circles represent the supplemental EVQs selected from SDSS spectroscopy alone. The \MgII\ broad line luminosity increases when the continuum luminosity increases, {but with a slope shallower than 1}. Right: the relation between the equivalent width of broad \MgII\ and the continuum luminosity. {In both panels, the red dot-dashed line shows the 1-to-1 relation between the (fractional) variability of the line and the continuum, and the blue dot-dashed line corresponds to constant line flux independent of continuum changes. These two cases are the two extreme cases where the \MgII\ line either fully responds to the continuum changes or does not respond at all. On average these quasars are between these two extreme cases. The black dashed line and gray dotted line are the best-fit linear regression results for the main sample and the combined sample.} \label{fig:L_EW}}
  \end{figure*}

\section{Results} \label{sec:results}
We summarize the photometric variability of these EVQs in Table \ref{tab:photometry}. $\Delta g_{\rm max}$ is the maximum variability in the $g$ band from the SDSS, PS1, and DES observations. All of them have $\Delta g_{\rm max}$ larger than 1.5 mag. $\Delta g_{\rm phot}$ is the $g$-band magnitude difference between imaging photometry closest to the Gemini and SDSS spectra epochs. The majority of them rebrightened since they were intially selected in the first 3 years of DES (see their light curves in the left panels of Figure \ref{fig:evq_example} and Figure \ref{fig:cl_others}). There are only four with $\Delta g_{\rm phot}>1.5$ mag, and 9 with $\Delta g_{\rm phot}>1.0$ mag. 
$\Delta g_{\rm spec}$ is the magnitude difference between the Gemini and SDSS spectra convolved with the SDSS $g$-band filter curve. $\Delta g_{\rm back}$ is the $g$-band magnitude difference between the latest DES imaging epoch and the faintest epoch in earlier DES data, describing how much they have brightened since. Among them, 9 EVQs did not vary much with $-0.3 \leq \Delta g_{\rm back} \leq 0$, and 7 EVQs brightened with $-1.7 < \Delta g_{\rm back} < -0.6$. Hence nearly half of them have significantly brightened (by more than 0.5 mag) since the faintest epoch, which is consistent with the findings in \citet{MacLeod2019}. The light curves show that the extreme variability mostly occurs over a few years, but can also occur on much shorter timescale. For example, J0140$+$0052 brightened by $\sim 2$ mag from the GMOS spectral epoch to the last DES epoch (less than one year in the observed frame), and J2209$-$0055 brightened by $\sim 1$ mag and then again dimmed by $\sim 1$ mag from 2016 to 2018. This is consistent with the type transition timescale of CLQs spanning from 0.9 to 13 years in \citet{Yang2018}.

J2159$+$0005 is an EVQ that has been dimming continuously since 2001 as seen from its light curves in the $g$ and $r$ bands (shown in Figure \ref{fig:evq_example}). The latest DES photometry of J2159$+$0005 is $>2$ mag fainter than the earlier SDSS photometry in 2001. The right panel in Figure \ref{fig:evq_example} shows that its continuum became much fainter and redder. Figure \ref{fig:fitting} shows our fitting results around \MgII\ and \hbeta. While there is no detectable broad \hbeta\ flux in the GMOS spectrum, broad \MgII\ is still visible (albeit at reduced flux; see below). The narrow \MgII\ emission on top of the broad-line emission is more distinctive in the faint state spectrum. We witness a normal blue quasar changing to a quasar with a broad \MgII\ emission line but no broad \hbeta\ emission line, which bears some similarities to the objects presented in \citet{Roig2014}, and similar to those CLQs presented in \citet{MacLeod2019}. Although the SDSS spectrum does not cover \hbeta, this quasar is possibly a CLQ with the highest redshift of $z=0.936$ known to date. It is plausible that some broad-line quasars at redshift $z>1$, where \hbeta\ moves out of the optical window, with a red continuum are similar to J2159$+$0005 seen at redshift $z<1$.

EVQ J2343$+$0038 is a CLQ at $z=0.667$, based on the behaviour of \hbeta. It has broad \hbeta\ emission with ${\rm L_{H\beta}} = 10^{42.32 \pm 0.09}$ erg s$^{-1}$ and ${\rm FWHM_{H\beta} = 4068 \pm 1980}$ km s$^{-1}$ in the bright state from the SDSS spectrum. In the dim state, in the GMOS spectrum, the broad \hbeta\ emission has disappeared and we can clearly see \hbeta\ in absorption instead. 

The light curves and spectra of the other 14 EVQs are listed in the Appendix. \hbeta\ is available in eight GMOS spectra, in four of them \hbeta\ is broad in the faint state (J0140$+$0052, J0048$-$0113, J2209$-$0055, and J2350$+$0025), in one it is narrow (J2159$+$0005), in one it is absorption (J2343$+$0038), and in other two \hbeta\ is absent (J2252$+$0004 and J2228$-$0032). The \hbeta\ and [O III] lines in J2252$+$0004 and J2228$-$0032 are marginally present probably because these lines lie beyond $10000$ ${\rm \AA}$, where the signal-to-noise ratio (S/N) is too low to verify the existence of these lines. 
\hbeta\ is only available in three of the SDSS spectra, including J2343$+$0038, J2335$-$0049, and J2209$-$0055. Unfortunately, \hbeta\ of J2335$-$0049 lands within a CCD gap, and we accidentally failed to obtain one exposure for it with the central wavelength at 6000 ${\rm \AA}$. The \hbeta\ emission of J2209$-$0055 is broad in both epochs, and its light curves show that it brightened before its GMOS spectrum was taken.
The continuum of J2141$-$0016, at $\lambda<2200$ ${\rm \AA}$ in the rest-frame, becomes much redder when it fades, similar to cases in \citet{Guo2016, Ross2018}. But the S/N of the GMOS spectra near 4000 ${\rm \AA}$ is low, and a higher quality spectrum is needed to confirm this result. 

Broad \MgII\ line remains visible in all 16 EVQs in both bright and faint epochs. We fit the SDSS and GMOS spectra in detail to analyse their continuum and emission line properties, and explore \MgII\ variability in \S\ref{sec:MgIIline} and \S\ref{sec:MgIIwidth}.
  
\subsection{A Supplemental Sample of EVQs}
To increase sample statistics, we add a supplemental sample of EVQs with multi-epoch spectra covering \MgII\ from the SDSS quasar catalog\footnote{https://data.sdss.org/sas/dr14/eboss/qso/DR14Q/DR14Q\_v4\_4.fits} in the 14th Data Release \citep[DR14Q, ][]{Paris2018}. The selection criteria are (1) the quality flag ``zWarning" in the quasar catalog is 0; (2) the median spectral S/N per pixel is higher than 2; (3) the spectrum covers \MgII; (4) the median spectral S/N around \MgII\ and the continuum at rest-frame 3000 ${\rm \AA}$, specifically from 2700 to 3100 ${\rm \AA}$, is higher than 5 to ensure the good quality of our spectral fitting; (5) multi-epoch spectra of the same quasar must all satisfy the criteria from (1) to (4); and (6) for the same quasar, the continuum flux ratio at rest-frame 3000 ${\rm \AA}$ between the brightest and faintest epochs from SDSS is larger than 3, which corresponds to a magnitude difference of 1.2 mag. These criteria result in 33 EVQs at $0.44<z<2.33$ selected from SDSS spectroscopy alone, nearly triple our EVQ sample with multi-epoch spectra.

\subsection{\MgII\ Line Variation \label{sec:MgIIline}} 

Our EVQ sample reveals that the broad \MgII\ line does vary in the same direction as the continuum. We summarize the spectral fitting of the main EVQs sample in Table \ref{tab:fitting}. Figure \ref{fig:L_EW} shows the variability of broad \MgII\ line luminosity between the two spectroscopic epochs, $\Delta {\rm log L(Mg II)}$, as a function of the 3000 ${\rm \AA}$ continuum luminosity variability between the same two epochs, $\Delta {\rm log L(3000\AA)}$ (in the left panel). The filled black dots are the 16 EVQs selected from SDSS and DES and followed-up with Gemini, and the open gray circles represent the additional EVQs selected from SDSS. The broad \MgII\ line luminosity increases when the continuum luminosity increases, as naively expected from photoionization. {We perform least-square fits to the main EVQ sample of 16 objects, forcing the line to cross $[0,0]$. We obtain}
\begin{equation}
    \Delta {\rm log L(Mg II)} = (0.39 \pm 0.07) \times \Delta {\rm log L(3000\AA)}.
\end{equation}
The result for the combined sample, including the main sample and the supplemental sample, is 
\begin{equation}
    \Delta {\rm log L(Mg II)} = (0.47 \pm 0.05) \times \Delta {\rm log L(3000\AA)}.
\end{equation}

The variation of broad \MgII\ is indeed smaller than that in the continuum, but still echos the continuum variation to some extent. This result suggests that at least some part of broad \MgII\ reverberates to continuum changes, or that the continuum flux that ionizes broad \MgII\ varies less than the continuum at rest-frame 3000\AA. It is interesting to note that \citet{Bruce2017} found that \MgII\ flux does not seem to respond to continuum variations in two extreme variability quasars, where the extreme variability can be caused by rare, high-amplitude microlensing events.

The weaker \MgII\ variations relative to the nearby continuum variations also manifests as an anti-correlation between the equivalent width ($W_{\lambda}$ hereafter) of \MgII, ${\rm log W_{\lambda}(Mg II)}$, with continuum luminosity, ${\rm log L(3000\AA)}$, which is the well-known Baldwin effect \citep[e.g.,][]{Baldwin1977, Green2001}. In the right panel of Figure \ref{fig:L_EW}, we fit linear regressions and obtain
\begin{equation}
    {\rm log W_{\lambda}(Mg II)} = (-0.47 \pm 0.07) \times {\rm log L(3000\AA)} + (23.01 \pm 2.97),
\end{equation}
for the main sample, and
\begin{equation}
    {\rm log W_{\lambda}(Mg II)} = (-0.33 \pm 0.03) \times {\rm log L(3000\AA)} + (16.91 \pm 1.31),
\end{equation}
for the combined sample. For the same object, the $W_{\lambda}$ increases when the continuum becomes fainter (listed in Table \ref{tab:fitting}). For the most variable EVQ, J2159$+$0005, in this sample, the log$W_{\lambda}{\rm (Mg II)}$ increases from $10^{1.76\pm0.03}$ to $10^{2.33\pm0.02}$ ${\rm \AA}$ when the continuum ${\rm L(3000\AA)}$ dims from $10^{45.9\pm0.01}$ to $10^{44.21\pm0.01}$ erg s$^{-1}$. The variable $W_{\lambda}$ indicates that the dramatic decrease in continuum and \MgII\ is not caused by variable dust reddening.

\begin{table*}
	\centering
	\rotatebox{90}{
		\begin{minipage}{\textheight}
		\centering
		\caption{Variability of EVQs}\label{tab:photometry}
		
		\begin{tabular}{cccccccrrrr}
		\hline
		Name & R.A. & Decl. & Redshift & SDSS & Gemini & ExpTime & $\Delta g_{\rm max}$ & $\Delta g_{\rm phot}$ & $\Delta g_{\rm spec}$ & $\Delta g_{\rm back}$ \\
		 & & & & & 2018 & (s) & (mag) & (mag) & (mag) & (mag) \\
		\hline
		J2159$+$0005 & 21:59:44.32 & $+$00:05:27.8 & 0.936 & 2001-09-21 & 05-12 & 675$\times$4 & 2.590$\pm$0.068 & 2.207$\pm$0.058 & 2.442$\pm$0.041 & $-$0.050$\pm$0.088 \\
        J0140$+$0052 & 01:40:27.89 & $+$00:52:12.5 & 1.020 & 2004-11-06 & 09-05 & 675$\times$4 & 2.272$\pm$0.067 & 1.647$\pm$0.061 & 0.312$\pm$0.013 & $-$0.078$\pm$0.086 \\
        J2213$-$0037 & 22:13:12.08 & $-$00:37:25.6 & 2.063 & 2003-09-30 & 07-10 & 675$\times$4 & 2.258$\pm$0.096 & 1.585$\pm$0.086 & 1.148$\pm$0.063 & $-$0.176$\pm$0.123 \\
        J2252$+$0004 & 22:52:50.73 & $+$00:04:18.3 & 1.001 & 2001-09-26 & 07-10 & 675$\times$4 & 2.137$\pm$0.042 & 1.517$\pm$0.050 & 0.708$\pm$0.021 & $-$0.121$\pm$0.062 \\
        J2217$+$0029 & 22:17:39.23 & $+$00:29:04.4 & 1.643 & 2003-08-22 & 06-17 & 450$\times$2 & 1.953$\pm$0.038 & 1.388$\pm$0.037 & 0.974$\pm$0.027 & $-$0.297$\pm$0.048 \\
        J2249$+$0047 & 22:49:24.01 & $+$00:47:50.4 & 1.360 & 2001-09-26 & 06-17 & 675$\times$4 & 2.419$\pm$0.090 & 1.249$\pm$0.072 & 1.202$\pm$0.043 & $-$0.344$\pm$0.110 \\
        J2228$-$0032 & 22:28:36.23 & $-$00:32:02.9 & 1.032 & 2001-08-22 & 07-10 & 900$\times$2 & 1.800$\pm$0.033 & 1.188$\pm$0.029 & 0.871$\pm$0.039 & $-$0.216$\pm$0.039 \\
        J2343$+$0038 & 23:43:07.38 & $+$00:38:54.7 & 0.667 & 2002-09-08 & 07-14 & 450$\times$1 & 1.841$\pm$0.042 & 1.104$\pm$0.047 & 0.840$\pm$0.037 & 0.000$\pm$0.056 \\
        J2328$-$0053 & 23:28:38.04 & $-$00:53:13.6 & 1.551 & 2001-10-17 & 07-10 & 675$\times$4 & 1.971$\pm$0.048 & 1.020$\pm$0.034 & 0.623$\pm$0.019 & $-$0.859$\pm$0.054 \\
        J2338$-$0101 & 23:38:53.44 & $-$01:01:19.4 & 1.483 & 2002-09-08 & 07-12 & 900$\times$2 & 2.024$\pm$0.056 & 0.949$\pm$0.052 & 0.635$\pm$0.025 & $-$0.682$\pm$0.069 \\
        J2335$-$0049 & 23:35:17.82 & $-$00:49:27.0 & 0.671 & 2001-10-17 & 07-12 & 450$\times$2 & 1.753$\pm$0.082 & 0.949$\pm$0.032 & 0.459$\pm$0.096 & $-$0.319$\pm$0.081 \\
        J0140$-$0035 & 01:40:48.62 & $-$00:35:00.9 & 1.383 & 2003-01-05 & 09-05 & 675$\times$4 & 2.255$\pm$0.068 & 0.863$\pm$0.045 & 0.115$\pm$0.014 & $-$1.067$\pm$0.076 \\
        J2141$-$0016 & 21:41:30.61 & $-$00:16:49.1 & 1.282 & 2004-08-10 & 05-12 & 675$\times$4 & 1.768$\pm$0.039 & 0.622$\pm$0.083 & 1.074$\pm$0.025 & $-$1.035$\pm$0.088 \\
        J0048$-$0113 & 00:48:26.09 & $-$01:13:10.7 & 1.034 & 2003-09-01 & 07-09 & 675$\times$4 & 1.695$\pm$0.052 & 0.479$\pm$0.040 & 0.453$\pm$0.018 & $-$0.836$\pm$0.052 \\
        J2209$-$0055 & 22:09:08.24 & $-$00:55:58.8 & 0.528 & 2000-09-01 & 05-12 & 450$\times$2 & 1.966$\pm$0.075 & 0.251$\pm$0.017 & 1.065$\pm$0.007 & $-$1.715$\pm$0.076 \\
        J2350$+$0025 & 23:50:40.09 & $+$00:25:58.9 & 1.062 & 2002-09-06 & 07-14 & 900$\times$2 & 2.101$\pm$0.078 & 0.141$\pm$0.018 & $-$0.217$\pm$0.005 & $-$1.890$\pm$0.078 \\
        \hline
		\end{tabular}
		
		\begin{flushleft}
		The table is ranked by $\Delta g_{\rm phot}$, which is the $g$-band magnitude difference between imaging photometry closest to the Gemini and SDSS spectra epochs. $\Delta g_{\rm max}$ is the maximum variability in the $g$ band. $\Delta g_{\rm spec}$ is the spectrophotometry magnitude difference with the SDSS and Gemini spectra convolved with the SDSS $g$-band filter. $\Delta g_{\rm back}$ is the $g$-band magnitude difference between the latest DES imaging epoch and the faintest DES epoch.
		\end{flushleft}
		\end{minipage}}
     \end{table*}

 \begin{figure}
 \subfigure{
  \includegraphics[width=3.3in]{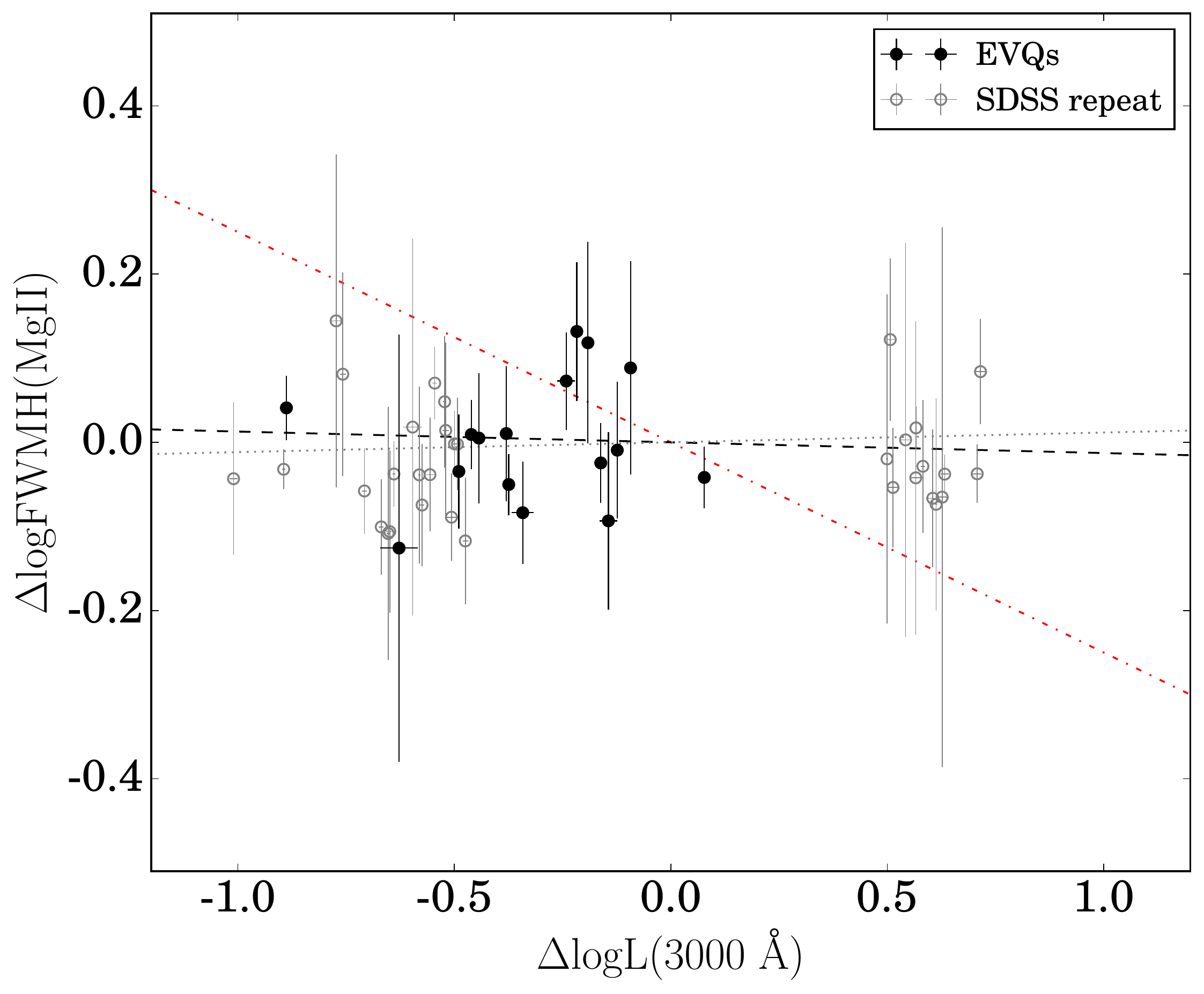}}
  \caption{Broad \MgII\ FWHM variability versus continuum variability. The red dotted line indicates the expected virial relation of $\Delta {\rm log W} = -0.25 \Delta {\rm log L}$. The FWHM of broad \MgII\ does not vary accordingly in response to changes in the continuum (see \S\ref{sec:MgIIwidth} for details).}  \label{fig:FWHM}
  \end{figure}

\subsection{\MgII\ Line Width} \label{sec:MgIIwidth}

In contrast to the flux, we found that the width (in terms of FWHM) of the broad \MgII\ line barely changes with luminosity for the bulk of the sample. For J2159$+$0005, which has the most dramatic change in continuum flux, the broad \MgII\ component has a FWHM of $9881 \pm 693$ km s$^{-1}$ for the bright state and a FWHM of $10858 \pm 579$ km s$^{-1}$ for the faint state, resulting in a change of FWHM of $977 \pm 903$ km s$^{-1}$. Assuming that the gas clouds emitting broad \MgII\ are virialized, the BH mass is 
\begin{equation}
    M_{\rm BH} = \frac{V^2 R}{G} = f \frac{W^2 R}{G},
\end{equation}
where $V$ is the virial velocity, $R$ is the BLR radius, $f$ is a scale factor that accounts for the orientation, kinematics, and structure of the BLR, and $W$ is the width of the broad line assuming that the broad line is Doppler broadened by virial motion. If the BLR size increases as continuum luminosity increases, as observed for broad \hbeta\ as ${R \propto L^{0.5}}$ as found in local RM studies or based on the simple photoionization argument, then a relation between the EW and luminosity of the form $\Delta {\rm log W} = -0.25 \Delta {\rm log L}$ is expected. Figure \ref{fig:FWHM} shows the changes of broad \MgII\ FWHM between the bright and faint epochs, $\Delta {\rm log FWHM (MgII)}$, as a function of $\Delta {\rm log L(3000\AA)}$, the corresponding changes of the continuum luminosity. Our results show that the broad \MgII\ width does not vary as expected from the simple virial relation above, in sharp contrast to the behaviour of broad \hbeta\ \citep[e.g.,][]{Park2012,Shen2013}. Fitting a linear regression model to our main EVQ sample, we obtain \\
\begin{equation}
    \Delta {\rm log FWHM (MgII)} = (-0.013 \pm 0.030) \Delta {\rm log L(3000 \AA)}.
\end{equation}
For the combined EVQ sample we obtain 
\begin{equation}
    \Delta {\rm log FWHM (MgII)} = (0.012 \pm 0.012) \Delta {\rm log L(3000 \AA)}.
\end{equation}
In both cases the broad \MgII\ FWHM remains more or less constant despite the large changes in the continuum luminosity, which is consistent with the findings in \cite{Shen2013}. 

To eliminate a possibility that variable fitting of \FeII\ emission compromises the line width variation of \MgII, we do a test by fixing the \FeII\ parameters for the two epochs of the same object. We get \\
\begin{equation}
    \Delta {\rm log FWHM (MgII)} = (-0.016 \pm 0.030) \Delta {\rm log L(3000 \AA)}
\end{equation}
for the main EVQ sample, and \\
\begin{equation}
    \Delta {\rm log FWHM (MgII)} = (0.019 \pm 0.012) \Delta {\rm log L(3000 \AA)}
\end{equation}
for the combined sample. The result shows that the line width of \MgII\ barely changes, irrespective of our approach.

We discuss the potential causes of the non-variable broad \MgII\ FWHM and its implications for virial BH mass estimation using \MgII\ in \S\ref{sec:discussion}.

\section{Discussion} \label{sec:discussion}
The coordinated variations in the continuum flux and in the broad \MgII\ flux suggest that at least part of the \MgII\ is photoionized by the continuum. This result offers some support to use the RM technique to infer the distances of the \MgII\ broad-line clouds. However, the associated \MgII\ variability is reduced compared to the level of continuum changes, indicating some part of the \MgII\ line might not be photoionized, thus complicating the interpretation of the \MgII\ RM results. In addition, the weaker \MgII\ variability, as observed for a large sample of quasars in a dedicated RM program \citep{Sun2015}, makes it generally more difficult to detect a RM lag for the broad \MgII\ line.

On the other hand, for the bulk of EVQs in our sample, the FWHM of broad \MgII\ does not vary with luminosity as expected from the simple virial assumption, in contrast to the case of broad \hbeta. This result suggests that the single-epoch virial BH mass based on \MgII, which depends both on the continuum luminosity and single-epoch \MgII\ width, will have an additional scatter in individual objects when luminosity varies while \MgII\ width remains the same. To demonstrate this point, we estimate single-epoch \MgII-based black hole masses using the bright and faint-state spectra and the recipe from \citet[][{listed in Table \ref{tab:fitting}}]{McLure2004}. Apart from some quasars with small continuum changes (since some EVQs brightened again in our late GMOS spectroscopy) and some quasars with bad spectral quality, the black hole mass estimates in the bright and faint states are inconsistent (at $>1\sigma$ of measurement errors) for 8 of the EVQs in our main sample, including J2159$+$0005, J2252$+$0004, J2249$+$0047, J2228$-$0032, J2328$-$0053, J2335$-$0049, J2141$-$0016, J2209$-$0055. For example, the black hole mass estimates for J2159$+$0005 are $(1.48\pm0.17)\times 10^9 M_{\odot}$ in the bright state and $(5.03\pm1.18)\times10^8 M_{\odot}$ in the dim state. These two black hole masses are inconsistent at $>4 \sigma$ confidence level. Using the width of broad \MgII\ to estimate the black hole mass therefore introduces a luminosity-dependent bias \citep[see detailed discussion in][]{Shen2013} with the single-epoch mass technique that is most severe for quasars that undergo significant luminosity changes. For individual quasars, clearly the $R-L$ relation is different or even absent for broad \MgII, although there may still be a global $R-L$ relation for quasars over a broad range of luminosity. The latter case will justify the use of \MgII\ as an single-epoch mass estimator for the general population of quasars (albeit still suffering from the luminosity-dependent scatter in individual objects), and the existence of such a global $R-L$ relation for broad \MgII\ can be tested with upcoming RM results on large quasar samples \citep[e.g.,][]{Shen2015, Hoormann2019}.

The so-called ``breathing" effect of the BLR, an increase in ionizing continuum luminosity leads to an increase in BLR radius and hence a decrease in line width, is observed conclusively for \hbeta\
\citep{Korista2004, Cackett2006, Denney2009, Park2012, Barth2015, Runco2016}. Different from \hbeta\ as a recombination line, \MgII\ are resonance lines. The greater optical depth for \MgII, which results in a large number of scatterings before escape from the BLR, may cause the \MgII\ photons to be emitted over a larger radius than \hbeta\ \citep{Korista2004}. Thus the variation in ionizing continuum may be diluted at a larger radius, leading to a smaller amplitude of \MgII\ variability. The lack of a clear intrinsic $R-L$ relation for broad \MgII\ could mean that the broad \MgII\ emission is confined in a narrower range of radii than the broader distribution of gas that can potentially produce \hbeta\ emission. The transition layer between the H II zone and the H I zone, where most of the continuum photons between 13.6 and 55 eV are absorbed and a large portion of the line emission is produced, is geometrically thin \citep{Krolik1999}. This layer could emit a larger fraction of the total \MgII\ emission than it does for the total Balmer-line emission \citep[e.g.,][]{Kwan1981}.
Thus when the ionizing continuum changes, the flux of \MgII\ echos, but the radius of the gas emitting \MgII\ barely changes, leading to more or less constant broad \MgII\ width. Another possibility is that part of gas emitting \MgII\ is virialized and photoionized, while the other part is not. It has been shown that \MgII\ profile is complex \citep{Kovacevic2015, Jonic2016}, and can be affected by non-virial effects, such as outflows \citep{Tavares2013, Popovic2019}, especially for very broad \MgII\ (${\rm FWHM} > 6000\ {\rm km\ s^{-1}}$). Potential detailed shape change of \MgII\ profile can be analysed with dedicated \MgII\ RM data and used to test these scenarios.

\begin{table*}
	\centering
	\caption{\ Spectral Fitting to the Two-Epoch Spectra }
	\label{tab:fitting}
	\begin{tabular}{ccccccc} 
		\hline
		Name & MJD & logL(3000 ${\rm \AA}$) & logL(Mg II) & FWHM(Mg II) & logW$_{\lambda}$(Mg II) & ${\rm M_{BH}}$ \\
		 & (days) & (erg s$^{-1}$) & (erg s$^{-1}$) & (km s$^{-1}$) & (${\rm \AA}$) & ($10^8 M_{\odot}$)\\
		\hline
		J2159$+$0005 & 52173 & 45.09 $\pm$ 0.01 & 43.43 $\pm$ 0.02 & 9847 $\pm$ 590 & 1.76 $\pm$ 0.03 & 14.76 $\pm$ 1.74 \\
         & 58250 & 44.21 $\pm$ 0.01 & 43.09 $\pm$ 0.02 & 10826 $\pm$ 1334 & 2.33 $\pm$ 0.02 & 5.03 $\pm$ 1.18 \\
        J0140$+$0052 & 53315 & 45.12 $\pm$ 0.01 & 43.18 $\pm$ 0.03 & 4384 $\pm$ 914 & 1.49 $\pm$ 0.03 & 3.05 $\pm$ 1.25 \\
         & 58366 & 44.90 $\pm$ 0.00 & 43.35 $\pm$ 0.02 & 6104 $\pm$ 245 & 1.88 $\pm$ 0.02 & 4.33 $\pm$ 0.35 \\
        J2213$-$0037 & 52912 & 45.61 $\pm$ 0.04 & 43.91 $\pm$ 0.11 & 9633 $\pm$ 5678 & 1.77 $\pm$ 0.12 & 29.74 $\pm$ 31.98 \\
         & 58309 & 44.99 $\pm$ 0.02 & 43.55 $\pm$ 0.02 & 7630 $\pm$ 671 & 1.96 $\pm$ 0.03 & 7.63 $\pm$ 1.30 \\
        J2252$+$0004 & 52178 & 45.24 $\pm$ 0.01 & 43.37 $\pm$ 0.03 & 7918 $\pm$ 1256 & 1.55 $\pm$ 0.03 & 11.85 $\pm$ 3.76 \\
         & 58309 & 44.80 $\pm$ 0.00 & 43.25 $\pm$ 0.01 & 7979 $\pm$ 506 & 1.86 $\pm$ 0.01 & 6.38 $\pm$ 0.77 \\
        J2217$+$0029 & 52873 & 45.63 $\pm$ 0.01 & 43.86 $\pm$ 0.05 & 6500 $\pm$ 983 & 1.69 $\pm$ 0.05 & 13.92 $\pm$ 4.08 \\
         & 58286 & 45.25 $\pm$ 0.00 & 43.55 $\pm$ 0.04 & 6431 $\pm$ 710 & 1.71 $\pm$ 0.03 & 7.90 $\pm$ 1.77 \\
        J2249$+$0047 & 52178 & 45.09 $\pm$ 0.01 & 43.45 $\pm$ 0.04 & 12230 $\pm$ 992 & 1.79 $\pm$ 0.04 & 22.73 $\pm$ 3.59 \\
         & 58286 & 44.72 $\pm$ 0.00 & 43.39 $\pm$ 0.01 & 10863 $\pm$ 362 & 2.09 $\pm$ 0.01 & 10.51 $\pm$ 0.68 \\
        J2228$-$0032 & 52143 & 45.36 $\pm$ 0.00 & 43.69 $\pm$ 0.02 & 13727 $\pm$ 943 & 1.76 $\pm$ 0.02 & 42.08 $\pm$ 5.60 \\
         & 58309 & 44.90 $\pm$ 0.01 & 43.59 $\pm$ 0.01 & 14021 $\pm$ 775 & 2.11 $\pm$ 0.01 & 22.81 $\pm$ 2.74 \\
        J2343$+$0038 & 52525 & 44.38 $\pm$ 0.01 & 43.13 $\pm$ 0.08 & 4907 $\pm$ 433 & 2.20 $\pm$ 0.07 & 1.32 $\pm$ 0.21 \\
         & 58313 & 44.13 $\pm$ 0.01 & 42.72 $\pm$ 0.03 & 6009 $\pm$ 666 & 2.06 $\pm$ 0.05 & 1.40 $\pm$ 0.31 \\
        J2328$-$0053 & 52199 & 45.35 $\pm$ 0.02 & 43.74 $\pm$ 0.04 & 7468 $\pm$ 829 & 1.79 $\pm$ 0.05 & 12.28 $\pm$ 2.68 \\
         & 58309 & 45.01 $\pm$ 0.00 & 43.56 $\pm$ 0.01 & 6230 $\pm$ 466 & 1.97 $\pm$ 0.01 & 5.28 $\pm$ 0.82 \\
        J2338$-$0101 & 52525 & 45.38 $\pm$ 0.01 & 43.51 $\pm$ 0.04 & 5090 $\pm$ 1117 & 1.56 $\pm$ 0.05 & 5.96 $\pm$ 2.49 \\
         & 58311 & 45.19 $\pm$ 0.00 & 43.49 $\pm$ 0.03 & 6781 $\pm$ 929 & 1.73 $\pm$ 0.03 & 8.05 $\pm$ 2.09 \\
        J2335$-$0049 & 52199 & 44.53 $\pm$ 0.01 & 43.18 $\pm$ 0.05 & 6360 $\pm$ 919 & 2.10 $\pm$ 0.07 & 2.76 $\pm$ 0.81 \\
         & 58311 & 44.39 $\pm$ 0.02 & 42.85 $\pm$ 0.05 & 5079 $\pm$ 715 & 1.87 $\pm$ 0.05 & 1.43 $\pm$ 0.40 \\
        J0140$-$0035 & 52644 & 45.22 $\pm$ 0.01 & 43.51 $\pm$ 0.03 & 4004 $\pm$ 935 & 1.71 $\pm$ 0.04 & 2.91 $\pm$ 1.44 \\
         & 58366 & 45.12 $\pm$ 0.00 & 43.55 $\pm$ 0.04 & 4978 $\pm$ 1034 & 1.85 $\pm$ 0.04 & 3.94 $\pm$ 1.46 \\
        J2141$-$0016 & 53227 & 45.19 $\pm$ 0.01 & 43.62 $\pm$ 0.03 & 4844 $\pm$ 439 & 1.86 $\pm$ 0.03 & 4.08 $\pm$ 0.73 \\
         & 58250 & 45.02 $\pm$ 0.00 & 43.36 $\pm$ 0.01 & 4643 $\pm$ 267 & 1.76 $\pm$ 0.01 & 2.98 $\pm$ 0.36 \\
        J0048$-$0113 & 52883 & 44.86 $\pm$ 0.01 & 43.26 $\pm$ 0.07 & 5000 $\pm$ 821 & 1.83 $\pm$ 0.08 & 2.75 $\pm$ 0.87 \\
         & 58308 & 44.74 $\pm$ 0.00 & 43.18 $\pm$ 0.02 & 4671 $\pm$ 516 & 1.89 $\pm$ 0.02 & 2.01 $\pm$ 0.45 \\
        J2209$-$0055 & 51788 & 45.02 $\pm$ 0.00 & 43.37 $\pm$ 0.02 & 6184 $\pm$ 393 & 1.78 $\pm$ 0.02 & 5.24 $\pm$ 0.68 \\
         & 58250 & 44.53 $\pm$ 0.01 & 43.00 $\pm$ 0.03 & 5643 $\pm$ 787 & 1.89 $\pm$ 0.03 & 2.16 $\pm$ 0.64 \\
        J2350$+$0025 & 52523 & 45.31 $\pm$ 0.01 & 43.62 $\pm$ 0.02 & 4968 $\pm$ 478 & 1.73 $\pm$ 0.02 & 5.11 $\pm$ 0.98 \\
         & 58313 & 45.38 $\pm$ 0.00 & 43.69 $\pm$ 0.01 & 4521 $\pm$ 138 & 1.72 $\pm$ 0.01 & 4.71 $\pm$ 0.29 \\
		\hline
	\end{tabular}
	
\begin{flushleft}
M$_{\rm BH}$ is the black hole mass estimated based on the single-epoch recipe based on \MgII\ in \citet{McLure2004}. This table is ranked in the same order as Table \ref{tab:photometry}, from the most dimmed to the least dimmed quasars.
\end{flushleft}
\end{table*}

\section{Summary}\label{sec:summary}
We have presented Gemini/GMOS spectroscopy of 16 EVQs, with $>1.5$ mag maximum variability in the $g$ band from the combined DES and SDSS photometric light curves \citep[][]{Rumbaugh2018}, and studied their spectral changes between the GMOS spectra and the earlier SDSS spectra. 

Our main results are:

(1) About half of the EVQs brightened again a few years since reaching the minimum flux. This is consistent with the conclusions in \citet{Rumbaugh2018} that extreme variability events are common among the general population of quasars on multi-year timescales, as well as the findings in the recent spectroscopic work by \citet{MacLeod2019}. The extreme variability can also occur, albeit much less frequently, on much shorter timescales of $<1$ year, consistent with the findings in previous studies \citep{Gezari2017,Yang2018,Trakhtenbrot2019}.\\
(2) We identified two quasars that apparently changed their spectroscopic appearance. J2159$+$0005 is a possible CLQ. If confirmed, it would be the most distant CLQ so far discovered at $z=0.936$. J2343$+$0038 is a CLQ, at $z=0.667$, based on the disappearance of of \hbeta.\\
(3) The \MgII\ broad line flux varies accordingly as the continuum changes, albeit with a smaller amplitude than that for the continuum, $\Delta {\rm log L(Mg II)} = (0.39 \pm 0.07) \times \Delta {\rm log L(3000\AA)}$. This suggests that at least part of the broad \MgII\ emission is photoionized, or the continuum flux that ionizes broad \MgII\ varies less than the continuum at rest-frame 3000\AA. \\
(4) The \MgII\ broad line FWHM remains roughly constant for the bulk of the objects, even though the continuum luminosity changed significantly, in contrast to the properties of broad \hbeta. We discussed potential causes of the different kinematic properties of the broad \MgII\ line, but to confirm these speculations require future dedicated \MgII\ RM data. Nevertheless, the different variability properties of broad \MgII\ compared to broad \hbeta\ provide some cautionary notes on the use of \MgII\ as a single-epoch BH mass estimator. In particular, the absence of a clear intrinsic $R-L$ relation for \MgII\ will inevitably lead to a luminosity-dependent bias in the BH mass estimates in individual systems as the continuum luminosity varies significantly. 

\section*{Acknowledgements}
QY and YS acknowledge support from an Alfred P. Sloan Research Fellowship (YS) and NSF grant AST-1715579. We thank Patrick Hall, Tamara Davis, Shu Wang, and Hengxiao Guo for useful discussions and suggestions. 

Funding for the DES Projects has been provided by the U.S. Department of Energy, the U.S. National Science Foundation, the Ministry of Science and Education of Spain, the Science and Technology Facilities Council of the United Kingdom, the Higher Education Funding Council for England, the National Center for Supercomputing Applications at the University of Illinois at Urbana-Champaign, the Kavli Institute of Cosmological Physics at the University of Chicago, the Center for Cosmology and Astro-Particle Physics at the Ohio State University, the Mitchell Institute for Fundamental Physics and Astronomy at Texas A\&M University, Financiadora de Estudos e Projetos, Funda{\c c}{\~a}o Carlos Chagas Filho de Amparo {\`a} Pesquisa do Estado do Rio de Janeiro, Conselho Nacional de Desenvolvimento Cient{\'i}fico e Tecnol{\'o}gico and 
the Minist{\'e}rio da Ci{\^e}ncia, Tecnologia e Inova{\c c}{\~a}o, the Deutsche Forschungsgemeinschaft and the Collaborating Institutions in the Dark Energy Survey. 

The Collaborating Institutions are Argonne National Laboratory, the University of California at Santa Cruz, the University of Cambridge, Centro de Investigaciones Energ{\'e}ticas, Medioambientales y Tecnol{\'o}gicas-Madrid, the University of Chicago, University College London, the DES-Brazil Consortium, the University of Edinburgh, the Eidgen{\"o}ssische Technische Hochschule (ETH) Z{\"u}rich, Fermi National Accelerator Laboratory, the University of Illinois at Urbana-Champaign, the Institut de Ci{\`e}ncies de l'Espai (IEEC/CSIC), the Institut de F{\'i}sica d'Altes Energies, Lawrence Berkeley National Laboratory, the Ludwig-Maximilians Universit{\"a}t M{\"u}nchen and the associated Excellence Cluster Universe, 
the University of Michigan, the National Optical Astronomy Observatory, the University of Nottingham, The Ohio State University, the University of Pennsylvania, the University of Portsmouth, SLAC National Accelerator Laboratory, Stanford University, the University of Sussex, Texas A\&M University, and the OzDES Membership Consortium.

Based in part on observations at Cerro Tololo Inter-American Observatory, National Optical Astronomy Observatory, which is operated by the Association of Universities for Research in Astronomy (AURA) under a cooperative agreement with the National Science Foundation.

The DES data management system is supported by the National Science Foundation under Grant Numbers AST-1138766 and AST-1536171.
The DES participants from Spanish institutions are partially supported by MINECO under grants AYA2015-71825, ESP2015-66861, FPA2015-68048, SEV-2016-0588, SEV-2016-0597, and MDM-2015-0509, some of which include ERDF funds from the European Union. IFAE is partially funded by the CERCA program of the Generalitat de Catalunya. Research leading to these results has received funding from the European Research Council under the European Union's Seventh Framework Program (FP7/2007-2013) including ERC grant agreements 240672, 291329, and 306478. We acknowledge support from the Brazilian Instituto Nacional de Ci\^enciae Tecnologia (INCT) e-Universe (CNPq grant 465376/2014-2).

This manuscript has been authored by Fermi Research Alliance, LLC under Contract No. DE-AC02-07CH11359 with the U.S. Department of Energy, Office of Science, Office of High Energy Physics. The United States Government retains and the publisher, by accepting the article for publication, acknowledges that the United States Government retains a non-exclusive, paid-up, irrevocable, world-wide license to publish or reproduce the published form of this manuscript, or allow others to do so, for United States Government purposes.

Funding for SDSS-III has been provided by the Alfred P. Sloan Foundation, the Participating Institutions, the National Science Foundation, and the U.S. Department of Energy Office of Science. The SDSS-III website is \url{http://www.sdss3.org/}. SDSS-III is managed by the Astrophysical Research Consortium for the Participating Institutions of the SDSS-III Collaboration including the University of Arizona, the Brazilian Participation Group, Brookhaven National Laboratory, Carnegie Mellon University, University of Florida, the French Participation Group, the German Participation Group, Harvard University, the Instituto de Astrofisica de Canarias, the Michigan State/Notre Dame/JINA Participation Group, Johns Hopkins University, Lawrence Berkeley National Laboratory, Max Planck Institute for Astrophysics, Max Planck Institute for Extraterrestrial Physics, New Mexico State University, New York University, Ohio State University, Pennsylvania State University, University of Portsmouth, Princeton University, the Spanish Participation Group, University of Tokyo, University of Utah, Vanderbilt University, University of Virginia, University of Washington, and Yale University.

The PS1 has been made possible through contributions by the Institute for Astronomy, the University of Hawaii, the Pan-STARRS Project Office, the Max-Planck Society and its participating institutes, the Max Planck Institute for Astronomy, Heidelberg and the Max Planck Institute for Extraterrestrial Physics, Garching, The Johns Hopkins University, Durham University, the University of Edinburgh, Queen's University Belfast, the Harvard-Smithsonian Center for Astrophysics, the Las Cumbres Observatory Global Telescope Network Incorporated, the National Central University of Taiwan, the Space Telescope Science Institute, the National Aeronautics and Space Administration under Grant No. NNX08AR22G issued through the Planetary Science Division of the NASA Science Mission Directorate, the National Science Foundation under Grant No. AST-1238877, the University of Maryland, and Eotvos Lorand University (ELTE). 



\appendix
\renewcommand{\thefigure}{A.\arabic{figure}}

The light curves and spectra of the remaining 14 EVQs, in addition to the two shown in Figure \ref{fig:evq_example}, are provided in Figure \ref{fig:cl_others}, in the same order as that in Table  \ref{tab:photometry} in terms of the level of dimming. 

\begin{figure*}
  \centering
  \hspace{0cm}
  \subfigure{
   \hspace{-1.0cm}
  \includegraphics[width=3.8in]{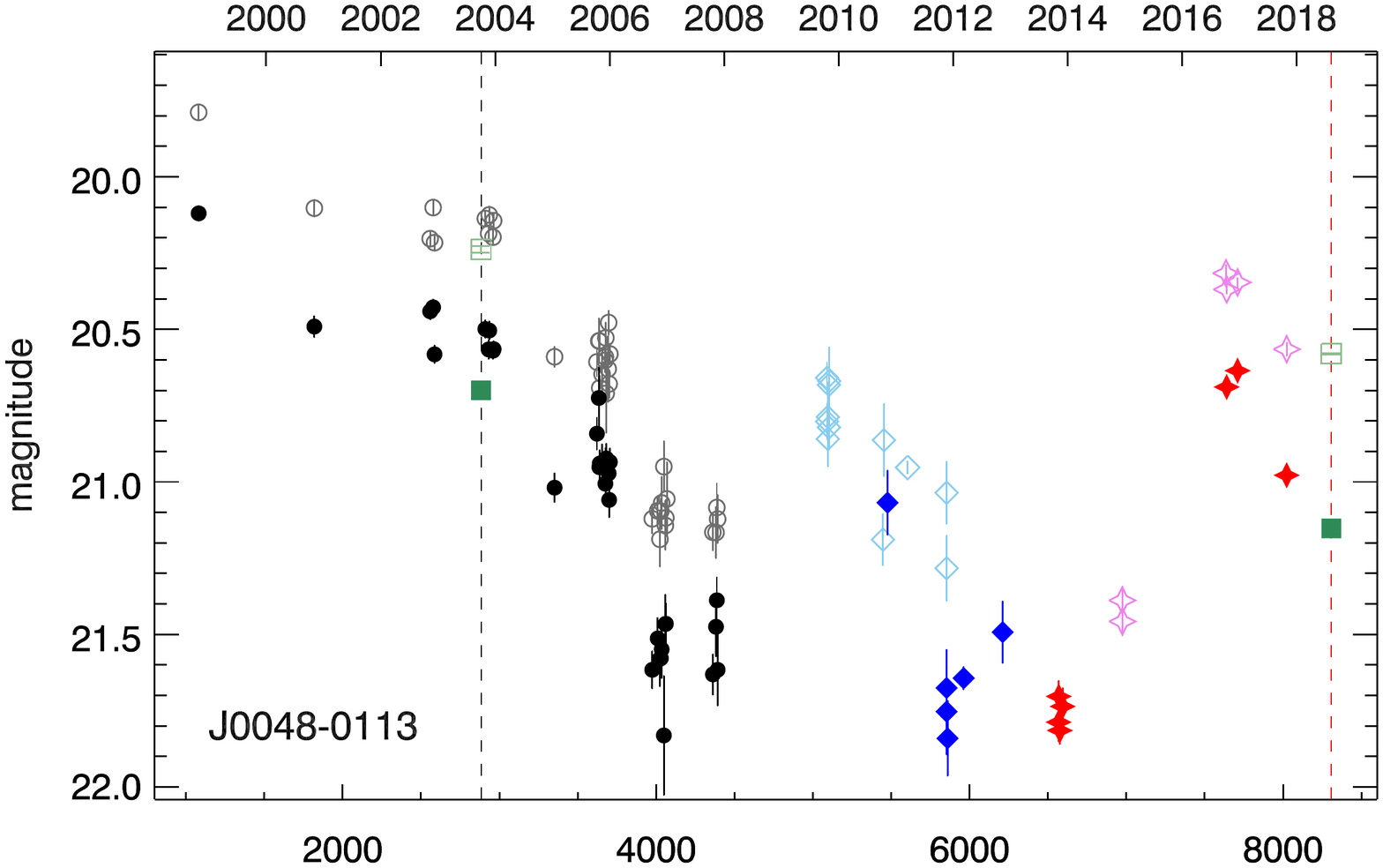}}
 \hspace{-1.4cm}
 \subfigure{
  \includegraphics[width=3.8in]{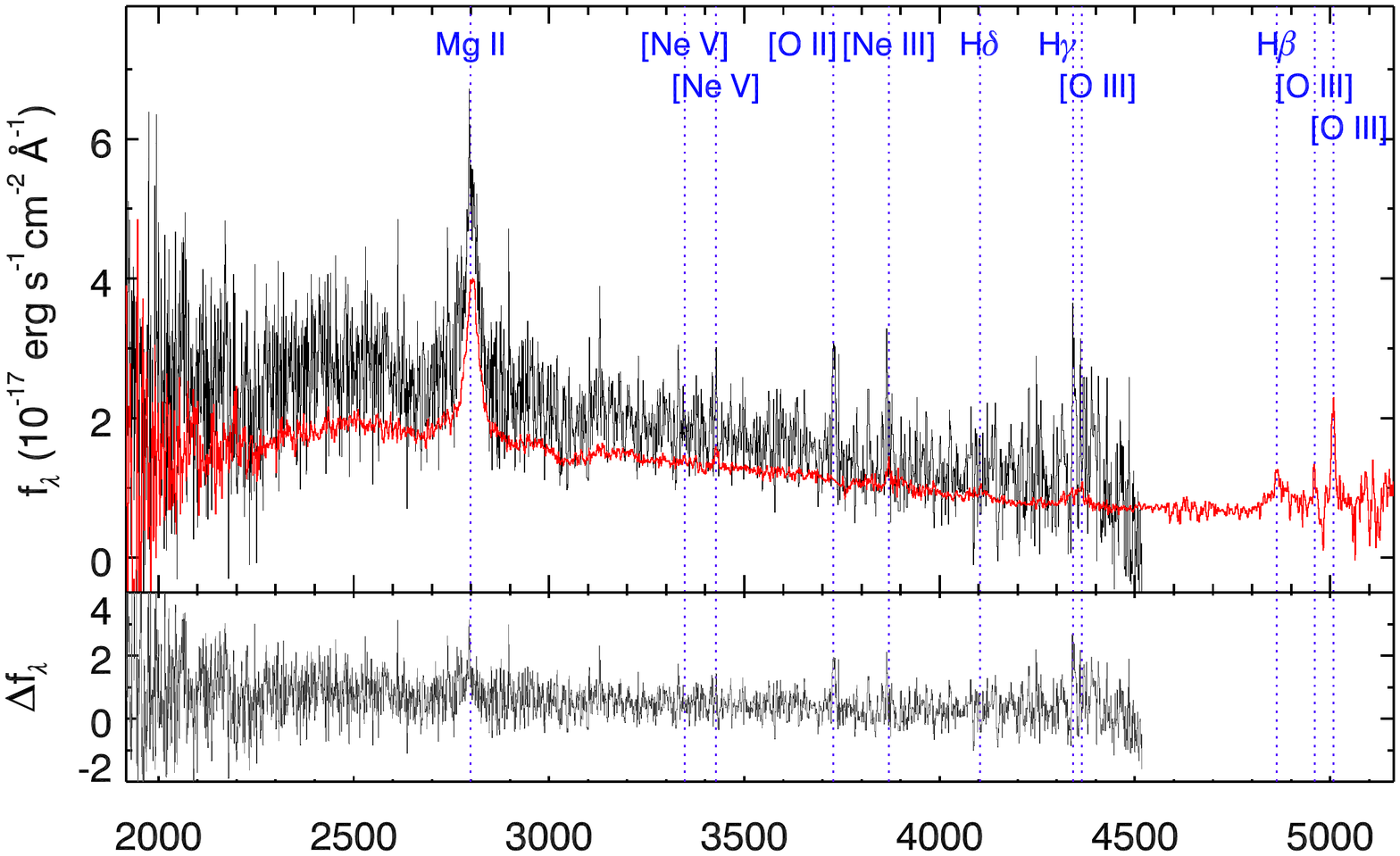}}\\
 \vspace{-2.5cm} 

 \centering
 \hspace{0cm}
 \subfigure{
  \hspace{-1.0cm}
  \includegraphics[width=3.8in]{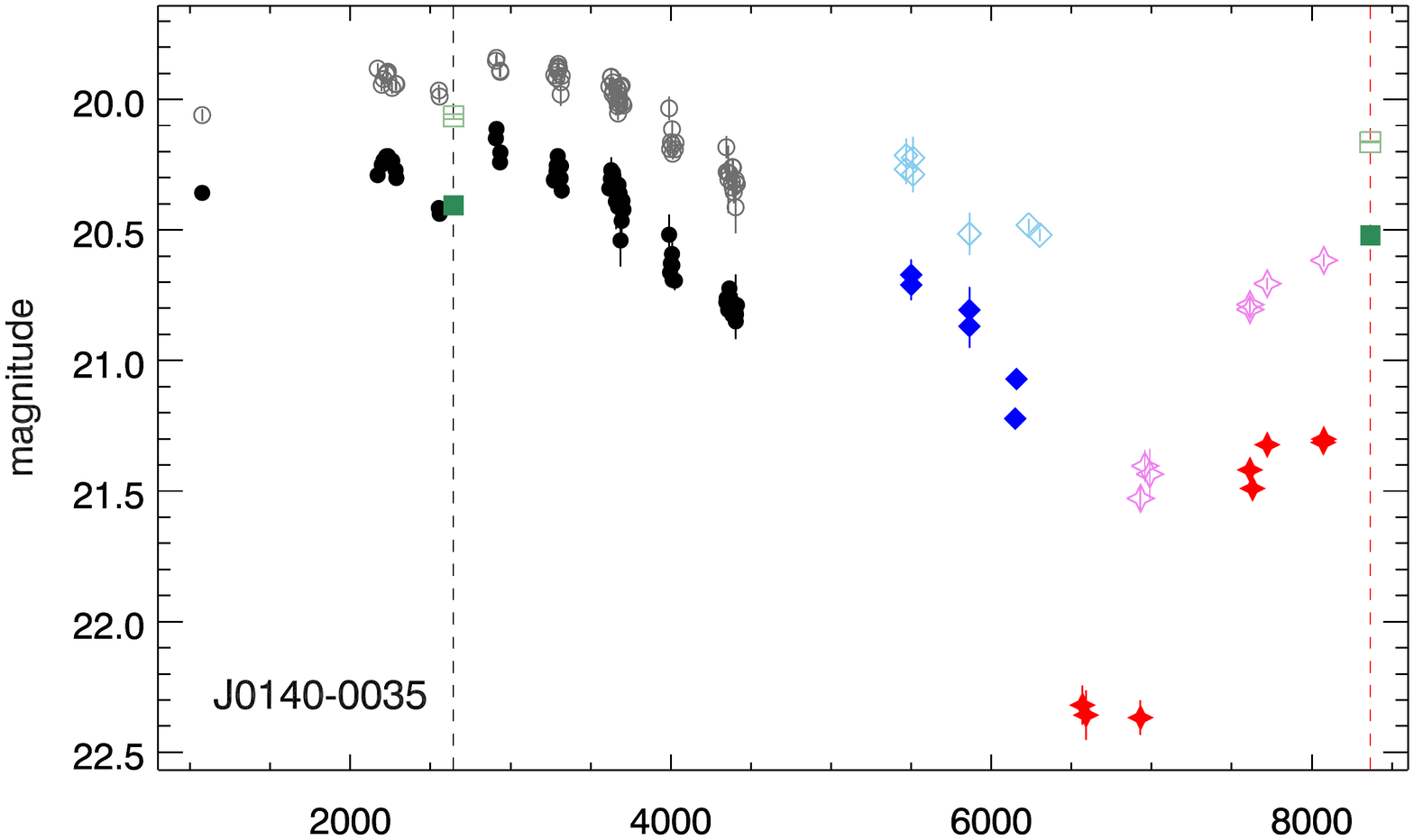}}
 \hspace{-1.4cm}
 \subfigure{
  \includegraphics[width=3.8in]{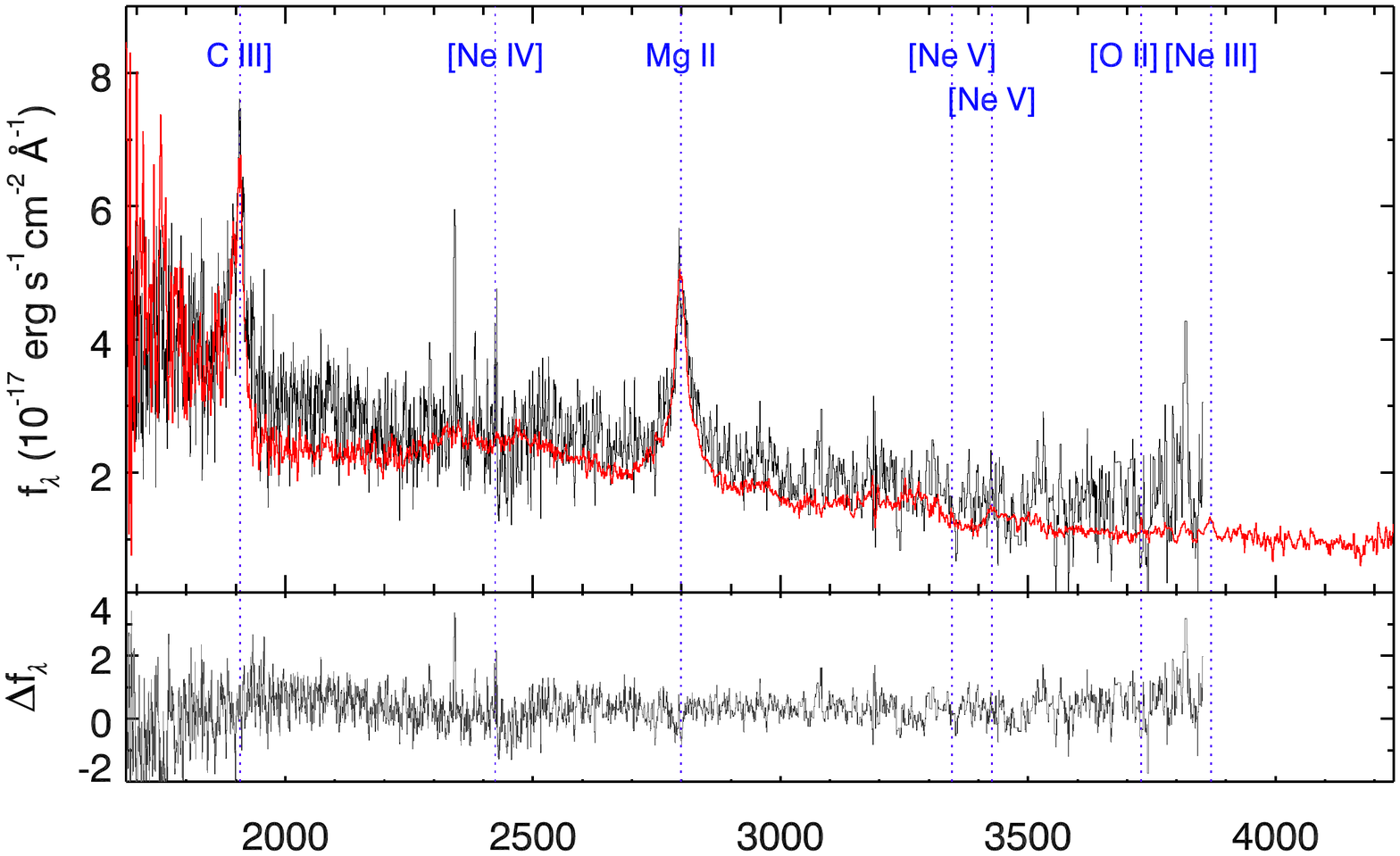}}\\
 \vspace{-2.5cm}

  \centering
  \hspace{0cm}
  \subfigure{
   \hspace{-1.0cm}
  \includegraphics[width=3.8in]{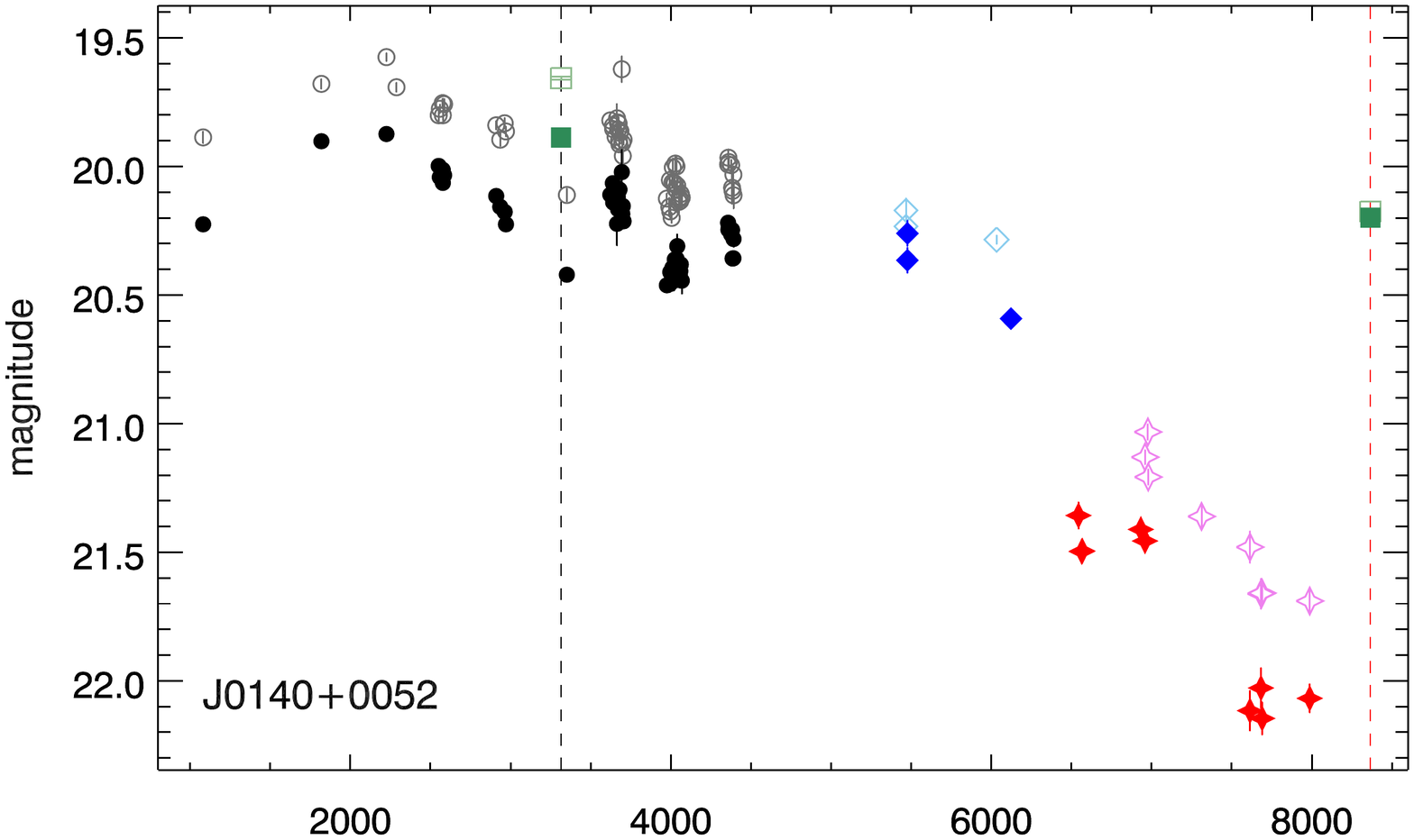}}
 \hspace{-1.4cm}
 \subfigure{
  \includegraphics[width=3.8in]{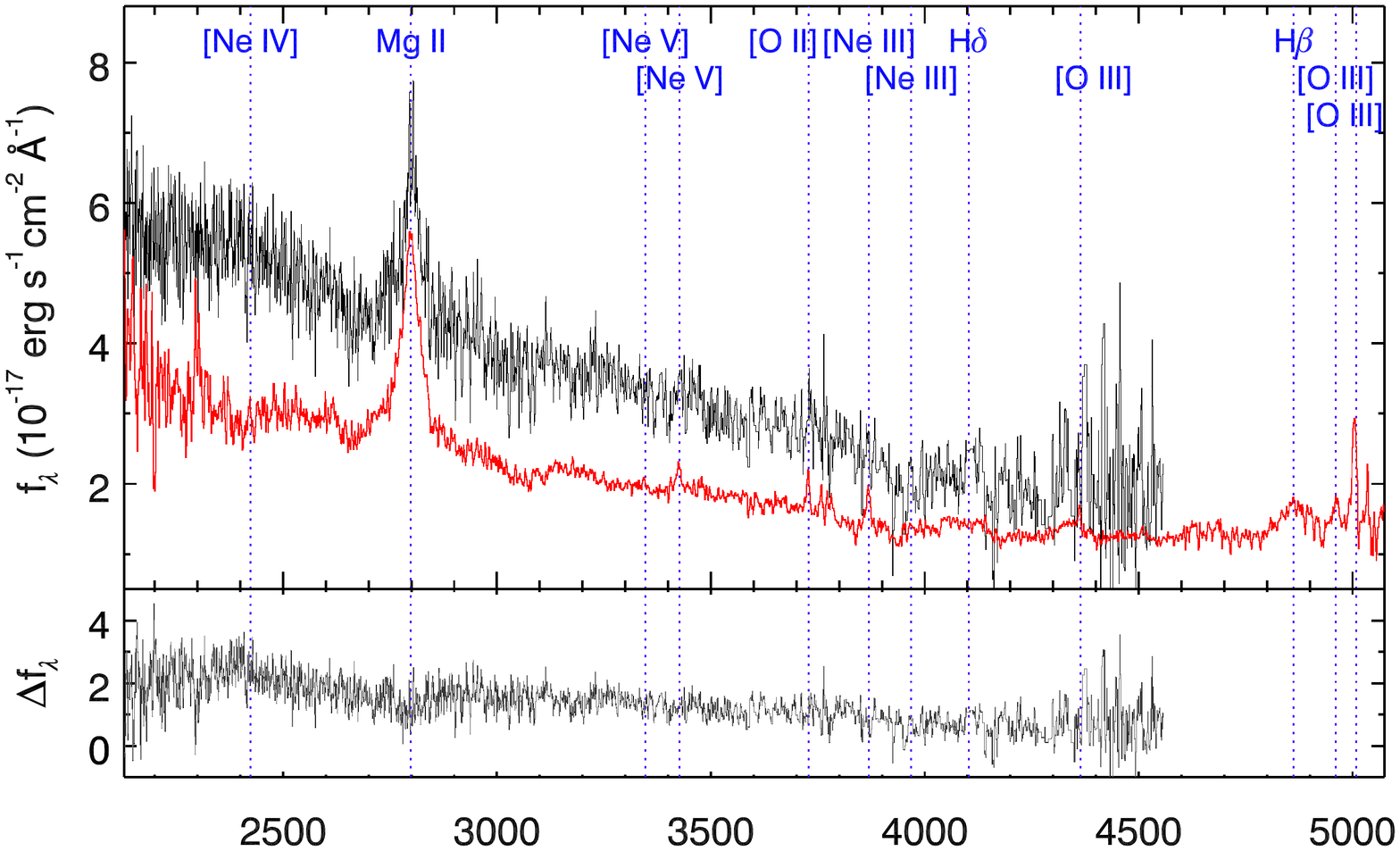}}\\
 \vspace{-2.5cm}
 
  \centering
 \hspace{0cm}
  \subfigure{
   \hspace{-1.0cm}
  \includegraphics[width=3.8in]{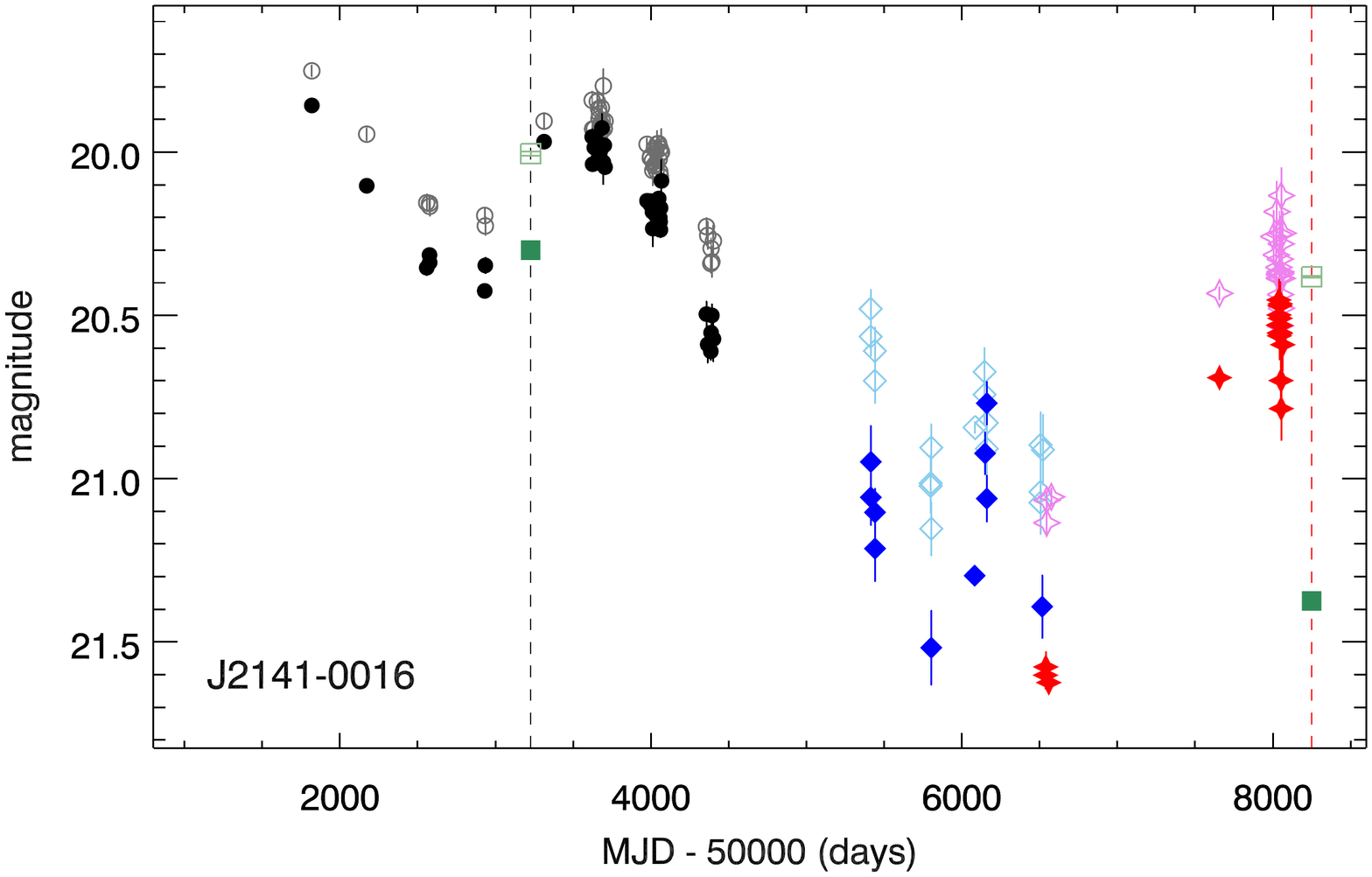}}
 \hspace{-1.4cm}
 \subfigure{
  \includegraphics[width=3.8in]{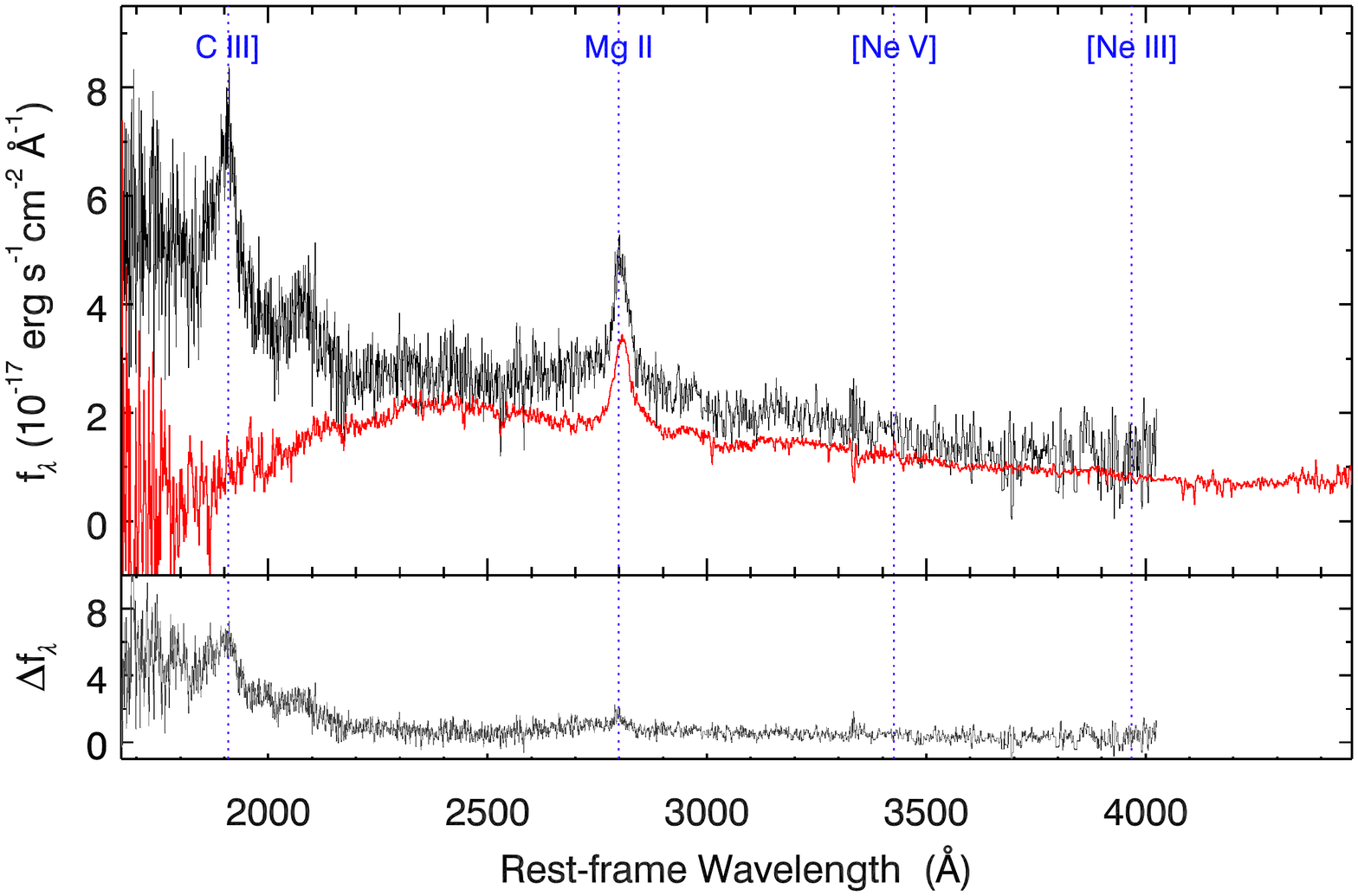}}\\
 \vspace{-1.0cm}
 
 \setcounter{figure}{0}
  \caption{\label{fig:cl_others} The rest of the EVQ sample observed with GMOS. Notations are the same as Figure \ref{fig:evq_example}. }
\end{figure*}

\renewcommand{\thefigure}{A.\arabic{figure}}
\addtocounter{figure}{-1}

\begin{figure*}

  \centering
  \hspace{0cm}
  \subfigure{
   \hspace{-1.0cm}
  \includegraphics[width=3.8in]{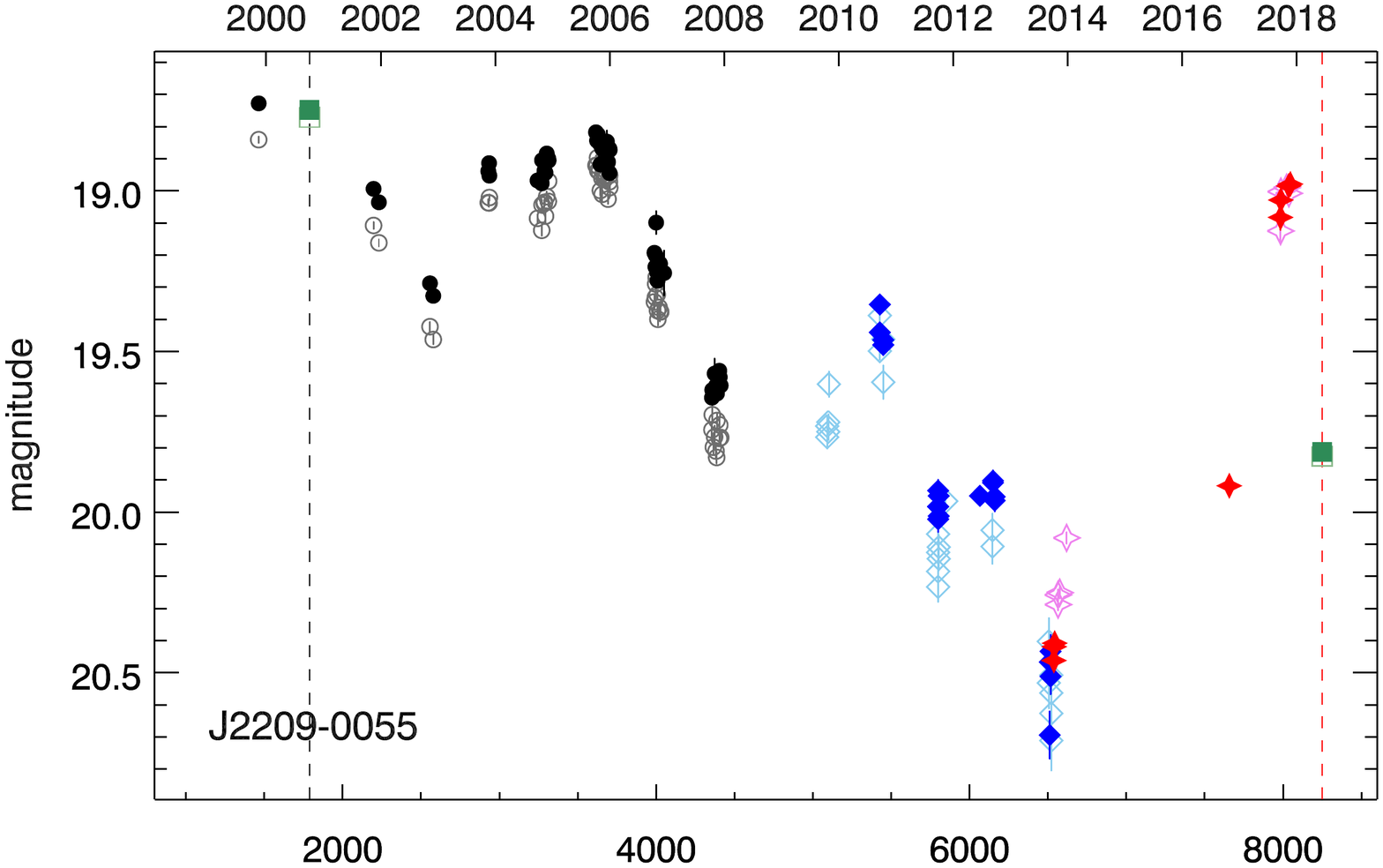}}
 \hspace{-1.4cm}
 \subfigure{
  \includegraphics[width=3.8in]{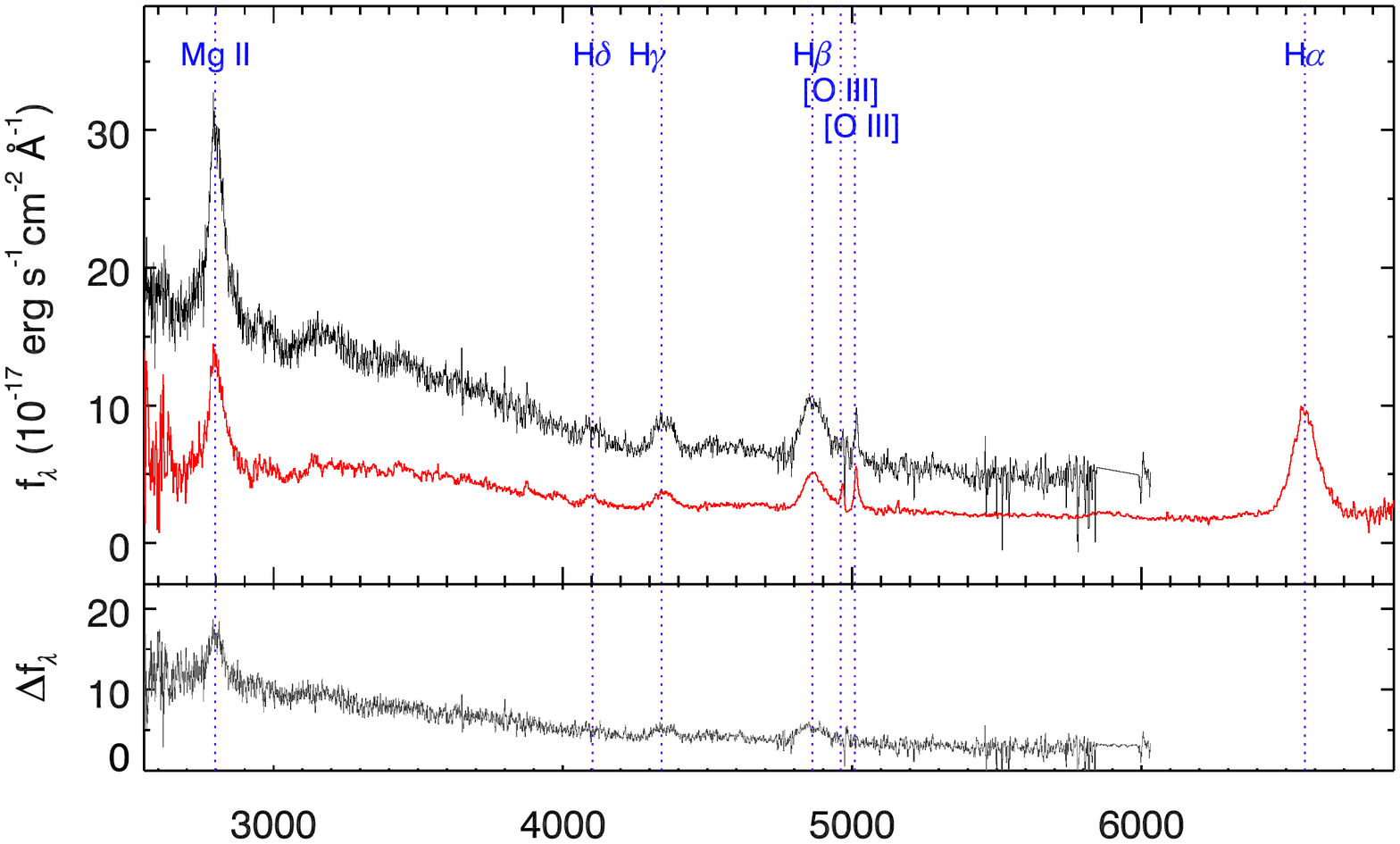}}\\
 \vspace{-2.5cm}

 \centering
 \hspace{0cm}
 \subfigure{
  \hspace{-1.0cm}
  \includegraphics[width=3.8in]{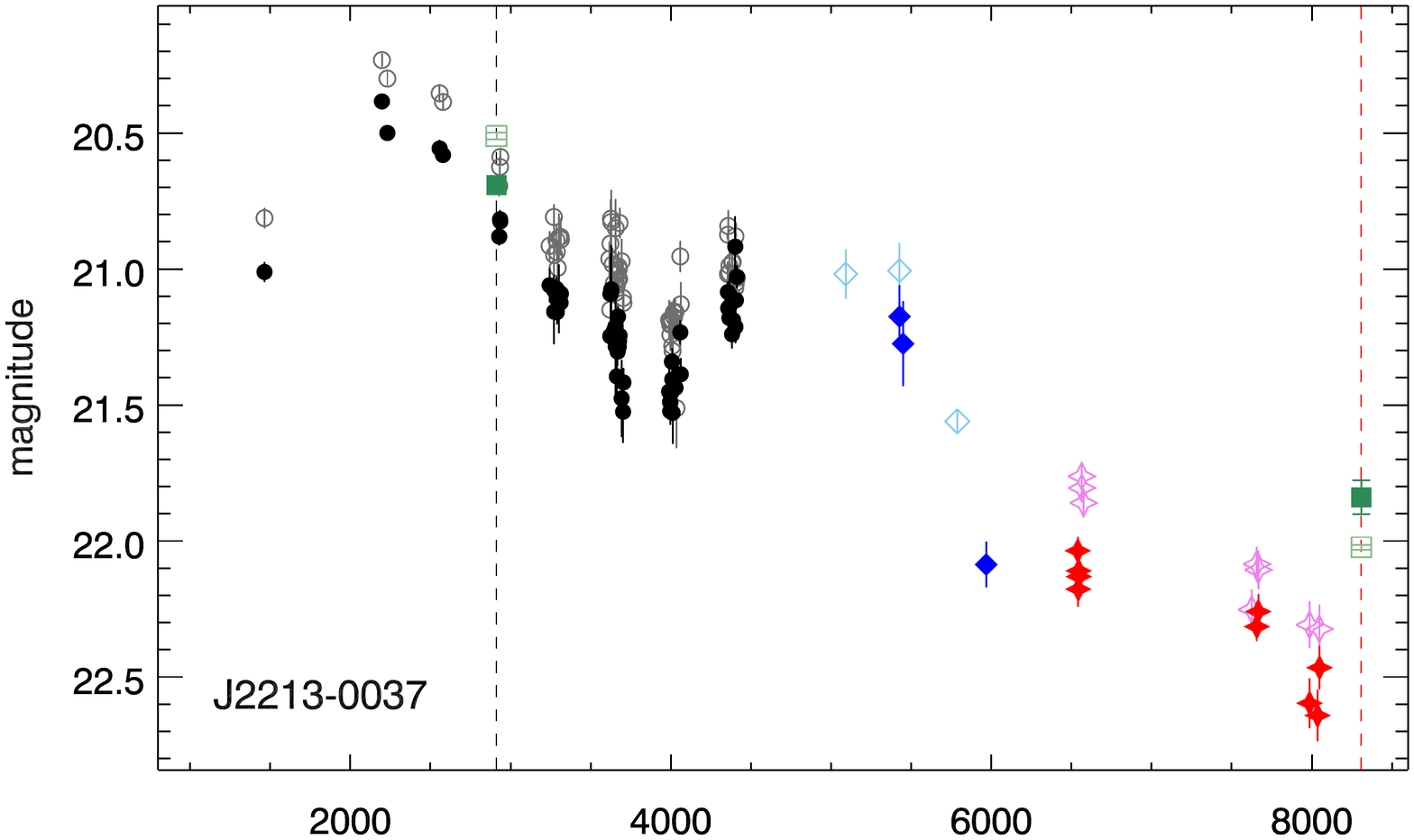}}
 \hspace{-1.4cm}
 \subfigure{
  \includegraphics[width=3.8in]{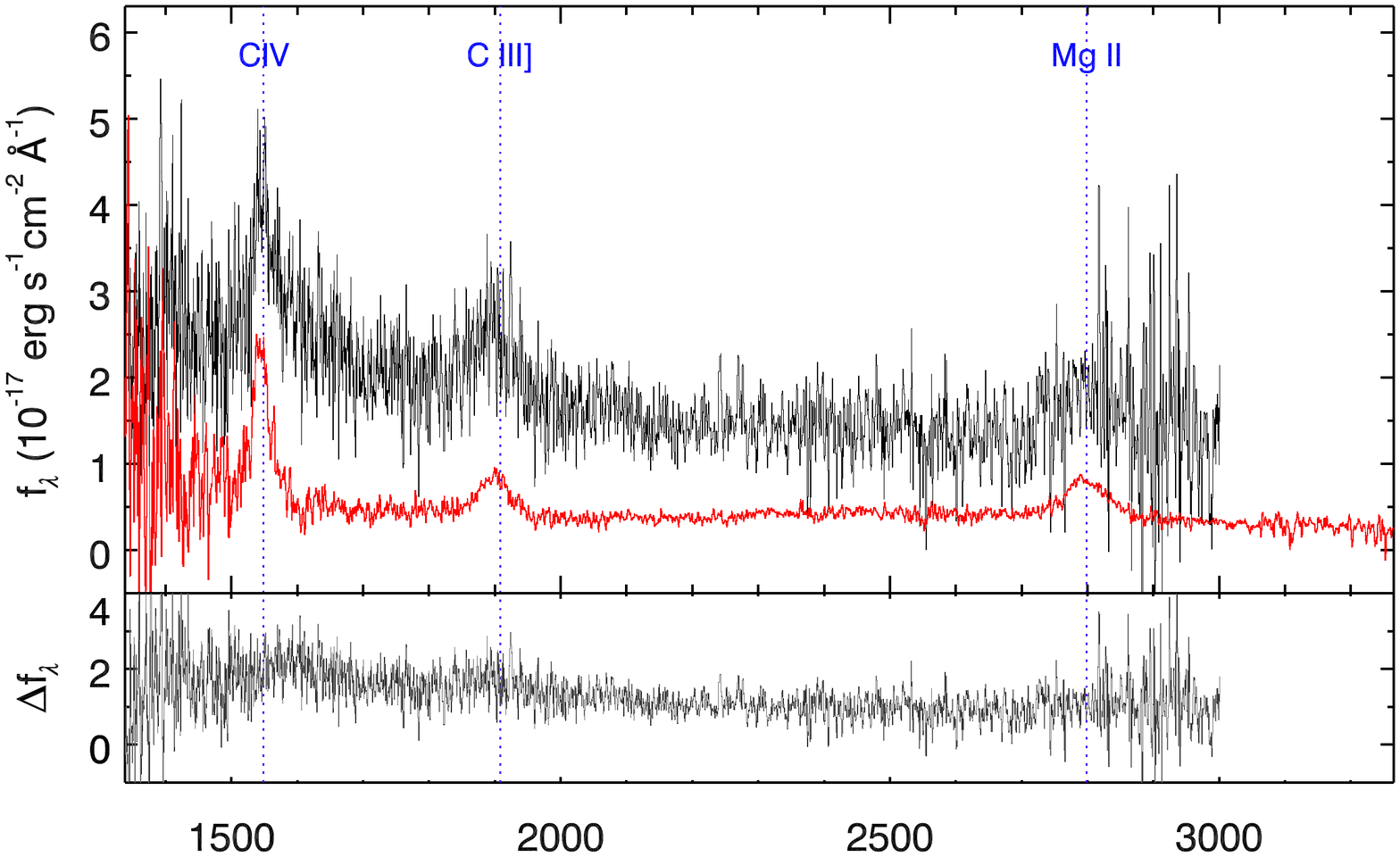}}\\
 \vspace{-2.5cm}
 
  \centering
 \hspace{0cm}
  \subfigure{
   \hspace{-1.0cm}
  \includegraphics[width=3.8in]{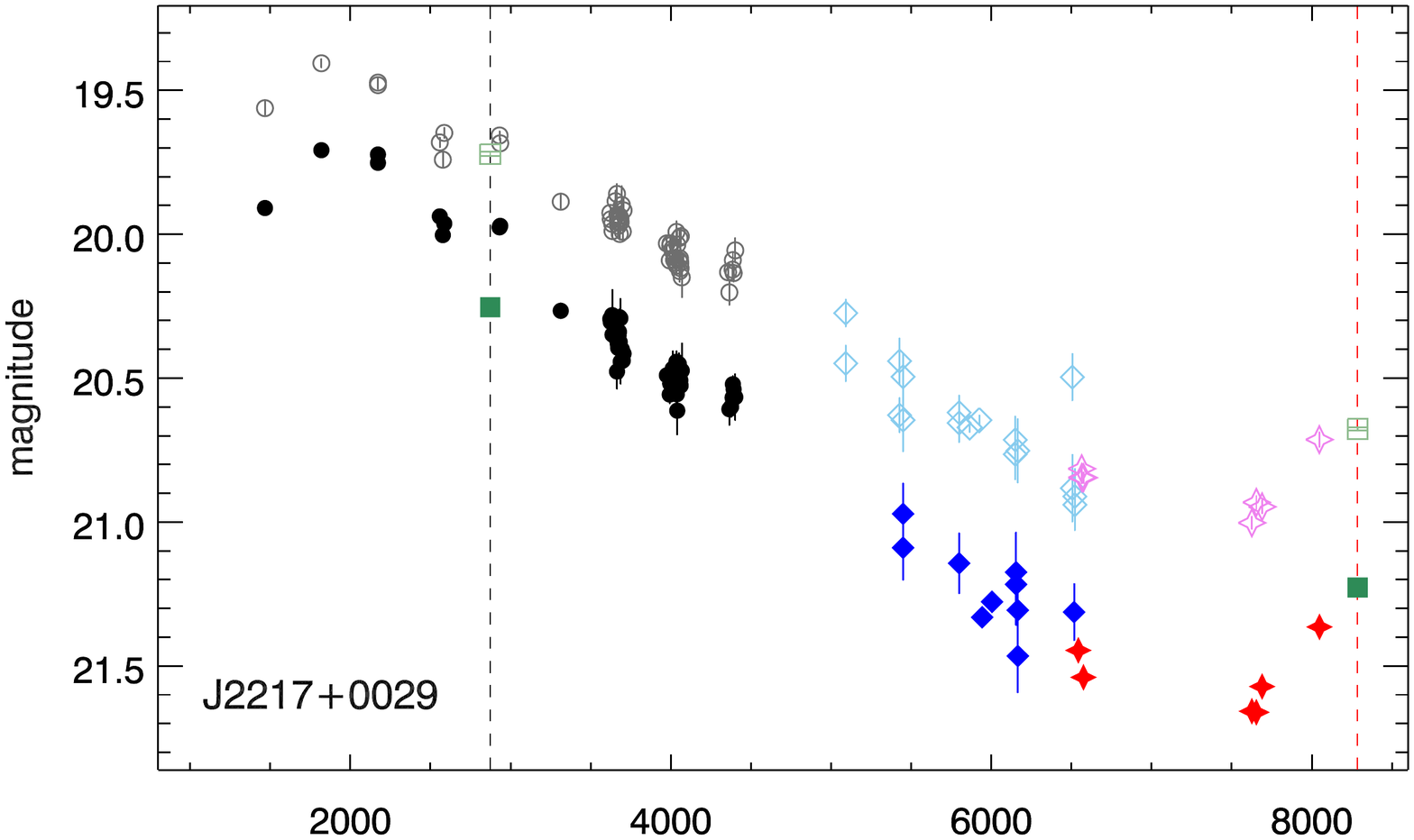}}
 \hspace{-1.4cm}
 \subfigure{
  \includegraphics[width=3.8in]{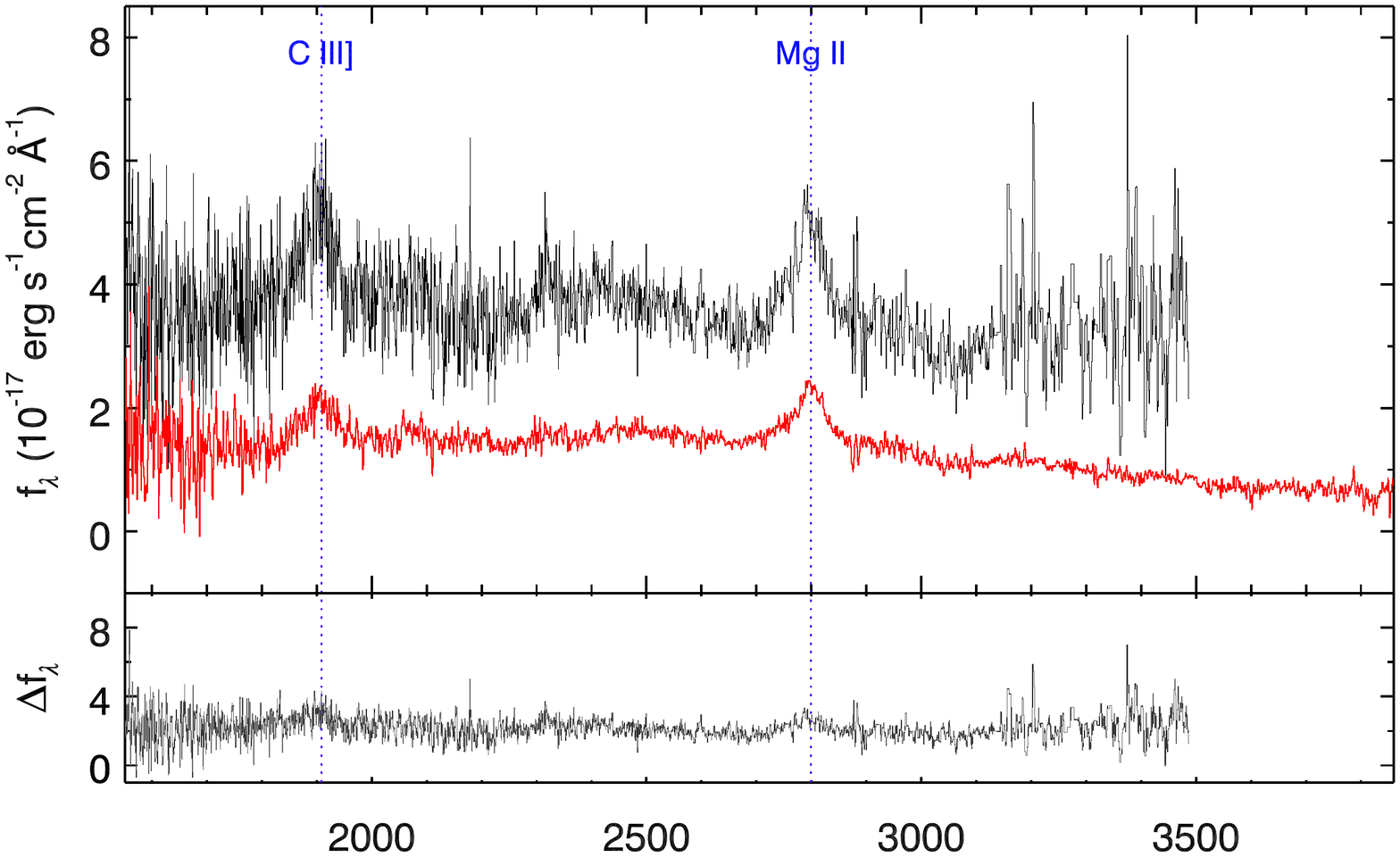}}\\
 \vspace{-2.5cm}
 
   \centering
  \hspace{0cm}
  \subfigure{
   \hspace{-1.0cm}
  \includegraphics[width=3.8in]{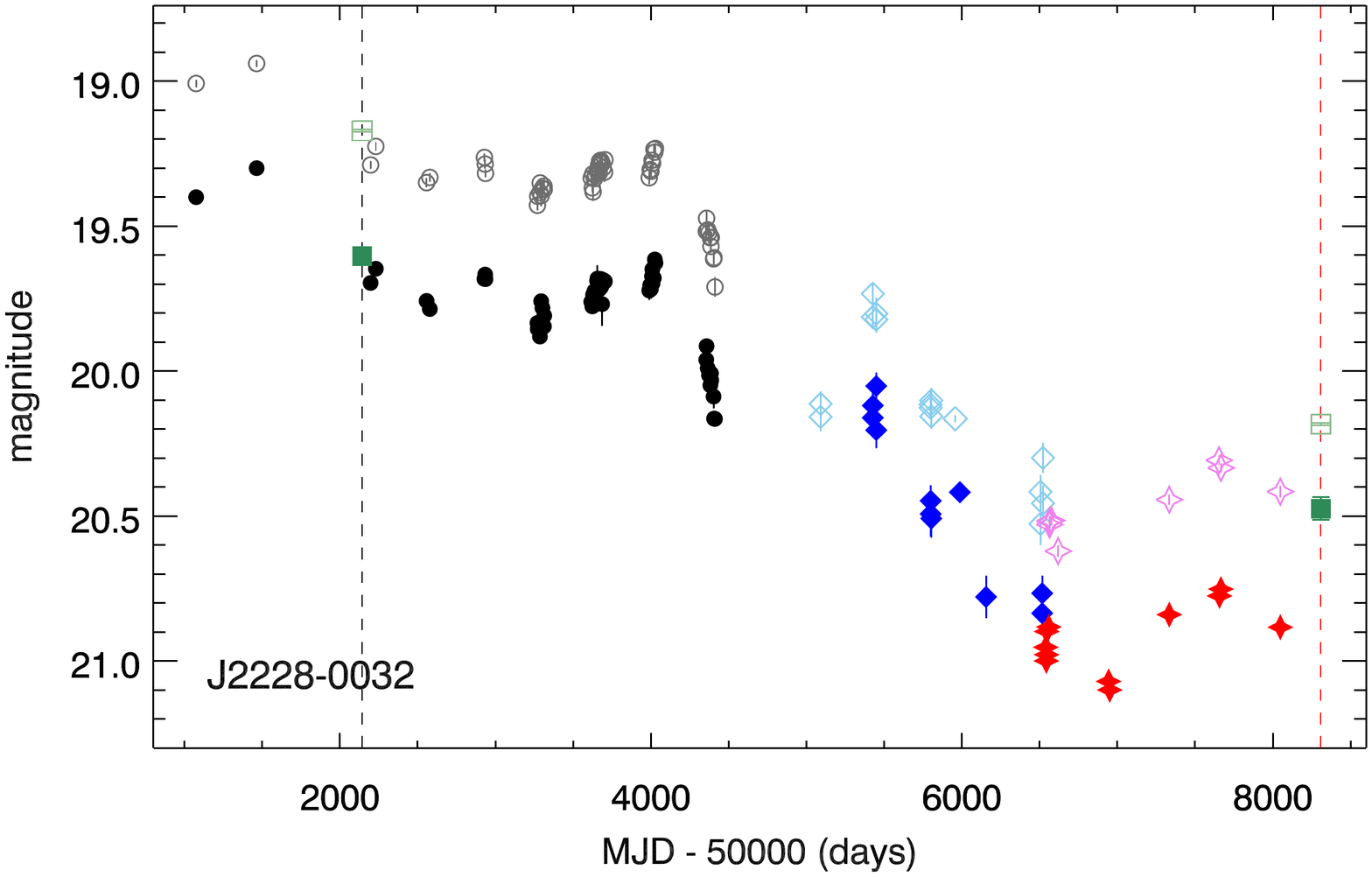}}
 \hspace{-1.4cm}
 \subfigure{
  \includegraphics[width=3.8in]{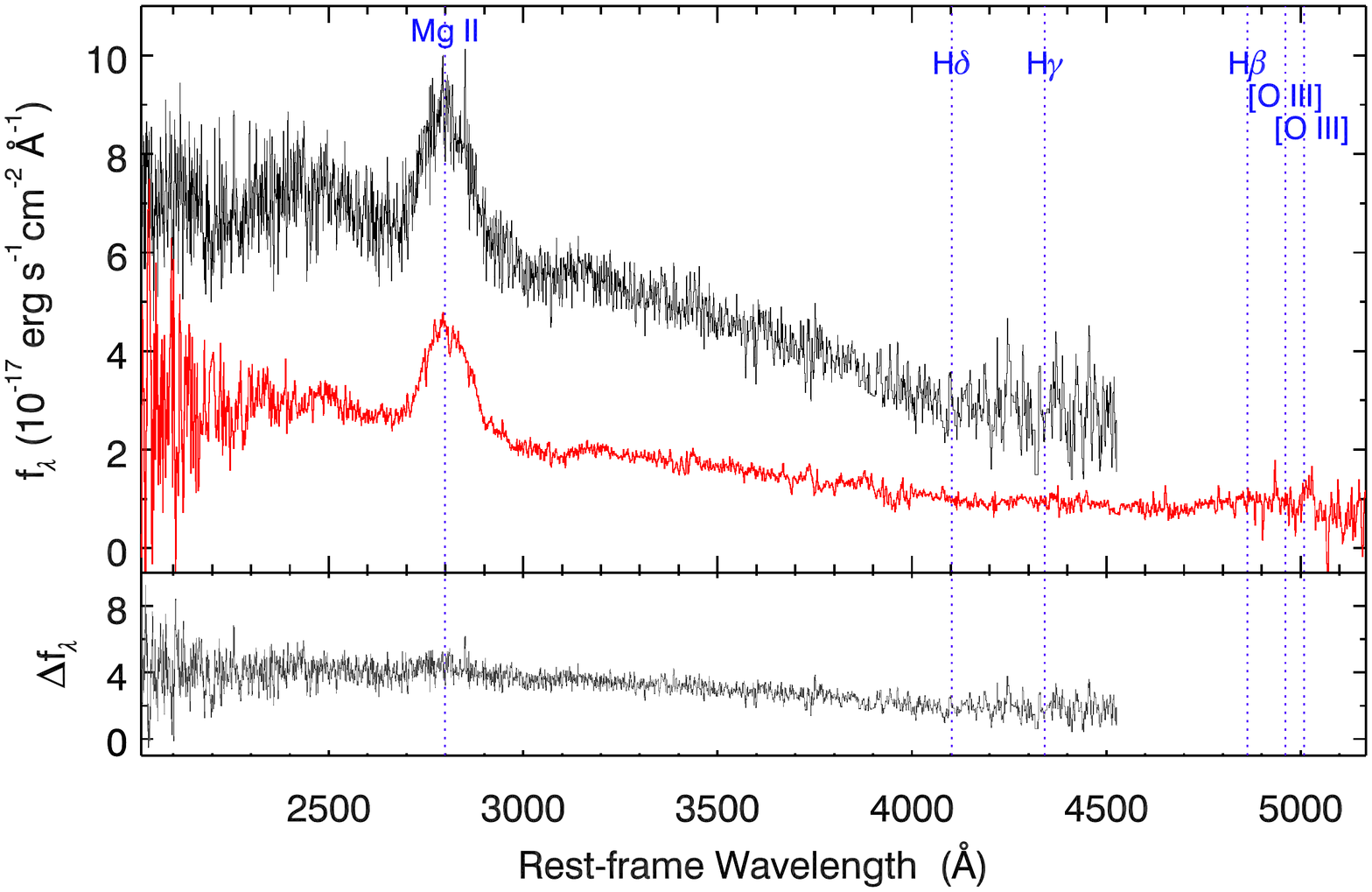}}\\
 \vspace{-1.0cm}
 
  \caption{ (Continued.)}
\end{figure*}

\renewcommand{\thefigure}{A.\arabic{figure}}
\addtocounter{figure}{-1}

\begin{figure*}
 \centering
 \hspace{0cm}
 \subfigure{
  \hspace{-1.0cm}
  \includegraphics[width=3.8in]{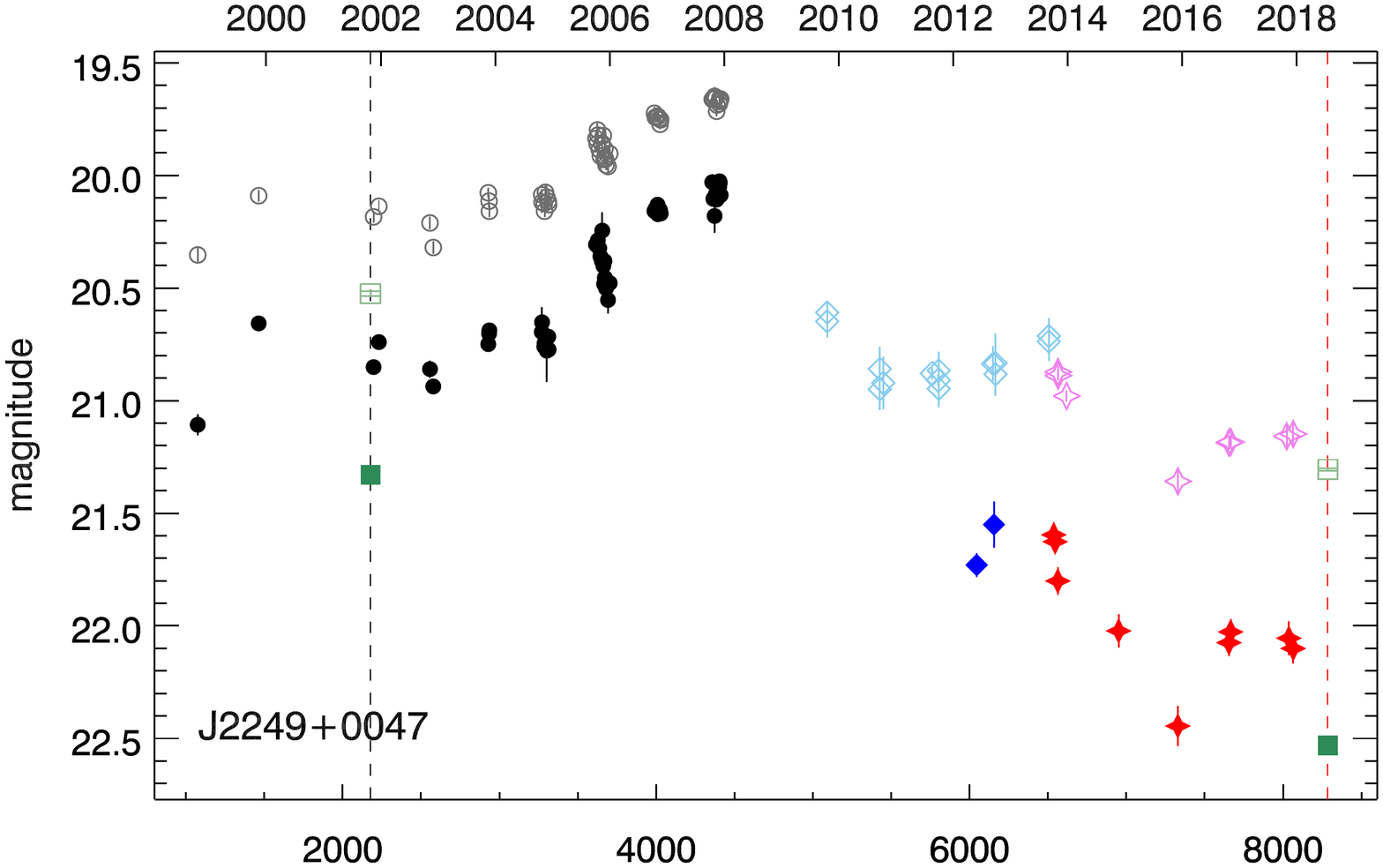}}
 \hspace{-1.4cm}
 \subfigure{
  \includegraphics[width=3.8in]{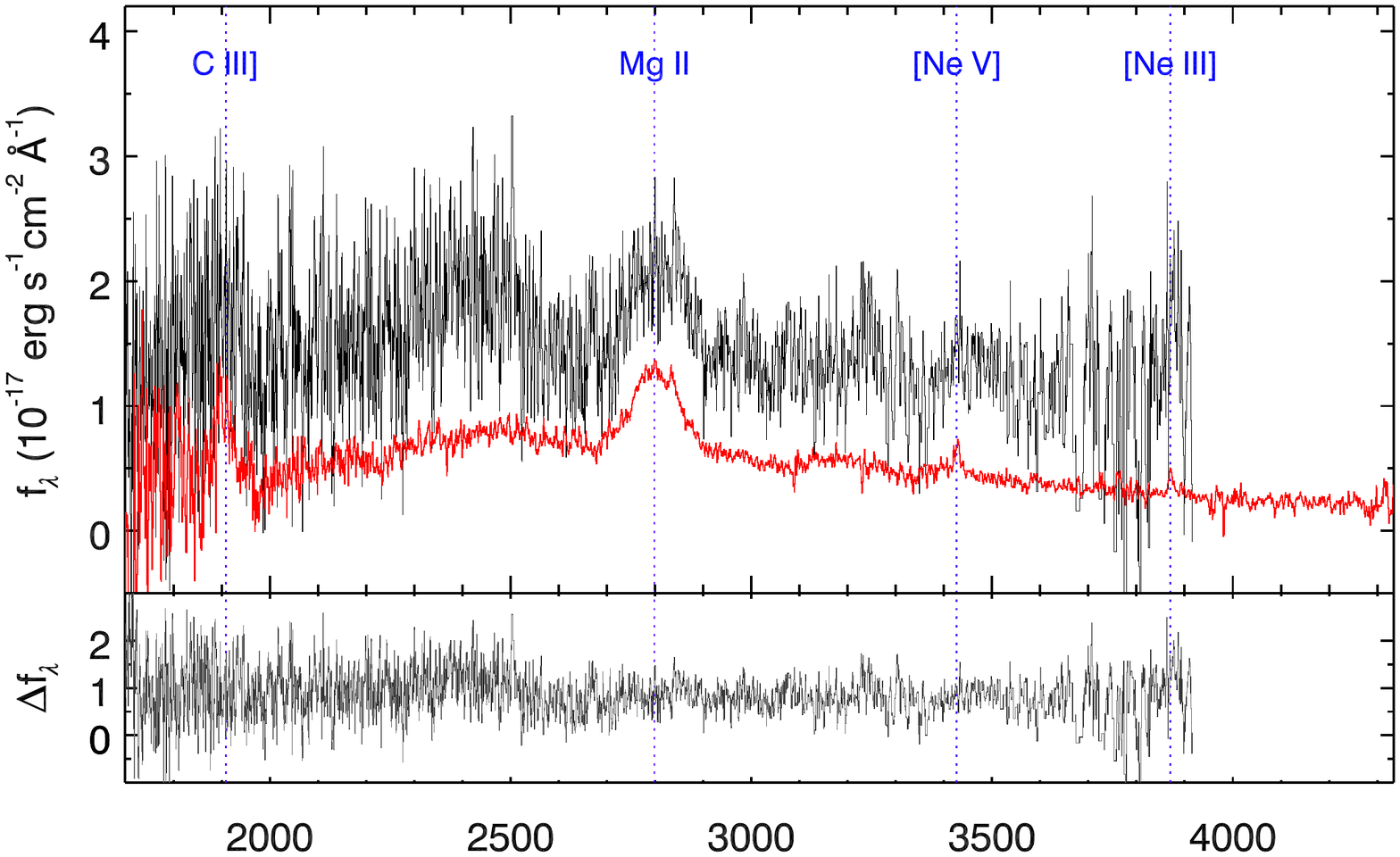}}\\
 \vspace{-2.5cm}

  \centering
  \hspace{0cm}
  \subfigure{
   \hspace{-1.0cm}
  \includegraphics[width=3.8in]{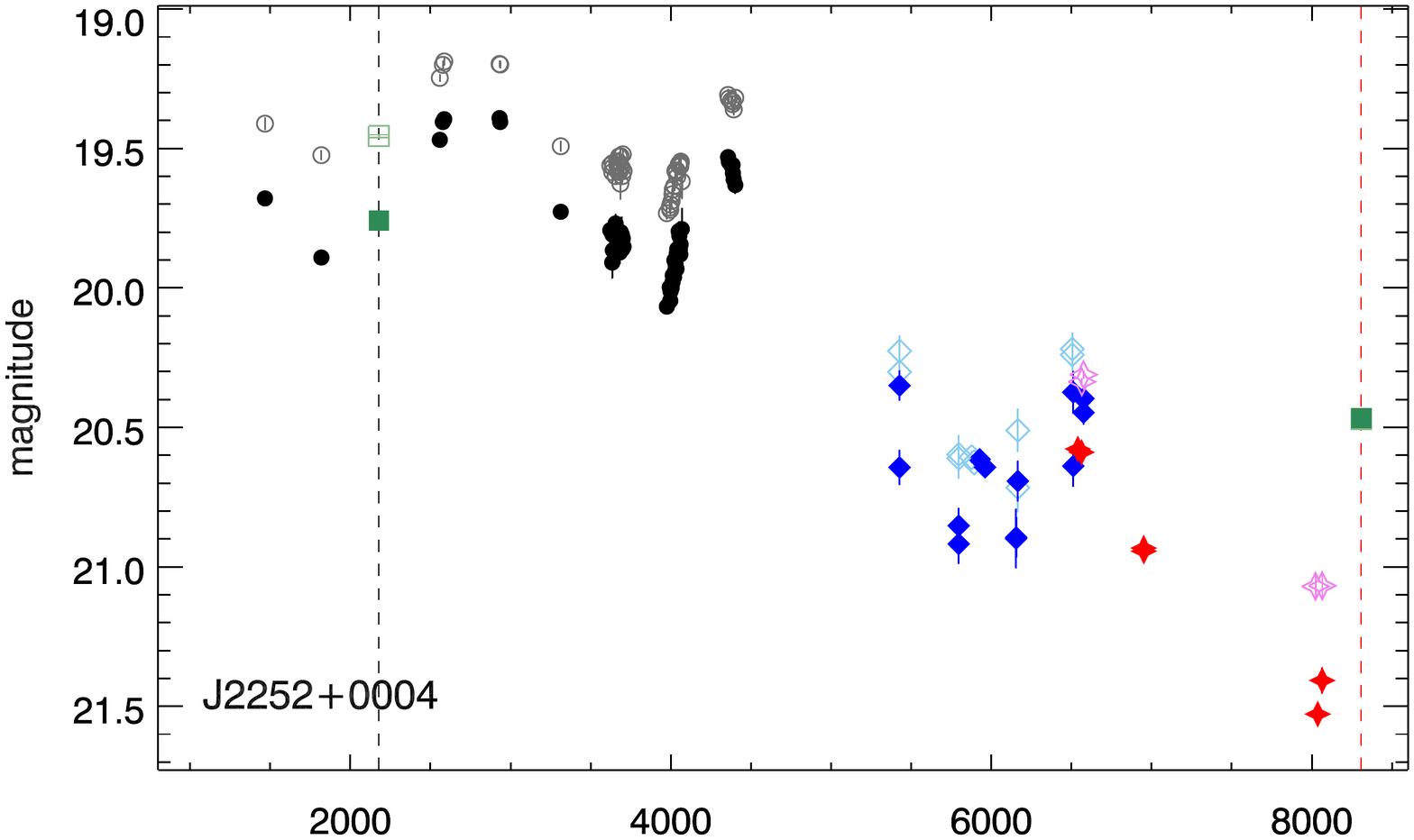}}
 \hspace{-1.4cm}
 \subfigure{
  \includegraphics[width=3.8in]{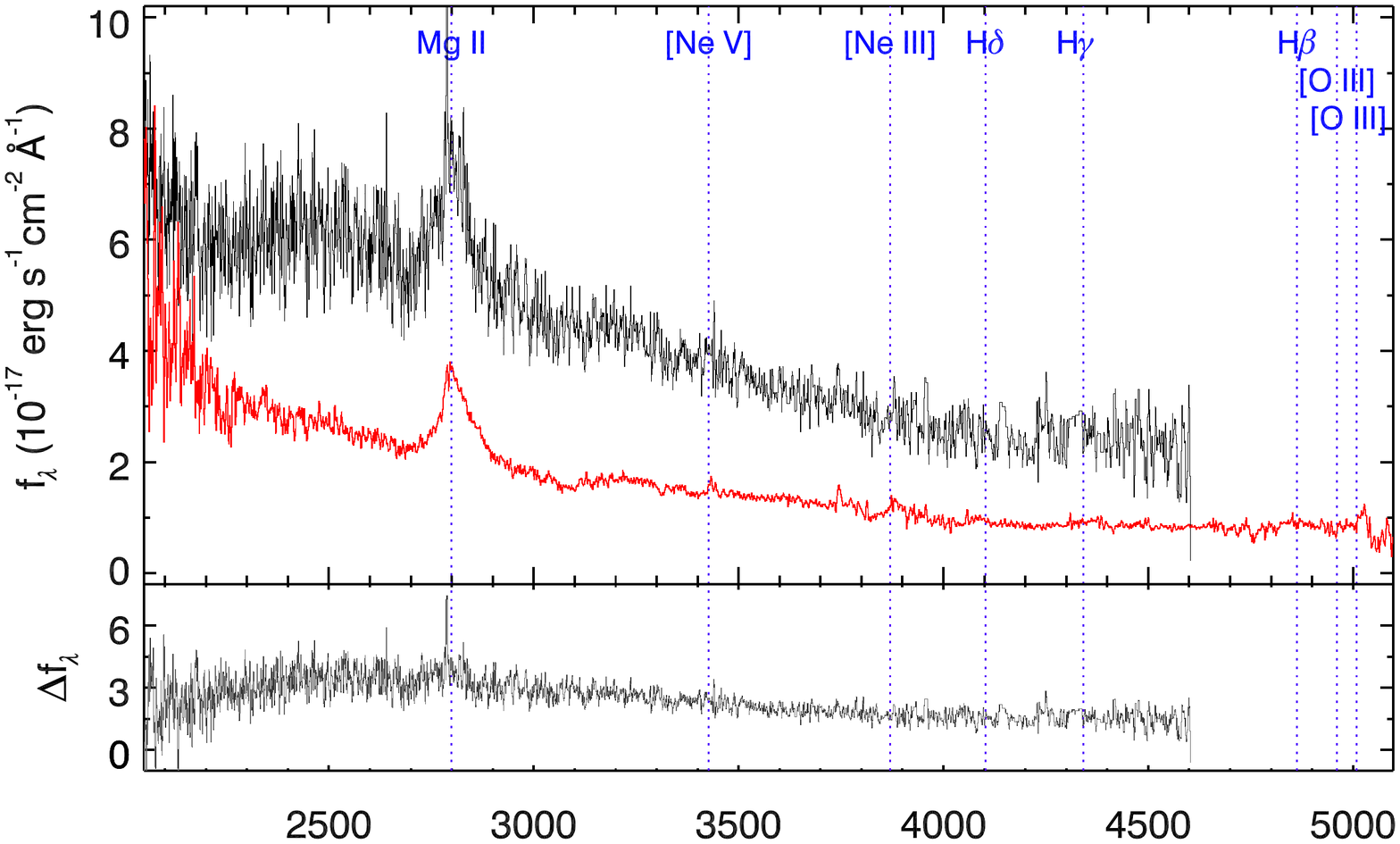}}\\
 \vspace{-2.5cm}
 
  \centering
 \hspace{0cm}
  \subfigure{
   \hspace{-1.0cm}
  \includegraphics[width=3.8in]{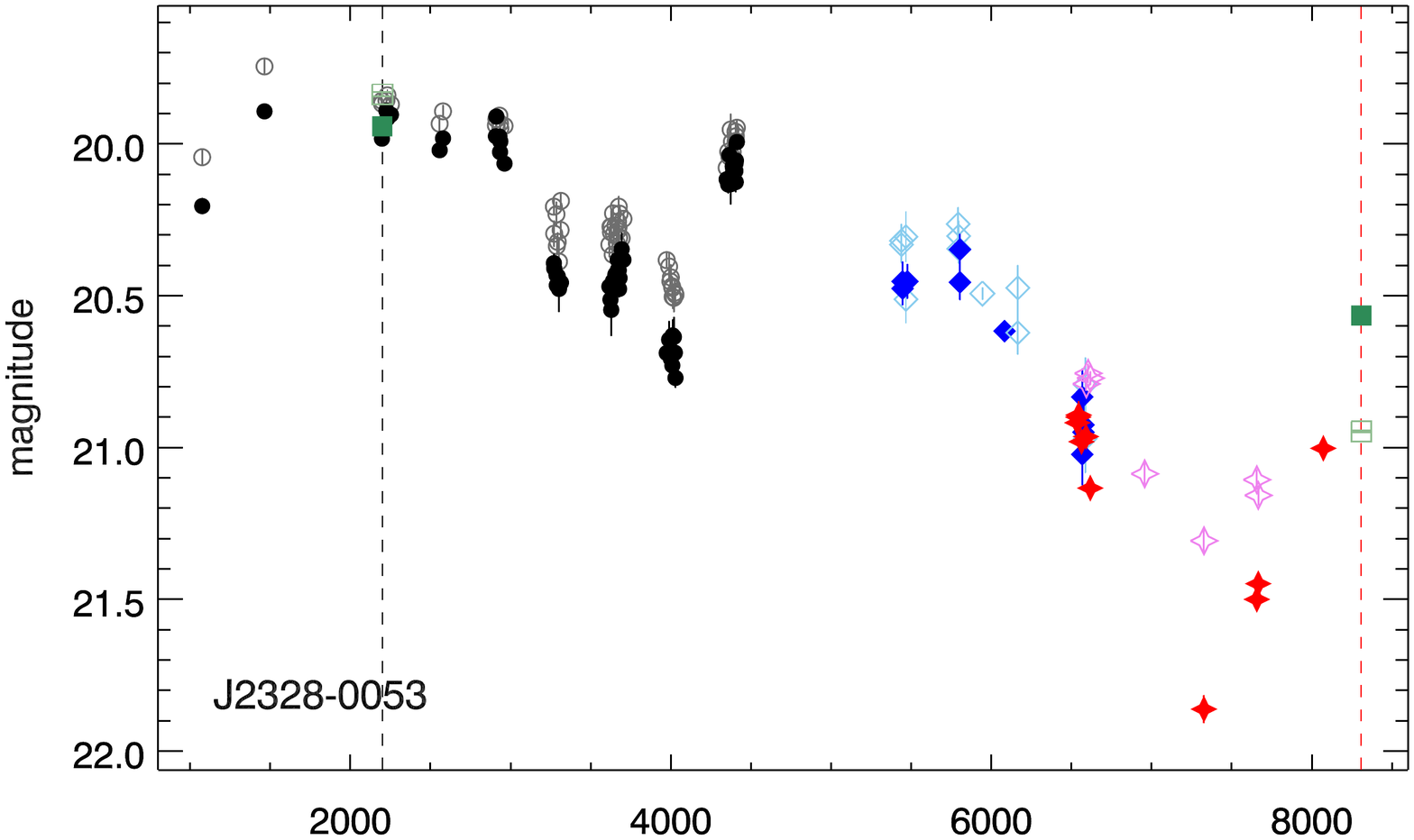}}
 \hspace{-1.4cm}
 \subfigure{
  \includegraphics[width=3.8in]{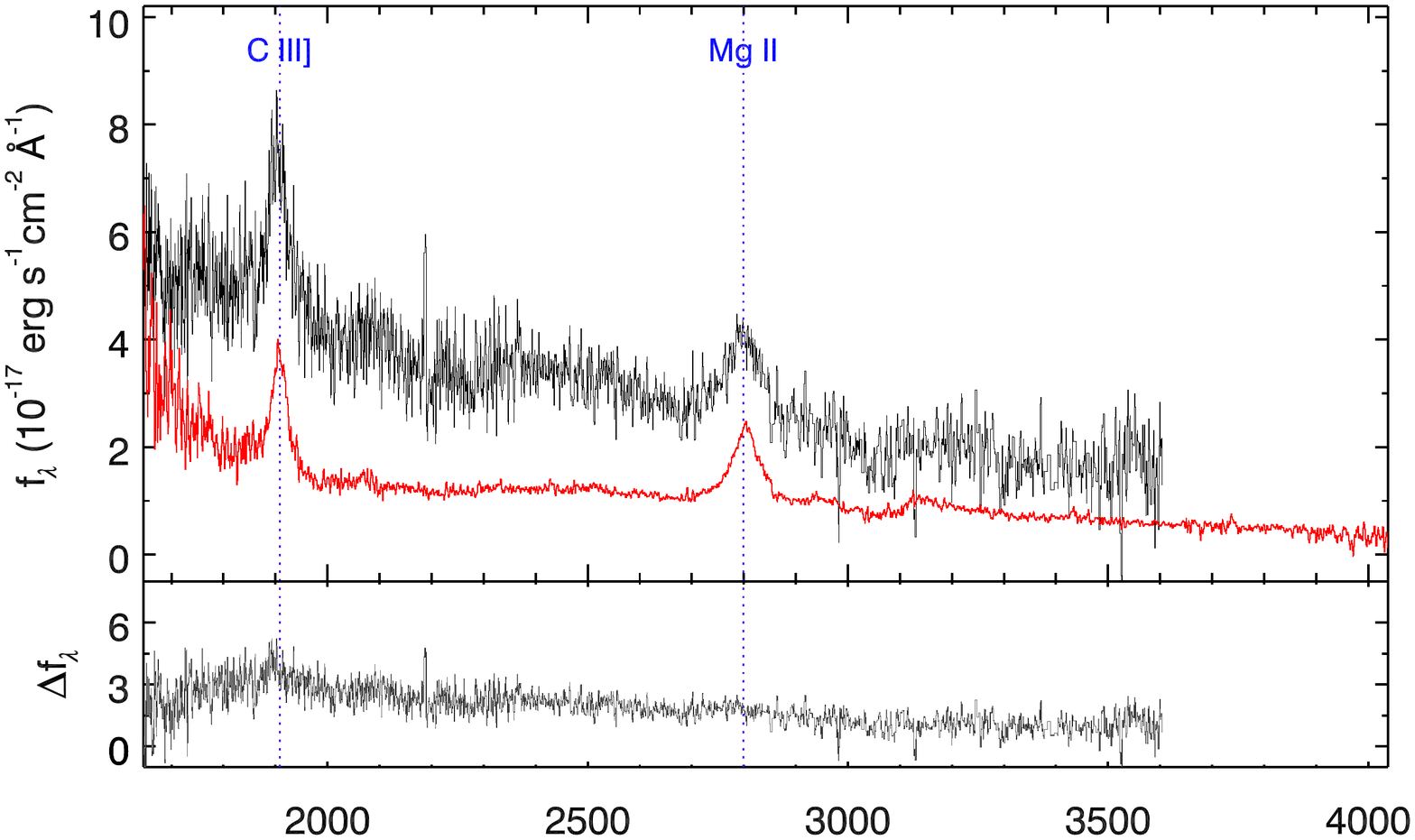}}\\
 \vspace{-2.5cm}
 
   \centering
  \hspace{0cm}
  \subfigure{
   \hspace{-1.0cm}
  \includegraphics[width=3.8in]{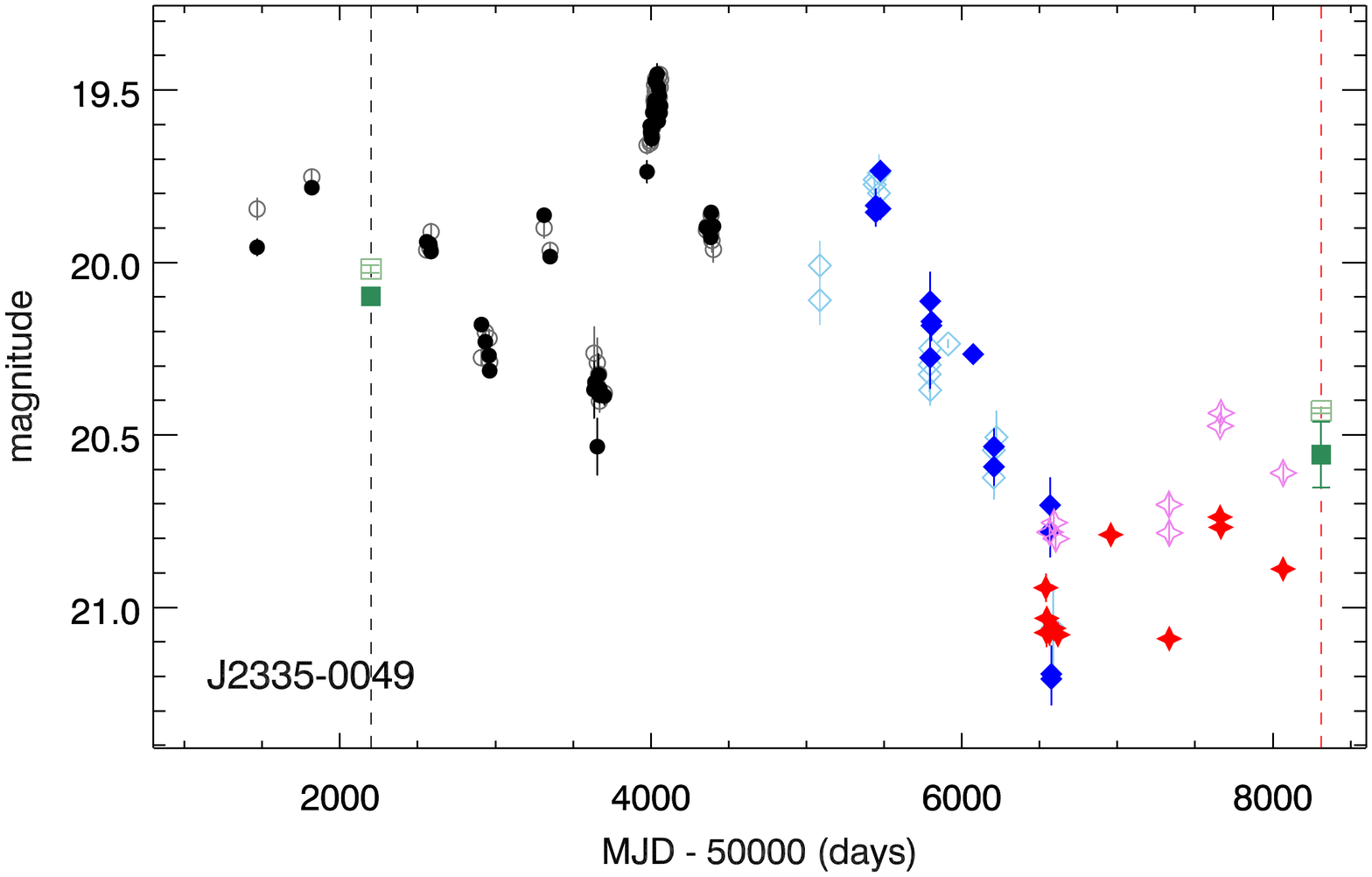}}
 \hspace{-1.4cm}
 \subfigure{
  \includegraphics[width=3.8in]{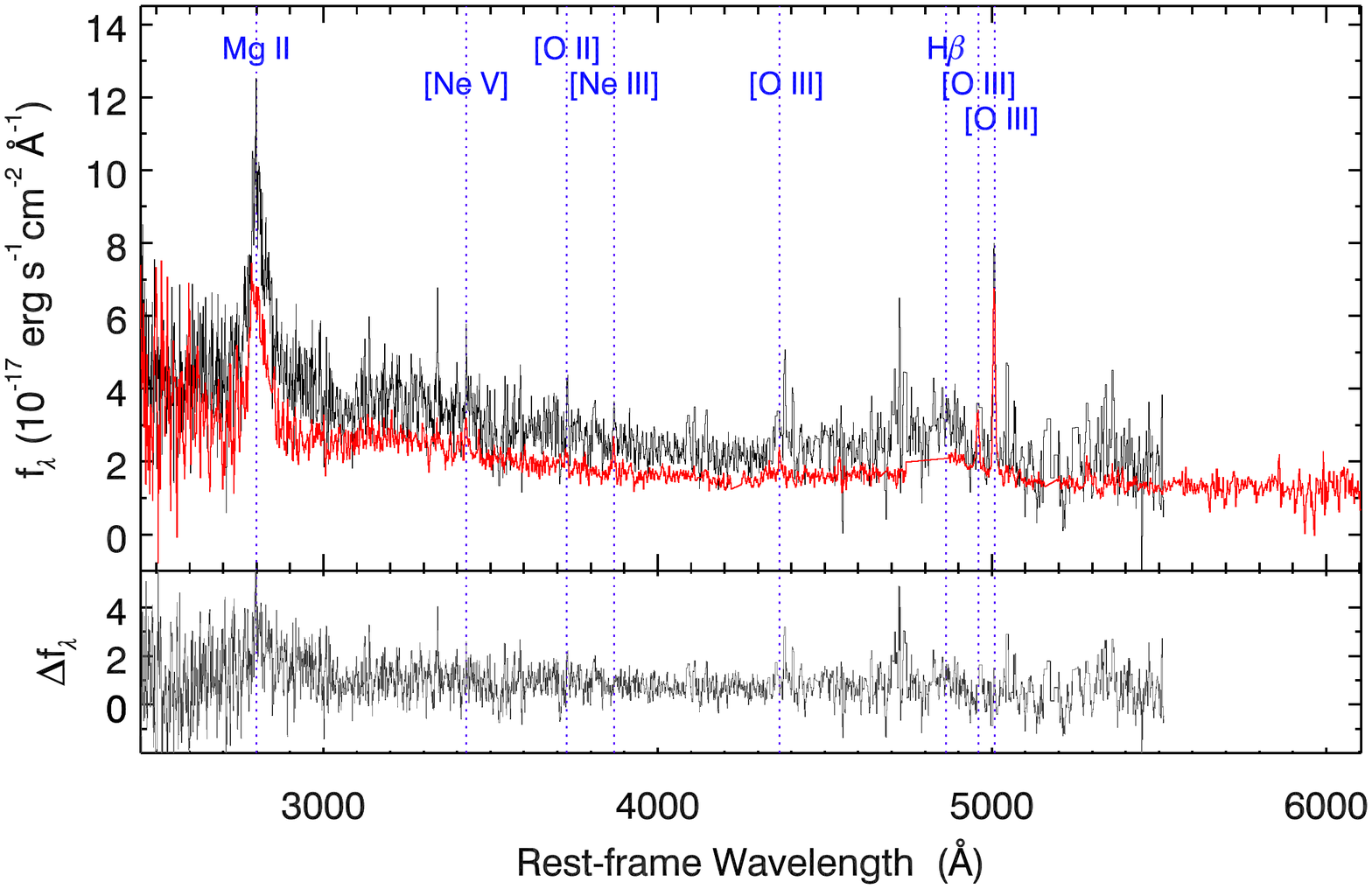}}\\
 \vspace{-1.0cm}
 
  \caption{ (Continued.) }
\end{figure*}

\renewcommand{\thefigure}{A.\arabic{figure}}
\addtocounter{figure}{-1}

\begin{figure*}
 \vspace{-1cm}
 \centering
 \hspace{0cm}
 \subfigure{
  \hspace{-1.0cm}
  \includegraphics[width=3.8in]{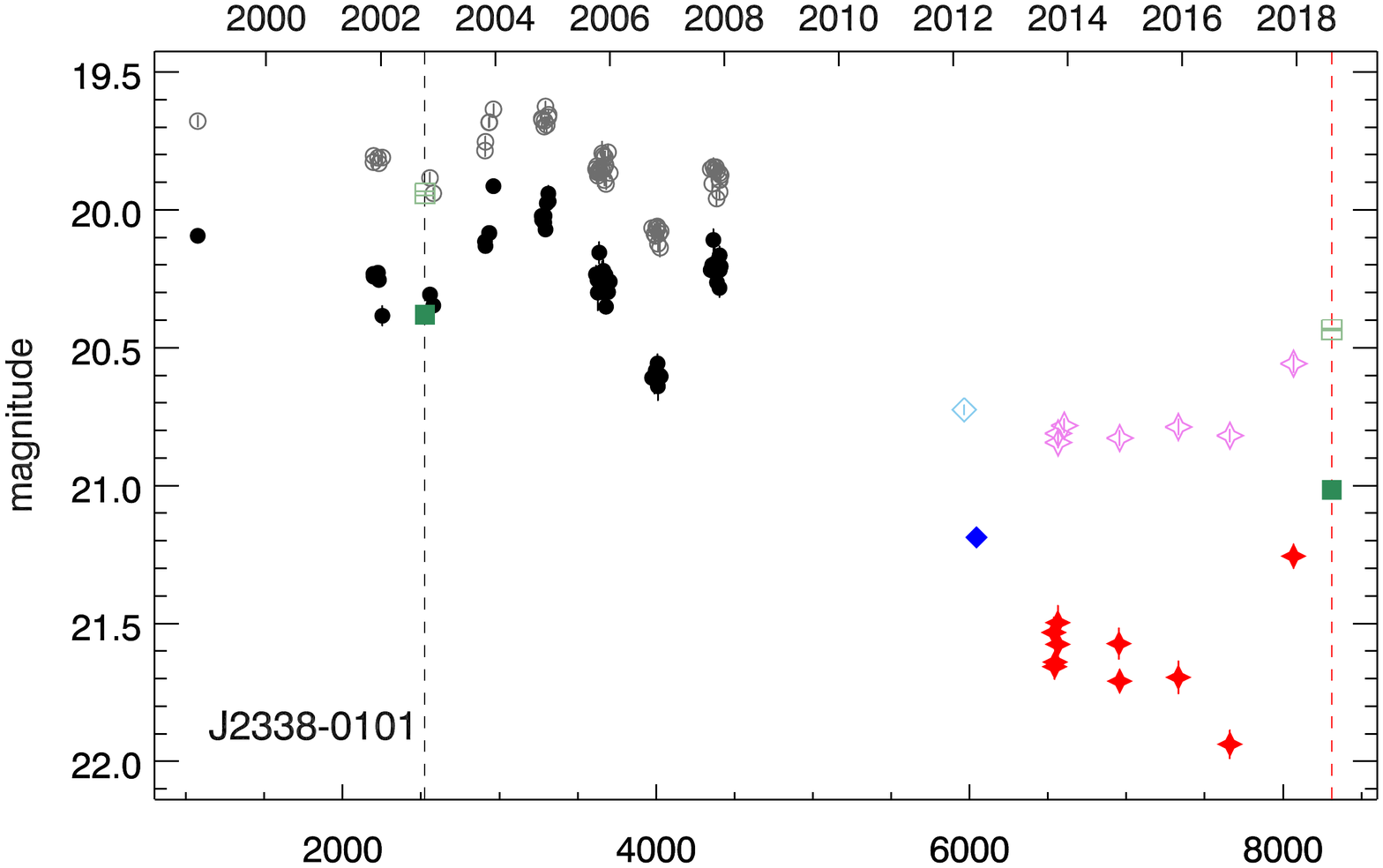}}
 \hspace{-1.4cm}
 \subfigure{
  \includegraphics[width=3.8in]{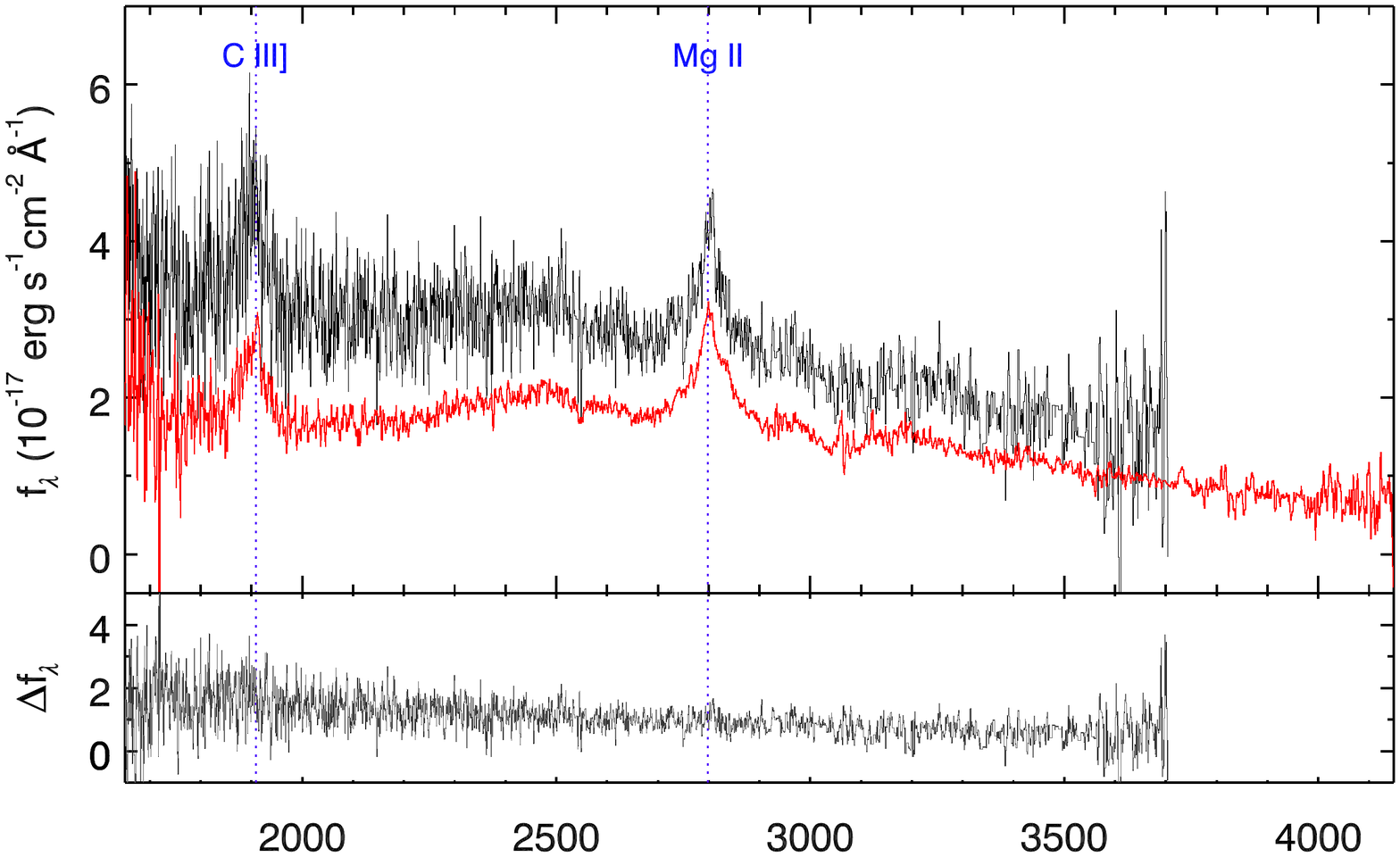}}\\
 \vspace{-2.5cm}

  \centering
  \hspace{0cm}
  \subfigure{
   \hspace{-1.0cm}
  \includegraphics[width=3.8in]{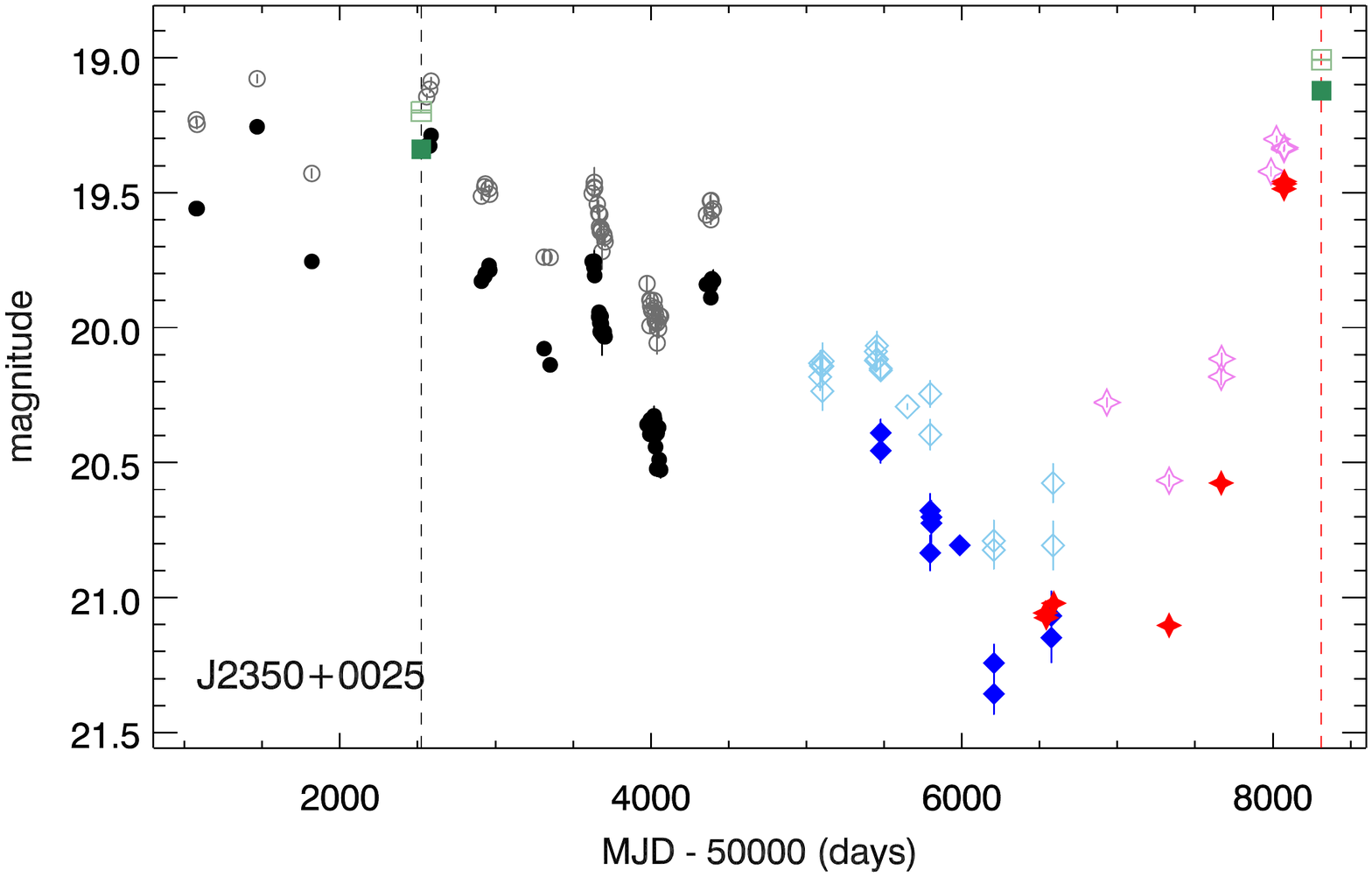}}
 \hspace{-1.4cm}
 \subfigure{
  \includegraphics[width=3.8in]{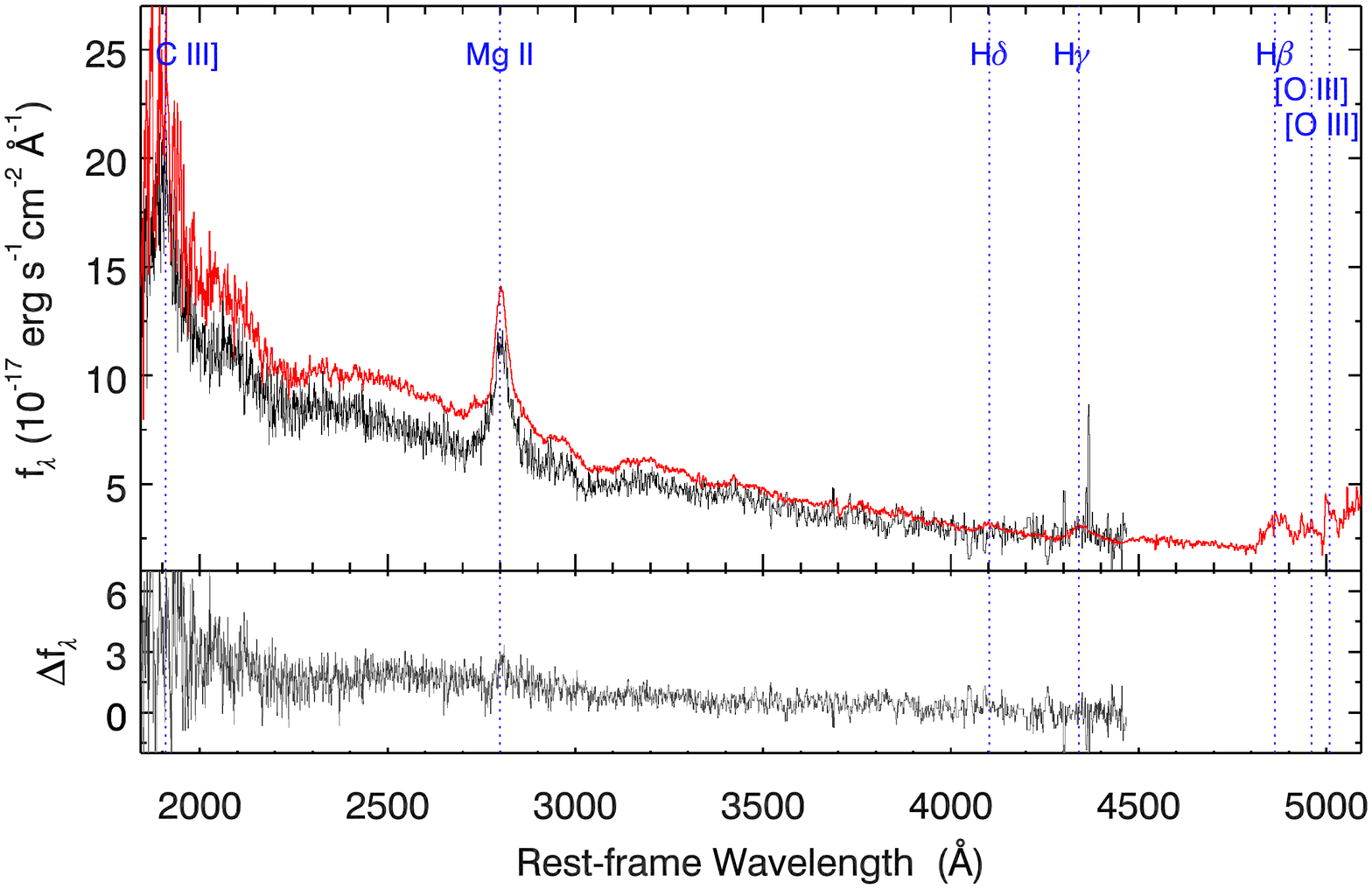}}\\
 \vspace{-1cm}
  \caption{ (Continued.) }
\end{figure*}


\bibliographystyle{aa_url}
\bibliography{ref} 

\begin{thebibliography}{77}
\expandafter\ifx\csname natexlab\endcsname\relax\def\natexlab#1{#1}\fi

\bibitem[{{Abazajian} {et~al.}(2009){Abazajian}, {Adelman-McCarthy},
  {Ag{\"u}eros}, {Allam}, {Allende Prieto}, {An}, {Anderson}, {Anderson},
  {Annis}, {Bahcall}, \& et~al.}]{Abazajian2009}
{Abazajian}, K.~N., {Adelman-McCarthy}, J.~K., {Ag{\"u}eros}, M.~A., {et~al.}
  2009,
  \href{http://dx.doi.org/10.1088/0067-0049/182/2/543}{\color{magenta}\apjs},
  \href{http://adsabs.harvard.edu/abs/2009ApJS..182..543A}{182, 543}

\bibitem[{{Abbott} {et~al.}(2018){Abbott}, {Abdalla}, {Allam}, {Amara},
  {Annis}, {Asorey}, {Avila}, {Ballester}, {Banerji}, {Barkhouse}, {Baruah},
  {Baumer}, {Bechtol}, {Becker}, {Benoit-L{\'e}vy}, {Bernstein}, {Bertin},
  {Blazek}, {Bocquet}, {Brooks}, {Brout}, {Buckley-Geer}, {Burke}, {Busti},
  {Campisano}, {Cardiel-Sas}, {Carnero Rosell}, {Carrasco Kind}, {Carretero},
  {Castander}, {Cawthon}, {Chang}, {Chen}, {Conselice}, {Costa}, {Crocce},
  {Cunha}, {D'Andrea}, {da Costa}, {Das}, {Daues}, {Davis}, {Davis}, {De
  Vicente}, {DePoy}, {DeRose}, {Desai}, {Diehl}, {Dietrich}, {Dodelson},
  {Doel}, {Drlica-Wagner}, {Eifler}, {Elliott}, {Evrard}, {Farahi}, {Fausti
  Neto}, {Fernandez}, {Finley}, {Flaugher}, {Foley}, {Fosalba}, {Friedel},
  {Frieman}, {Garc{\'{\i}}a-Bellido}, {Gaztanaga}, {Gerdes}, {Giannantonio},
  {Gill}, {Glazebrook}, {Goldstein}, {Gower}, {Gruen}, {Gruendl}, {Gschwend},
  {Gupta}, {Gutierrez}, {Hamilton}, {Hartley}, {Hinton}, {Hislop}, {Hollowood},
  {Honscheid}, {Hoyle}, {Huterer}, {Jain}, {James}, {Jeltema}, {Johnson},
  {Johnson}, {Kacprzak}, {Kent}, {Khullar}, {Klein}, {Kovacs}, {Koziol},
  {Krause}, {Kremin}, {Kron}, {Kuehn}, {Kuhlmann}, {Kuropatkin}, {Lahav},
  {Lasker}, {Li}, {Li}, {Liddle}, {Lima}, {Lin}, {L{\'o}pez-Reyes}, {MacCrann},
  {Maia}, {Maloney}, {Manera}, {March}, {Marriner}, {Marshall}, {Martini},
  {McClintock}, {McKay}, {McMahon}, {Melchior}, {Menanteau}, {Miller},
  {Miquel}, {Mohr}, {Morganson}, {Mould}, {Neilsen}, {Nichol}, {Nogueira},
  {Nord}, {Nugent}, {Nunes}, {Ogando}, {Old}, {Pace}, {Palmese},
  {Paz-Chinch{\'o}n}, {Peiris}, {Percival}, {Petravick}, {Plazas}, {Poh},
  {Pond}, {Porredon}, {Pujol}, {Refregier}, {Reil}, {Ricker}, {Rollins},
  {Romer}, {Roodman}, {Rooney}, {Ross}, {Rykoff}, {Sako}, {Sanchez}, {Sanchez},
  {Santiago}, {Saro}, {Scarpine}, {Scolnic}, {Serrano}, {Sevilla-Noarbe},
  {Sheldon}, {Shipp}, {Silveira}, {Smith}, {Smith}, {Smith}, {Soares-Santos},
  {Sobreira}, {Song}, {Stebbins}, {Suchyta}, {Sullivan}, {Swanson}, {Tarle},
  {Thaler}, {Thomas}, {Thomas}, {Troxel}, {Tucker}, {Vikram}, {Vivas},
  {Walker}, {Wechsler}, {Weller}, {Wester}, {Wolf}, {Wu}, {Yanny}, {Zenteno},
  {Zhang}, {Zuntz}, {DES Collaboration}, {Juneau}, {Fitzpatrick}, {Nikutta},
  {Nidever}, {Olsen}, {Scott}, \& {Data Lab}}]{Abbott2018}
{Abbott}, T.~M.~C., {Abdalla}, F.~B., {Allam}, S., {et~al.} 2018,
  \href{http://dx.doi.org/10.3847/1538-4365/aae9f0}{\color{magenta}\apjs},
  \href{http://adsabs.harvard.edu/abs/2018ApJS..239...18A}{239, 18}

\bibitem[{{Antonucci}(1993)}]{Antonucci1993}
{Antonucci}, R. 1993,
  \href{http://dx.doi.org/10.1146/annurev.aa.31.090193.002353}{\color{magenta}\araa},
  \href{http://adsabs.harvard.edu/abs/1993ARA%26A..31..473A}{31, 473}

\bibitem[{{Baldwin}(1977)}]{Baldwin1977}
{Baldwin}, J.~A. 1977,
  \href{http://dx.doi.org/10.1086/155294}{\color{magenta}\apj},
  \href{http://adsabs.harvard.edu/abs/1977ApJ...214..679B}{214, 679}

\bibitem[{{Barth} {et~al.}(2015){Barth}, {Bennert}, {Canalizo}, {Filippenko},
  {Gates}, {Greene}, {Li}, {Malkan}, {Pancoast}, {Sand}, {Stern}, {Treu},
  {Woo}, {Assef}, {Bae}, {Brewer}, {Cenko}, {Clubb}, {Cooper},
  {Diamond-Stanic}, {Hiner}, {H{\"o}nig}, {Hsiao}, {Kandrashoff}, {Lazarova},
  {Nierenberg}, {Rex}, {Silverman}, {Tollerud}, \& {Walsh}}]{Barth2015}
{Barth}, A.~J., {Bennert}, V.~N., {Canalizo}, G., {et~al.} 2015,
  \href{http://dx.doi.org/10.1088/0067-0049/217/2/26}{\color{magenta}\apjs},
  \href{http://adsabs.harvard.edu/abs/2015ApJS..217...26B}{217, 26}

\bibitem[{{Boroson} \& {Green}(1992)}]{Boroson1992}
{Boroson}, T.~A. \& {Green}, R.~F. 1992,
  \href{http://dx.doi.org/10.1086/191661}{\color{magenta}\apjs},
  \href{http://adsabs.harvard.edu/abs/1992ApJS...80..109B}{80, 109}

\bibitem[{{Bruce} {et~al.}(2017){Bruce}, {Lawrence}, {MacLeod}, {Elvis},
  {Ward}, {Collinson}, {Gezari}, {Marshall}, {Lam}, {Kotak}, {Inserra},
  {Polshaw}, {Kaiser}, {Kudritzki}, {Magnier}, \& {Waters}}]{Bruce2017}
{Bruce}, A., {Lawrence}, A., {MacLeod}, C., {et~al.} 2017,
  \href{http://dx.doi.org/10.1093/mnras/stx168}{\color{magenta}\mnras},
  \href{http://adsabs.harvard.edu/abs/2017MNRAS.467.1259B}{467, 1259}

\bibitem[{{Cackett} {et~al.}(2015){Cackett}, {G{\"u}ltekin}, {Bentz},
  {Fausnaugh}, {Peterson}, {Troyer}, \& {Vestergaard}}]{Cackett2015}
{Cackett}, E.~M., {G{\"u}ltekin}, K., {Bentz}, M.~C., {et~al.} 2015,
  \href{http://dx.doi.org/10.1088/0004-637X/810/2/86}{\color{magenta}\apj},
  \href{http://adsabs.harvard.edu/abs/2015ApJ...810...86C}{810, 86}

\bibitem[{{Cackett} \& {Horne}(2006)}]{Cackett2006}
{Cackett}, E.~M. \& {Horne}, K. 2006,
  \href{http://dx.doi.org/10.1111/j.1365-2966.2005.09795.x}{\color{magenta}\mnras},
  \href{http://adsabs.harvard.edu/abs/2006MNRAS.365.1180C}{365, 1180}

\bibitem[{{Cardelli} {et~al.}(1989){Cardelli}, {Clayton}, \&
  {Mathis}}]{Cardelli1989}
{Cardelli}, J.~A., {Clayton}, G.~C., \& {Mathis}, J.~S. 1989,
  \href{http://dx.doi.org/10.1086/167900}{\color{magenta}\apj},
  \href{http://adsabs.harvard.edu/abs/1989ApJ...345..245C}{345, 245}

\bibitem[{{Chambers} {et~al.}(2016){Chambers}, {Magnier}, {Metcalfe},
  {Flewelling}, {Huber}, {Waters}, {Denneau}, {Draper}, {Farrow}, {Finkbeiner},
  {Holmberg}, {Koppenhoefer}, {Price}, {Saglia}, {Schlafly}, {Smartt},
  {Sweeney}, {Wainscoat}, {Burgett}, {Grav}, {Heasley}, {Hodapp}, {Jedicke},
  {Kaiser}, {Kudritzki}, {Luppino}, {Lupton}, {Monet}, {Morgan}, {Onaka},
  {Stubbs}, {Tonry}, {Banados}, {Bell}, {Bender}, {Bernard}, {Botticella},
  {Casertano}, {Chastel}, {Chen}, {Chen}, {Cole}, {Deacon}, {Frenk},
  {Fitzsimmons}, {Gezari}, {Goessl}, {Goggia}, {Goldman}, {Grebel}, {Hambly},
  {Hasinger}, {Heavens}, {Heckman}, {Henderson}, {Henning}, {Holman}, {Hopp},
  {Ip}, {Isani}, {Keyes}, {Koekemoer}, {Kotak}, {Long}, {Lucey}, {Liu},
  {Martin}, {McLean}, {Morganson}, {Murphy}, {Nieto-Santisteban}, {Norberg},
  {Peacock}, {Pier}, {Postman}, {Primak}, {Rae}, {Rest}, {Riess}, {Riffeser},
  {Rix}, {Roser}, {Schilbach}, {Schultz}, {Scolnic}, {Szalay}, {Seitz},
  {Shiao}, {Small}, {Smith}, {Soderblom}, {Taylor}, {Thakar}, {Thiel},
  {Thilker}, {Urata}, {Valenti}, {Walter}, {Watters}, {Werner}, {White},
  {Wood-Vasey}, \& {Wyse}}]{Chambers2016}
{Chambers}, K.~C., {Magnier}, E.~A., {Metcalfe}, N., {et~al.} 2016, arXiv
  e-prints
  \href{http://adsabs.harvard.edu/abs/2016arXiv161205560C}{[\eprint[arXiv]{1612.05560}]}

\bibitem[{{Clavel} {et~al.}(1991){Clavel}, {Reichert}, {Alloin}, {Crenshaw},
  {Kriss}, {Krolik}, {Malkan}, {Netzer}, {Peterson}, {Wamsteker}, {Altamore},
  {Baribaud}, {Barr}, {Beck}, {Binette}, {Bromage}, {Brosch}, {Diaz},
  {Filippenko}, {Fricke}, {Gaskell}, {Giommi}, {Glass}, {Gondhalekar},
  {Hackney}, {Halpern}, {Hutter}, {Joersaeter}, {Kinney}, {Kollatschny},
  {Koratkar}, {Korista}, {Laor}, {Lasota}, {Leibowitz}, {Maoz}, {Martin},
  {Mazeh}, {Meurs}, {Nair}, {O'Brien}, {Pelat}, {Perez}, {Perola}, {Ptak},
  {Rodriguez-Pascual}, {Rosenblatt}, {Sadun}, {Santos-Lleo}, {Shaw}, {Smith},
  {Stirpe}, {Stoner}, {Sun}, {Ulrich}, {van Groningen}, \&
  {Zheng}}]{Clavel1991}
{Clavel}, J., {Reichert}, G.~A., {Alloin}, D., {et~al.} 1991,
  \href{http://dx.doi.org/10.1086/169540}{\color{magenta}\apj},
  \href{http://adsabs.harvard.edu/abs/1991ApJ...366...64C}{366, 64}

\bibitem[{{Denney} {et~al.}(2014){Denney}, {De Rosa}, {Croxall}, {Gupta},
  {Bentz}, {Fausnaugh}, {Grier}, {Martini}, {Mathur}, {Peterson}, {Pogge}, \&
  {Shappee}}]{Denney2014}
{Denney}, K.~D., {De Rosa}, G., {Croxall}, K., {et~al.} 2014,
  \href{http://dx.doi.org/10.1088/0004-637X/796/2/134}{\color{magenta}\apj},
  \href{http://adsabs.harvard.edu/abs/2014ApJ...796..134D}{796, 134}

\bibitem[{{Denney} {et~al.}(2009){Denney}, {Peterson}, {Dietrich},
  {Vestergaard}, \& {Bentz}}]{Denney2009}
{Denney}, K.~D., {Peterson}, B.~M., {Dietrich}, M., {Vestergaard}, M., \&
  {Bentz}, M.~C. 2009,
  \href{http://dx.doi.org/10.1088/0004-637X/692/1/246}{\color{magenta}\apj},
  \href{http://adsabs.harvard.edu/abs/2009ApJ...692..246D}{692, 246}

\bibitem[{{Diehl} {et~al.}(2018){Diehl}, {Neilsen}, {Gruendl}, {Abbott},
  {Allam}, {Alvarez}, {Annis}, {Balbinot}, {Bhargava}, {Bechtol}, {Bernstein},
  {Bhatawdekar}, {Bocquet}, {Brout}, {Capasso}, {Cawthon}, {Chang}, {Cook},
  {Conselice}, {Cruz}, {D'Andrea}, {da Costa}, {Das}, {DePoy}, {Drlica-Wagner},
  {Elliott}, {Everett}, {Frieman}, {Fausti Neto}, {Fert{\'e}}, {Friswell},
  {Furnell}, {Gelman}, {Gerdes}, {Gill}, {Goldstein}, {Gruen}, {Gulledge},
  {Hamilton}, {Hollowood}, {Honscheid}, {James}, {Johnson}, {Johnson}, {Kent},
  {Kessler}, {Khullar}, {Kovacs}, {Kremin}, {Kron}, {Kuropatkin}, {Lasker},
  {Lathrop}, {Li}, {Manera}, {March}, {Marshall}, {Medford}, {Menanteau},
  {Mohammed}, {Monroy}, {Moraes}, {Morganson}, {Muir}, {Murphy}, {Nord},
  {Pace}, {Palmese}, {Park}, {Paz-Chinch{\'o}n}, {Pereira}, {Petravick},
  {Plazas}, {Poh}, {Prochaska}, {Romer}, {Reil}, {Roodman}, {Sako}, {Sauseda},
  {Scolnic}, {Secco}, {Sevilla-Noarbe}, {Shipp}, {Smith}, {Soares-Santos},
  {Soergel}, {Stebbins}, {Story}, {Stringer}, {Tarsitano}, {Thomas}, {Tucker},
  {Vivas}, {Walker}, {Wang}, {Weaverdyck}, {Weaverdyck}, {Wester}, {Wethers},
  {Wilkenson}, {Wu}, {Yanny}, {Zenteno}, \& {Zhang}}]{Diehl2018}
{Diehl}, H.~T., {Neilsen}, E., {Gruendl}, R.~A., {et~al.} 2018, in Society of
  Photo-Optical Instrumentation Engineers (SPIE) Conference Series, Vol. 10704,
  Observatory Operations: Strategies, Processes, and Systems VII,
  \href{http://adsabs.harvard.edu/abs/2018SPIE10704E..0DD}{107040D}

\bibitem[{{Flaugher} {et~al.}(2015){Flaugher}, {Diehl}, {Honscheid}, {Abbott},
  {Alvarez}, {Angstadt}, {Annis}, {Antonik}, {Ballester}, {Beaufore},
  {Bernstein}, {Bernstein}, {Bigelow}, {Bonati}, {Boprie}, {Brooks},
  {Buckley-Geer}, {Campa}, {Cardiel-Sas}, {Castander}, {Castilla}, {Cease},
  {Cela-Ruiz}, {Chappa}, {Chi}, {Cooper}, {da Costa}, {Dede}, {Derylo},
  {DePoy}, {de Vicente}, {Doel}, {Drlica-Wagner}, {Eiting}, {Elliott}, {Emes},
  {Estrada}, {Fausti Neto}, {Finley}, {Flores}, {Frieman}, {Gerdes},
  {Gladders}, {Gregory}, {Gutierrez}, {Hao}, {Holland}, {Holm}, {Huffman},
  {Jackson}, {James}, {Jonas}, {Karcher}, {Karliner}, {Kent}, {Kessler},
  {Kozlovsky}, {Kron}, {Kubik}, {Kuehn}, {Kuhlmann}, {Kuk}, {Lahav}, {Lathrop},
  {Lee}, {Levi}, {Lewis}, {Li}, {Mandrichenko}, {Marshall}, {Martinez},
  {Merritt}, {Miquel}, {Mu{\~n}oz}, {Neilsen}, {Nichol}, {Nord}, {Ogando},
  {Olsen}, {Palaio}, {Patton}, {Peoples}, {Plazas}, {Rauch}, {Reil}, {Rheault},
  {Roe}, {Rogers}, {Roodman}, {Sanchez}, {Scarpine}, {Schindler}, {Schmidt},
  {Schmitt}, {Schubnell}, {Schultz}, {Schurter}, {Scott}, {Serrano}, {Shaw},
  {Smith}, {Soares-Santos}, {Stefanik}, {Stuermer}, {Suchyta}, {Sypniewski},
  {Tarle}, {Thaler}, {Tighe}, {Tran}, {Tucker}, {Walker}, {Wang}, {Watson},
  {Weaverdyck}, {Wester}, {Woods}, {Yanny}, \& {DES
  Collaboration}}]{Flaugher2015}
{Flaugher}, B., {Diehl}, H.~T., {Honscheid}, K., {et~al.} 2015,
  \href{http://dx.doi.org/10.1088/0004-6256/150/5/150}{\color{magenta}\aj},
  \href{http://adsabs.harvard.edu/abs/2015AJ....150..150F}{150, 150}

\bibitem[{{Frieman} {et~al.}(2008){Frieman}, {Bassett}, {Becker}, {Choi},
  {Cinabro}, {DeJongh}, {Depoy}, {Dilday}, {Doi}, {Garnavich}, {Hogan},
  {Holtzman}, {Im}, {Jha}, {Kessler}, {Konishi}, {Lampeitl}, {Marriner},
  {Marshall}, {McGinnis}, {Miknaitis}, {Nichol}, {Prieto}, {Riess}, {Richmond},
  {Romani}, {Sako}, {Schneider}, {Smith}, {Takanashi}, {Tokita}, {van der
  Heyden}, {Yasuda}, {Zheng}, {Adelman-McCarthy}, {Annis}, {Assef},
  {Barentine}, {Bender}, {Blandford}, {Boroski}, {Bremer}, {Brewington},
  {Collins}, {Crotts}, {Dembicky}, {Eastman}, {Edge}, {Edmondson}, {Elson},
  {Eyler}, {Filippenko}, {Foley}, {Frank}, {Goobar}, {Gueth}, {Gunn},
  {Harvanek}, {Hopp}, {Ihara}, {Ivezi{\'c}}, {Kahn}, {Kaplan}, {Kent},
  {Ketzeback}, {Kleinman}, {Kollatschny}, {Kron}, {Krzesi{\'n}ski}, {Lamenti},
  {Leloudas}, {Lin}, {Long}, {Lucey}, {Lupton}, {Malanushenko}, {Malanushenko},
  {McMillan}, {Mendez}, {Morgan}, {Morokuma}, {Nitta}, {Ostman}, {Pan},
  {Rockosi}, {Romer}, {Ruiz-Lapuente}, {Saurage}, {Schlesinger}, {Snedden},
  {Sollerman}, {Stoughton}, {Stritzinger}, {Subba Rao}, {Tucker}, {Vaisanen},
  {Watson}, {Watters}, {Wheeler}, {Yanny}, \& {York}}]{Frieman2008}
{Frieman}, J.~A., {Bassett}, B., {Becker}, A., {et~al.} 2008,
  \href{http://dx.doi.org/10.1088/0004-6256/135/1/338}{\color{magenta}\aj},
  \href{http://adsabs.harvard.edu/abs/2008AJ....135..338F}{135, 338}

\bibitem[{{Gezari} {et~al.}(2017){Gezari}, {Hung}, {Cenko}, {Blagorodnova},
  {Yan}, {Kulkarni}, {Mooley}, {Kong}, {Cantwell}, {Yu}, {Cao}, {Fremling},
  {Neill}, {Ngeow}, {Nugent}, \& {Wozniak}}]{Gezari2017}
{Gezari}, S., {Hung}, T., {Cenko}, S.~B., {et~al.} 2017,
  \href{http://dx.doi.org/10.3847/1538-4357/835/2/144}{\color{magenta}\apj},
  \href{http://adsabs.harvard.edu/abs/2017ApJ...835..144G}{835, 144}

\bibitem[{{Green} {et~al.}(2001){Green}, {Forster}, \&
  {Kuraszkiewicz}}]{Green2001}
{Green}, P.~J., {Forster}, K., \& {Kuraszkiewicz}, J. 2001,
  \href{http://dx.doi.org/10.1086/321600}{\color{magenta}\apj},
  \href{http://adsabs.harvard.edu/abs/2001ApJ...556..727G}{556, 727}

\bibitem[{{Gunn} {et~al.}(2006){Gunn}, {Siegmund}, {Mannery}, {Owen}, {Hull},
  {Leger}, {Carey}, {Knapp}, {York}, {Boroski}, {Kent}, {Lupton}, {Rockosi},
  {Evans}, {Waddell}, {Anderson}, {Annis}, {Barentine}, {Bartoszek}, {Bastian},
  {Bracker}, {Brewington}, {Briegel}, {Brinkmann}, {Brown}, {Carr},
  {Czarapata}, {Drennan}, {Dombeck}, {Federwitz}, {Gillespie}, {Gonzales},
  {Hansen}, {Harvanek}, {Hayes}, {Jordan}, {Kinney}, {Klaene}, {Kleinman},
  {Kron}, {Kresinski}, {Lee}, {Limmongkol}, {Lindenmeyer}, {Long}, {Loomis},
  {McGehee}, {Mantsch}, {Neilsen}, {Neswold}, {Newman}, {Nitta}, {Peoples},
  {Pier}, {Prieto}, {Prosapio}, {Rivetta}, {Schneider}, {Snedden}, \&
  {Wang}}]{Gunn2006}
{Gunn}, J.~E., {Siegmund}, W.~A., {Mannery}, E.~J., {et~al.} 2006,
  \href{http://dx.doi.org/10.1086/500975}{\color{magenta}\aj},
  \href{http://adsabs.harvard.edu/abs/2006AJ....131.2332G}{131, 2332}

\bibitem[{{Guo} \& {Gu}(2016)}]{Guo2016}
{Guo}, H. \& {Gu}, M. 2016,
  \href{http://dx.doi.org/10.3847/0004-637X/822/1/26}{\color{magenta}\apj},
  \href{http://adsabs.harvard.edu/abs/2016ApJ...822...26G}{822, 26}

\bibitem[{{Hoormann} {et~al.}(2019){Hoormann}, {Martini}, {Davis}, {King},
  {Lidman}, {Mudd}, {Sharp}, {Sommer}, {Tucker}, {Yu}, {Allam}, {Asorey},
  {Avila}, {Banerji}, {Brooks}, {Buckley-Geer}, {Burke}, {Calcino}, {Carnero
  Rosell}, {Carollo}, {Carrasco Kind}, {Carretero}, {Castander}, {Childress},
  {De Vicente}, {Desai}, {Diehl}, {Doel}, {Flaugher}, {Fosalba}, {Frieman},
  {Garc{\'{\i}}a-Bellido}, {Gerdes}, {Gruen}, {Gutierrez}, {Hartley}, {Hinton},
  {Hollowood}, {Honscheid}, {Hoyle}, {James}, {Krause}, {Kuehn}, {Kuropatkin},
  {Lewis}, {Lima}, {Macaulay}, {Maia}, {Menanteau}, {Miller}, {Miquel},
  {M{\"o}ller}, {Plazas}, {Romer}, {Roodman}, {Sanchez}, {Scarpine},
  {Schubnell}, {Serrano}, {Sevilla-Noarbe}, {Smith}, {Smith}, {Soares-Santos},
  {Sobreira}, {Suchyta}, {Swann}, {Swanson}, {Tarle}, {Uddin}, \& {DES
  Collaboration}}]{Hoormann2019}
{Hoormann}, J.~K., {Martini}, P., {Davis}, T.~M., {et~al.} 2019, arXiv e-prints
  \href{http://adsabs.harvard.edu/abs/2019arXiv190204206H}{[\eprint[arXiv]{1902.04206}]}

\bibitem[{{Hryniewicz} {et~al.}(2014){Hryniewicz}, {Czerny}, {Pych}, {Udalski},
  {Krupa}, {{\'S}wi{\c e}to{\'n}}, \& {Kaluzny}}]{Hryniewicz2014}
{Hryniewicz}, K., {Czerny}, B., {Pych}, W., {et~al.} 2014,
  \href{http://dx.doi.org/10.1051/0004-6361/201322487}{\color{magenta}\aap},
  \href{http://adsabs.harvard.edu/abs/2014A%26A...562A..34H}{562, A34}

\bibitem[{{Ivezi{\'c}} {et~al.}(2007){Ivezi{\'c}}, {Smith}, {Miknaitis}, {Lin},
  {Tucker}, {Lupton}, {Gunn}, {Knapp}, {Strauss}, {Sesar}, {Doi}, {Tanaka},
  {Fukugita}, {Holtzman}, {Kent}, {Yanny}, {Schlegel}, {Finkbeiner},
  {Padmanabhan}, {Rockosi}, {Juri{\'c}}, {Bond}, {Lee}, {Stoughton}, {Jester},
  {Harris}, {Harding}, {Morrison}, {Brinkmann}, {Schneider}, \&
  {York}}]{Ivezic2007}
{Ivezi{\'c}}, {\v Z}., {Smith}, J.~A., {Miknaitis}, G., {et~al.} 2007,
  \href{http://dx.doi.org/10.1086/519976}{\color{magenta}\aj},
  \href{http://adsabs.harvard.edu/abs/2007AJ....134..973I}{134, 973}

\bibitem[{{Joni{\'c}} {et~al.}(2016){Joni{\'c}}, {Kova{\v c}evi{\'c}-Doj{\v
  c}inovi{\'c}}, {Ili{\'c}}, \& {Popovi{\'c}}}]{Jonic2016}
{Joni{\'c}}, S., {Kova{\v c}evi{\'c}-Doj{\v c}inovi{\'c}}, J., {Ili{\'c}}, D.,
  \& {Popovi{\'c}}, L.~{\v C}. 2016,
  \href{http://dx.doi.org/10.1007/s10509-016-2680-9}{\color{magenta}\apss},
  \href{http://adsabs.harvard.edu/abs/2016Ap%26SS.361..101J}{361, 101}

\bibitem[{{Korista} \& {Goad}(2004)}]{Korista2004}
{Korista}, K.~T. \& {Goad}, M.~R. 2004,
  \href{http://dx.doi.org/10.1086/383193}{\color{magenta}\apj},
  \href{http://adsabs.harvard.edu/abs/2004ApJ...606..749K}{606, 749}

\bibitem[{{Kova{\v c}evi{\'c}-Doj{\v c}inovi{\'c}} \&
  {Popovi{\'c}}(2015)}]{Kovacevic2015}
{Kova{\v c}evi{\'c}-Doj{\v c}inovi{\'c}}, J. \& {Popovi{\'c}}, L.~{\v C}. 2015,
  \href{http://dx.doi.org/10.1088/0067-0049/221/2/35}{\color{magenta}\apjs},
  \href{http://adsabs.harvard.edu/abs/2015ApJS..221...35K}{221, 35}

\bibitem[{{Krolik}(1999)}]{Krolik1999}
{Krolik}, J.~H. 1999, {Active galactic nuclei : from the central black hole to
  the galactic environment} (Princeton University Press, c1999.)

\bibitem[{{Kwan} \& {Krolik}(1981)}]{Kwan1981}
{Kwan}, J. \& {Krolik}, J.~H. 1981,
  \href{http://dx.doi.org/10.1086/159395}{\color{magenta}\apj},
  \href{http://adsabs.harvard.edu/abs/1981ApJ...250..478K}{250, 478}

\bibitem[{{LaMassa} {et~al.}(2015){LaMassa}, {Cales}, {Moran}, {Myers},
  {Richards}, {Eracleous}, {Heckman}, {Gallo}, \& {Urry}}]{LaMassa2015}
{LaMassa}, S.~M., {Cales}, S., {Moran}, E.~C., {et~al.} 2015,
  \href{http://dx.doi.org/10.1088/0004-637X/800/2/144}{\color{magenta}\apj},
  \href{http://adsabs.harvard.edu/abs/2015ApJ...800..144L}{800, 144}

\bibitem[{{Lawrence}(2018)}]{Lawrence2018}
{Lawrence}, A. 2018,
  \href{http://dx.doi.org/10.1038/s41550-017-0372-1}{\color{magenta}Nature
  Astronomy}, \href{http://adsabs.harvard.edu/abs/2018NatAs...2..102L}{2, 102}

\bibitem[{{Le{\'o}n-Tavares} {et~al.}(2013){Le{\'o}n-Tavares}, {Chavushyan},
  {Pati{\~n}o-{\'A}lvarez}, {Valtaoja}, {Arshakian}, {Popovi{\'c}},
  {Tornikoski}, {Lobanov}, {Carrami{\~n}ana}, {Carrasco}, \&
  {L{\"a}hteenm{\"a}ki}}]{Tavares2013}
{Le{\'o}n-Tavares}, J., {Chavushyan}, V., {Pati{\~n}o-{\'A}lvarez}, V.,
  {et~al.} 2013,
  \href{http://dx.doi.org/10.1088/2041-8205/763/2/L36}{\color{magenta}\apjl},
  \href{http://adsabs.harvard.edu/abs/2013ApJ...763L..36L}{763, L36}

\bibitem[{{MacLeod} {et~al.}(2019){MacLeod}, {Green}, {Anderson}, {Bruce},
  {Eracleous}, {Graham}, {Homan}, {Lawrence}, {LeBleu}, {Ross}, {Ruan},
  {Runnoe}, {Stern}, {Burgett}, {Chambers}, {Kaiser}, {Magnier}, \&
  {Metcalfe}}]{MacLeod2019}
{MacLeod}, C.~L., {Green}, P.~J., {Anderson}, S.~F., {et~al.} 2019,
  \href{http://dx.doi.org/10.3847/1538-4357/ab05e2}{\color{magenta}\apj},
  \href{http://adsabs.harvard.edu/abs/2019ApJ...874....8M}{874, 8}

\bibitem[{{MacLeod} {et~al.}(2012){MacLeod}, {Ivezi{\'c}}, {Sesar}, {de Vries},
  {Kochanek}, {Kelly}, {Becker}, {Lupton}, {Hall}, {Richards}, {Anderson}, \&
  {Schneider}}]{MacLeod2012}
{MacLeod}, C.~L., {Ivezi{\'c}}, {\v Z}., {Sesar}, B., {et~al.} 2012,
  \href{http://dx.doi.org/10.1088/0004-637X/753/2/106}{\color{magenta}\apj},
  \href{http://adsabs.harvard.edu/abs/2012ApJ...753..106M}{753, 106}

\bibitem[{{MacLeod} {et~al.}(2016){MacLeod}, {Ross}, {Lawrence}, {Goad},
  {Horne}, {Burgett}, {Chambers}, {Flewelling}, {Hodapp}, {Kaiser}, {Magnier},
  {Wainscoat}, \& {Waters}}]{MacLeod2016}
{MacLeod}, C.~L., {Ross}, N.~P., {Lawrence}, A., {et~al.} 2016,
  \href{http://dx.doi.org/10.1093/mnras/stv2997}{\color{magenta}\mnras},
  \href{http://adsabs.harvard.edu/abs/2016MNRAS.457..389M}{457, 389}

\bibitem[{{McElroy} {et~al.}(2016){McElroy}, {Husemann}, {Croom}, {Davis},
  {Bennert}, {Busch}, {Combes}, {Eckart}, {Perez-Torres}, {Powell},
  {Scharw{\"a}chter}, {Tremblay}, \& {Urrutia}}]{McElroy2016}
{McElroy}, R.~E., {Husemann}, B., {Croom}, S.~M., {et~al.} 2016,
  \href{http://dx.doi.org/10.1051/0004-6361/201629102}{\color{magenta}\aap},
  \href{http://adsabs.harvard.edu/abs/2016A%26A...593L...8M}{593, L8}

\bibitem[{{McLure} \& {Dunlop}(2004)}]{McLure2004}
{McLure}, R.~J. \& {Dunlop}, J.~S. 2004,
  \href{http://dx.doi.org/10.1111/j.1365-2966.2004.08034.x}{\color{magenta}\mnras},
  \href{http://adsabs.harvard.edu/abs/2004MNRAS.352.1390M}{352, 1390}

\bibitem[{{McLure} \& {Jarvis}(2002)}]{McLure2002}
{McLure}, R.~J. \& {Jarvis}, M.~J. 2002,
  \href{http://dx.doi.org/10.1046/j.1365-8711.2002.05871.x}{\color{magenta}\mnras},
  \href{http://adsabs.harvard.edu/abs/2002MNRAS.337..109M}{337, 109}

\bibitem[{{Metzroth} {et~al.}(2006){Metzroth}, {Onken}, \&
  {Peterson}}]{Metzroth2006}
{Metzroth}, K.~G., {Onken}, C.~A., \& {Peterson}, B.~M. 2006,
  \href{http://dx.doi.org/10.1086/505525}{\color{magenta}\apj},
  \href{http://adsabs.harvard.edu/abs/2006ApJ...647..901M}{647, 901}

\bibitem[{{Morganson} {et~al.}(2014){Morganson}, {Burgett}, {Chambers},
  {Green}, {Kaiser}, {Magnier}, {Marshall}, {Morgan}, {Price}, {Rix},
  {Schlafly}, {Tonry}, \& {Walter}}]{Morganson2014}
{Morganson}, E., {Burgett}, W.~S., {Chambers}, K.~C., {et~al.} 2014,
  \href{http://dx.doi.org/10.1088/0004-637X/784/2/92}{\color{magenta}\apj},
  \href{http://adsabs.harvard.edu/abs/2014ApJ...784...92M}{784, 92}

\bibitem[{{P{\^a}ris} {et~al.}(2018){P{\^a}ris}, {Petitjean}, {Aubourg},
  {Myers}, {Streblyanska}, {Lyke}, {Anderson}, {Armengaud}, {Bautista},
  {Blanton}, {Blomqvist}, {Brinkmann}, {Brownstein}, {Brandt}, {Burtin},
  {Dawson}, {de la Torre}, {Georgakakis}, {Gil-Mar{\'{\i}}n}, {Green}, {Hall},
  {Kneib}, {LaMassa}, {Le Goff}, {MacLeod}, {Mariappan}, {McGreer}, {Merloni},
  {Noterdaeme}, {Palanque-Delabrouille}, {Percival}, {Ross}, {Rossi},
  {Schneider}, {Seo}, {Tojeiro}, {Weaver}, {Weijmans}, {Y{\`e}che}, {Zarrouk},
  \& {Zhao}}]{Paris2018}
{P{\^a}ris}, I., {Petitjean}, P., {Aubourg}, {\'E}., {et~al.} 2018,
  \href{http://dx.doi.org/10.1051/0004-6361/201732445}{\color{magenta}\aap},
  \href{http://adsabs.harvard.edu/abs/2018A%26A...613A..51P}{613, A51}

\bibitem[{{Park} {et~al.}(2012){Park}, {Woo}, {Treu}, {Barth}, {Bentz},
  {Bennert}, {Canalizo}, {Filippenko}, {Gates}, {Greene}, {Malkan}, \&
  {Walsh}}]{Park2012}
{Park}, D., {Woo}, J.-H., {Treu}, T., {et~al.} 2012,
  \href{http://dx.doi.org/10.1088/0004-637X/747/1/30}{\color{magenta}\apj},
  \href{http://adsabs.harvard.edu/abs/2012ApJ...747...30P}{747, 30}

\bibitem[{{Popovi{\'c}} {et~al.}(2019){Popovi{\'c}}, {Kova{\v c}evi{\'c}-Doj{\v
  c}inovi{\'c}}, \& {Mar{\v c}eta-Mandi{\'c}}}]{Popovic2019}
{Popovi{\'c}}, L.~{\v C}., {Kova{\v c}evi{\'c}-Doj{\v c}inovi{\'c}}, J., \&
  {Mar{\v c}eta-Mandi{\'c}}, S. 2019,
  \href{http://dx.doi.org/10.1093/mnras/stz157}{\color{magenta}\mnras}
  \href{http://adsabs.harvard.edu/abs/2019MNRAS.tmp..163P}{[\eprint[arXiv]{1901.03681}]}

\bibitem[{{Reichert} {et~al.}(1994){Reichert}, {Rodriguez-Pascual}, {Alloin},
  {Clavel}, {Crenshaw}, {Kriss}, {Krolik}, {Malkan}, {Netzer}, {Peterson},
  {Wamsteker}, {Altamore}, {Altieri}, {Anderson}, {Blackwell}, {Boisson},
  {Brosch}, {Carone}, {Dietrich}, {England}, {Evans}, {Filippenko}, {Gaskell},
  {Goad}, {Gondhalekar}, {Horne}, {Kazanas}, {Kollatschny}, {Koratkar},
  {Korista}, {MacAlpine}, {Maoz}, {Mazeh}, {McCollum}, {Miller}, {Mendes de
  Oliveira}, {O'Brien}, {Pastoriza}, {Pelat}, {Perez}, {Perola}, {Pogge},
  {Ptak}, {Recondo-Gonzalez}, {Rodriguez-Espinosa}, {Rosenblatt}, {Sadun},
  {Santos-Lleo}, {Shields}, {Shrader}, {Shull}, {Simkin}, {Sitko}, {Snijders},
  {Sparke}, {Stirpe}, {Stoner}, {Storchi-Bergmann}, {Sun}, {Wang}, {Welsh},
  {White}, {Winge}, \& {Zheng}}]{Reichert1994}
{Reichert}, G.~A., {Rodriguez-Pascual}, P.~M., {Alloin}, D., {et~al.} 1994,
  \href{http://dx.doi.org/10.1086/174007}{\color{magenta}\apj},
  \href{http://adsabs.harvard.edu/abs/1994ApJ...425..582R}{425, 582}

\bibitem[{{Roig} {et~al.}(2014){Roig}, {Blanton}, \& {Ross}}]{Roig2014}
{Roig}, B., {Blanton}, M.~R., \& {Ross}, N.~P. 2014,
  \href{http://dx.doi.org/10.1088/0004-637X/781/2/72}{\color{magenta}\apj},
  \href{http://adsabs.harvard.edu/abs/2014ApJ...781...72R}{781, 72}

\bibitem[{{Ross} {et~al.}(2018){Ross}, {Ford}, {Graham}, {McKernan}, {Stern},
  {Meisner}, {Assef}, {Dey}, {Drake}, {Jun}, \& {Lang}}]{Ross2018}
{Ross}, N.~P., {Ford}, K.~E.~S., {Graham}, M., {et~al.} 2018,
  \href{http://dx.doi.org/10.1093/mnras/sty2002}{\color{magenta}\mnras},
  \href{http://adsabs.harvard.edu/abs/2018MNRAS.480.4468R}{480, 4468}

\bibitem[{{Ruan} {et~al.}(2016){Ruan}, {Anderson}, {Cales}, {Eracleous},
  {Green}, {Morganson}, {Runnoe}, {Shen}, {Wilkinson}, {Blanton}, {Dwelly},
  {Georgakakis}, {Greene}, {LaMassa}, {Merloni}, \& {Schneider}}]{Ruan2016}
{Ruan}, J.~J., {Anderson}, S.~F., {Cales}, S.~L., {et~al.} 2016,
  \href{http://dx.doi.org/10.3847/0004-637X/826/2/188}{\color{magenta}\apj},
  \href{http://adsabs.harvard.edu/abs/2016ApJ...826..188R}{826, 188}

\bibitem[{{Rumbaugh} {et~al.}(2018){Rumbaugh}, {Shen}, {Morganson}, {Liu},
  {Banerji}, {McMahon}, {Abdalla}, {Benoit-L{\'e}vy}, {Bertin}, {Brooks},
  {Buckley-Geer}, {Capozzi}, {Carnero Rosell}, {Carrasco Kind}, {Carretero},
  {Cunha}, {D'Andrea}, {da Costa}, {DePoy}, {Desai}, {Doel}, {Frieman},
  {Garc{\'{\i}}a-Bellido}, {Gruen}, {Gruendl}, {Gschwend}, {Gutierrez},
  {Honscheid}, {James}, {Kuehn}, {Kuhlmann}, {Kuropatkin}, {Lima}, {Maia},
  {Marshall}, {Martini}, {Menanteau}, {Plazas}, {Reil}, {Roodman}, {Sanchez},
  {Scarpine}, {Schindler}, {Schubnell}, {Sheldon}, {Smith}, {Soares-Santos},
  {Sobreira}, {Suchyta}, {Swanson}, {Walker}, {Wester}, \& {(DES
  Collaboration}}]{Rumbaugh2018}
{Rumbaugh}, N., {Shen}, Y., {Morganson}, E., {et~al.} 2018,
  \href{http://dx.doi.org/10.3847/1538-4357/aaa9b6}{\color{magenta}\apj},
  \href{http://adsabs.harvard.edu/abs/2018ApJ...854..160R}{854, 160}

\bibitem[{{Runco} {et~al.}(2016){Runco}, {Cosens}, {Bennert}, {Scott},
  {Komossa}, {Malkan}, {Lazarova}, {Auger}, {Treu}, \& {Park}}]{Runco2016}
{Runco}, J.~N., {Cosens}, M., {Bennert}, V.~N., {et~al.} 2016,
  \href{http://dx.doi.org/10.3847/0004-637X/821/1/33}{\color{magenta}\apj},
  \href{http://adsabs.harvard.edu/abs/2016ApJ...821...33R}{821, 33}

\bibitem[{{Runnoe} {et~al.}(2016){Runnoe}, {Cales}, {Ruan}, {Eracleous},
  {Anderson}, {Shen}, {Green}, {Morganson}, {LaMassa}, {Greene}, {Dwelly},
  {Schneider}, {Merloni}, {Georgakakis}, \& {Roman-Lopes}}]{Runnoe2016}
{Runnoe}, J.~C., {Cales}, S., {Ruan}, J.~J., {et~al.} 2016,
  \href{http://dx.doi.org/10.1093/mnras/stv2385}{\color{magenta}\mnras},
  \href{http://adsabs.harvard.edu/abs/2016MNRAS.455.1691R}{455, 1691}

\bibitem[{{Salviander} {et~al.}(2007){Salviander}, {Shields}, {Gebhardt}, \&
  {Bonning}}]{Salviander2007}
{Salviander}, S., {Shields}, G.~A., {Gebhardt}, K., \& {Bonning}, E.~W. 2007,
  \href{http://dx.doi.org/10.1086/513086}{\color{magenta}\apj},
  \href{http://adsabs.harvard.edu/abs/2007ApJ...662..131S}{662, 131}

\bibitem[{{Schlegel} {et~al.}(1998){Schlegel}, {Finkbeiner}, \&
  {Davis}}]{Schlegel1998}
{Schlegel}, D.~J., {Finkbeiner}, D.~P., \& {Davis}, M. 1998,
  \href{http://dx.doi.org/10.1086/305772}{\color{magenta}\apj},
  \href{http://adsabs.harvard.edu/abs/1998ApJ...500..525S}{500, 525}

\bibitem[{{Schmidt} {et~al.}(2010){Schmidt}, {Marshall}, {Rix}, {Jester},
  {Hennawi}, \& {Dobler}}]{Schmidt2010}
{Schmidt}, K.~B., {Marshall}, P.~J., {Rix}, H.-W., {et~al.} 2010,
  \href{http://dx.doi.org/10.1088/0004-637X/714/2/1194}{\color{magenta}\apj},
  \href{http://adsabs.harvard.edu/abs/2010ApJ...714.1194S}{714, 1194}

\bibitem[{{Sesar} {et~al.}(2007){Sesar}, {Ivezi{\'c}}, {Lupton}, {Juri{\'c}},
  {Gunn}, {Knapp}, {DeLee}, {Smith}, {Miknaitis}, {Lin}, {Tucker}, {Doi},
  {Tanaka}, {Fukugita}, {Holtzman}, {Kent}, {Yanny}, {Schlegel}, {Finkbeiner},
  {Padmanabhan}, {Rockosi}, {Bond}, {Lee}, {Stoughton}, {Jester}, {Harris},
  {Harding}, {Brinkmann}, {Schneider}, {York}, {Richmond}, \& {Vanden
  Berk}}]{Sesar2007}
{Sesar}, B., {Ivezi{\'c}}, {\v Z}., {Lupton}, R.~H., {et~al.} 2007,
  \href{http://dx.doi.org/10.1086/521819}{\color{magenta}\aj},
  \href{http://adsabs.harvard.edu/abs/2007AJ....134.2236S}{134, 2236}

\bibitem[{{Shen}(2013)}]{Shen2013}
{Shen}, Y. 2013, Bulletin of the Astronomical Society of India,
  \href{http://adsabs.harvard.edu/abs/2013BASI...41...61S}{41, 61}

\bibitem[{{Shen} {et~al.}(2015){Shen}, {Brandt}, {Dawson}, {Hall}, {McGreer},
  {Anderson}, {Chen}, {Denney}, {Eftekharzadeh}, {Fan}, {Gao}, {Green},
  {Greene}, {Ho}, {Horne}, {Jiang}, {Kelly}, {Kinemuchi}, {Kochanek},
  {P{\^a}ris}, {Peters}, {Peterson}, {Petitjean}, {Ponder}, {Richards},
  {Schneider}, {Seth}, {Smith}, {Strauss}, {Tao}, {Trump}, {Wood-Vasey}, {Zu},
  {Eisenstein}, {Pan}, {Bizyaev}, {Malanushenko}, {Malanushenko}, \&
  {Oravetz}}]{Shen2015}
{Shen}, Y., {Brandt}, W.~N., {Dawson}, K.~S., {et~al.} 2015,
  \href{http://dx.doi.org/10.1088/0067-0049/216/1/4}{\color{magenta}\apjs},
  \href{http://adsabs.harvard.edu/abs/2015ApJS..216....4S}{216, 4}

\bibitem[{{Shen} {et~al.}(2008){Shen}, {Greene}, {Strauss}, {Richards}, \&
  {Schneider}}]{Shen2008}
{Shen}, Y., {Greene}, J.~E., {Strauss}, M.~A., {Richards}, G.~T., \&
  {Schneider}, D.~P. 2008,
  \href{http://dx.doi.org/10.1086/587475}{\color{magenta}\apj},
  \href{http://adsabs.harvard.edu/abs/2008ApJ...680..169S}{680, 169}

\bibitem[{{Shen} {et~al.}(2018){Shen}, {Hall}, {Horne}, {Zhu}, {McGreer},
  {Simm}, {Trump}, {Kinemuchi}, {Brandt}, {Green}, {Grier}, {Guo}, {Ho},
  {Homayouni}, {Jiang}, {I-Hsiu Li}, {Morganson}, {Petitjean}, {Richards},
  {Schneider}, {Starkey}, {Wang}, {Chambers}, {Kaiser}, {Kudritzki}, {Magnier},
  \& {Waters}}]{Shen2018}
{Shen}, Y., {Hall}, P.~B., {Horne}, K., {et~al.} 2018, arXiv e-prints
  \href{http://adsabs.harvard.edu/abs/2018arXiv181001447S}{[\eprint[arXiv]{1810.01447}]}

\bibitem[{{Shen} {et~al.}(2016){Shen}, {Horne}, {Grier}, {Peterson}, {Denney},
  {Trump}, {Sun}, {Brandt}, {Kochanek}, {Dawson}, {Green}, {Greene}, {Hall},
  {Ho}, {Jiang}, {Kinemuchi}, {McGreer}, {Petitjean}, {Richards}, {Schneider},
  {Strauss}, {Tao}, {Wood-Vasey}, {Zu}, {Pan}, {Bizyaev}, {Ge}, {Oravetz}, \&
  {Simmons}}]{Shen2016}
{Shen}, Y., {Horne}, K., {Grier}, C.~J., {et~al.} 2016,
  \href{http://dx.doi.org/10.3847/0004-637X/818/1/30}{\color{magenta}\apj},
  \href{http://adsabs.harvard.edu/abs/2016ApJ...818...30S}{818, 30}

\bibitem[{{Shen} {et~al.}(2011){Shen}, {Richards}, {Strauss}, {Hall},
  {Schneider}, {Snedden}, {Bizyaev}, {Brewington}, {Malanushenko},
  {Malanushenko}, {Oravetz}, {Pan}, \& {Simmons}}]{Shen2011}
{Shen}, Y., {Richards}, G.~T., {Strauss}, M.~A., {et~al.} 2011,
  \href{http://dx.doi.org/10.1088/0067-0049/194/2/45}{\color{magenta}\apjs},
  \href{http://adsabs.harvard.edu/abs/2011ApJS..194...45S}{194, 45}

\bibitem[{{Stern} {et~al.}(2018){Stern}, {McKernan}, {Graham}, {Ford}, {Ross},
  {Meisner}, {Assef}, {Balokovi{\'c}}, {Brightman}, {Dey}, {Drake},
  {Djorgovski}, {Eisenhardt}, \& {Jun}}]{Stern2018}
{Stern}, D., {McKernan}, B., {Graham}, M.~J., {et~al.} 2018,
  \href{http://dx.doi.org/10.3847/1538-4357/aac726}{\color{magenta}\apj},
  \href{http://adsabs.harvard.edu/abs/2018ApJ...864...27S}{864, 27}

\bibitem[{{Sun} {et~al.}(2015){Sun}, {Trump}, {Shen}, {Brandt}, {Dawson},
  {Denney}, {Hall}, {Ho}, {Horne}, {Jiang}, {Richards}, {Schneider}, {Bizyaev},
  {Kinemuchi}, {Oravetz}, {Pan}, \& {Simmons}}]{Sun2015}
{Sun}, M., {Trump}, J.~R., {Shen}, Y., {et~al.} 2015,
  \href{http://dx.doi.org/10.1088/0004-637X/811/1/42}{\color{magenta}\apj},
  \href{http://adsabs.harvard.edu/abs/2015ApJ...811...42S}{811, 42}

\bibitem[{{Tody}(1986)}]{Tody1986}
{Tody}, D. 1986, in \procspie, Vol. 627, Instrumentation in astronomy VI, ed.
  D.~L. {Crawford},
  \href{http://adsabs.harvard.edu/abs/1986SPIE..627..733T}{733}

\bibitem[{{Tody}(1993)}]{Tody1993}
{Tody}, D. 1993, in Astronomical Society of the Pacific Conference Series,
  Vol.~52, Astronomical Data Analysis Software and Systems II, ed. R.~J.
  {Hanisch}, R.~J.~V. {Brissenden}, \& J.~{Barnes},
  \href{http://adsabs.harvard.edu/abs/1993ASPC...52..173T}{173}

\bibitem[{{Trakhtenbrot} {et~al.}(2019){Trakhtenbrot}, {Arcavi}, {MacLeod},
  {Ricci}, {Kara}, {Graham}, {Stern}, {Harrison}, {Burke}, {Hiramatsu},
  {Hosseinzadeh}, {Howell}, {Smartt}, {Rest}, {Prieto}, {Shappee}, {Holoien},
  {Bersier}, {Filippenko}, {Brink}, {Zheng}, {Li}, {Remillard}, \&
  {Loewenstein}}]{Trakhtenbrot2019}
{Trakhtenbrot}, B., {Arcavi}, I., {MacLeod}, C.~L., {et~al.} 2019, arXiv
  e-prints
  \href{http://adsabs.harvard.edu/abs/2019arXiv190311084T}{[\eprint[arXiv]{1903.11084}]}

\bibitem[{{Trevese} {et~al.}(2007){Trevese}, {Paris}, {Stirpe}, {Vagnetti}, \&
  {Zitelli}}]{Trevese2007}
{Trevese}, D., {Paris}, D., {Stirpe}, G.~M., {Vagnetti}, F., \& {Zitelli}, V.
  2007,
  \href{http://dx.doi.org/10.1051/0004-6361:20077237}{\color{magenta}\aap},
  \href{http://adsabs.harvard.edu/abs/2007A%26A...470..491T}{470, 491}

\bibitem[{{Tsuzuki} {et~al.}(2006){Tsuzuki}, {Kawara}, {Yoshii}, {Oyabu},
  {Tanab{\'e}}, \& {Matsuoka}}]{Tsuzuki2006}
{Tsuzuki}, Y., {Kawara}, K., {Yoshii}, Y., {et~al.} 2006,
  \href{http://dx.doi.org/10.1086/506376}{\color{magenta}\apj},
  \href{http://adsabs.harvard.edu/abs/2006ApJ...650...57T}{650, 57}

\bibitem[{{Urry} \& {Padovani}(1995)}]{Urry1995}
{Urry}, C.~M. \& {Padovani}, P. 1995,
  \href{http://dx.doi.org/10.1086/133630}{\color{magenta}\pasp},
  \href{http://adsabs.harvard.edu/abs/1995PASP..107..803U}{107, 803}

\bibitem[{{Vanden Berk} {et~al.}(2004){Vanden Berk}, {Wilhite}, {Kron},
  {Anderson}, {Brunner}, {Hall}, {Ivezi{\'c}}, {Richards}, {Schneider}, {York},
  {Brinkmann}, {Lamb}, {Nichol}, \& {Schlegel}}]{VandenBerk2004}
{Vanden Berk}, D.~E., {Wilhite}, B.~C., {Kron}, R.~G., {et~al.} 2004,
  \href{http://dx.doi.org/10.1086/380563}{\color{magenta}\apj},
  \href{http://adsabs.harvard.edu/abs/2004ApJ...601..692V}{601, 692}

\bibitem[{{Vestergaard} \& {Wilkes}(2001)}]{Vestergaard2001}
{Vestergaard}, M. \& {Wilkes}, B.~J. 2001,
  \href{http://dx.doi.org/10.1086/320357}{\color{magenta}\apjs},
  \href{http://adsabs.harvard.edu/abs/2001ApJS..134....1V}{134, 1}

\bibitem[{{Wandel} {et~al.}(1999){Wandel}, {Peterson}, \&
  {Malkan}}]{Wandel1999}
{Wandel}, A., {Peterson}, B.~M., \& {Malkan}, M.~A. 1999,
  \href{http://dx.doi.org/10.1086/308017}{\color{magenta}\apj},
  \href{http://adsabs.harvard.edu/abs/1999ApJ...526..579W}{526, 579}

\bibitem[{{Wang} {et~al.}(2009){Wang}, {Dong}, {Wang}, {Ho}, {Yuan}, {Wang},
  {Zhang}, {Zhang}, \& {Zhou}}]{Wang2009}
{Wang}, J.-G., {Dong}, X.-B., {Wang}, T.-G., {et~al.} 2009,
  \href{http://dx.doi.org/10.1088/0004-637X/707/2/1334}{\color{magenta}\apj},
  \href{http://adsabs.harvard.edu/abs/2009ApJ...707.1334W}{707, 1334}

\bibitem[{{Wang} {et~al.}(2019){Wang}, {Shen}, {Jiang}, {Horne}, {Brandt},
  {Grier}, {Ho}, {Homayouni}, {I-Hsiu Li}, {Schneider}, \&
  {Trump}}]{WangShu2019}
{Wang}, S., {Shen}, Y., {Jiang}, L., {et~al.} 2019, arXiv e-prints
  \href{http://adsabs.harvard.edu/abs/2019arXiv190310015W}{[\eprint[arXiv]{1903.10015}]}

\bibitem[{{Wilhite} {et~al.}(2005){Wilhite}, {Vanden Berk}, {Kron},
  {Schneider}, {Pereyra}, {Brunner}, {Richards}, \& {Brinkmann}}]{Wilhite2005}
{Wilhite}, B.~C., {Vanden Berk}, D.~E., {Kron}, R.~G., {et~al.} 2005,
  \href{http://dx.doi.org/10.1086/430821}{\color{magenta}\apj},
  \href{http://adsabs.harvard.edu/abs/2005ApJ...633..638W}{633, 638}

\bibitem[{{Woo}(2008)}]{Woo2008}
{Woo}, J.-H. 2008,
  \href{http://dx.doi.org/10.1088/0004-6256/135/5/1849}{\color{magenta}\aj},
  \href{http://adsabs.harvard.edu/abs/2008AJ....135.1849W}{135, 1849}

\bibitem[{{Yang} {et~al.}(2018){Yang}, {Wu}, {Fan}, {Jiang}, {McGreer},
  {Shangguan}, {Yao}, {Wang}, {Joshi}, {Green}, {Wang}, {Feng}, {Fu}, {Yang},
  \& {Liu}}]{Yang2018}
{Yang}, Q., {Wu}, X.-B., {Fan}, X., {et~al.} 2018,
  \href{http://dx.doi.org/10.3847/1538-4357/aaca3a}{\color{magenta}\apj},
  \href{http://adsabs.harvard.edu/abs/2018ApJ...862..109Y}{862, 109}

\bibitem[{{Zhu} {et~al.}(2017){Zhu}, {Sun}, \& {Wang}}]{Zhu2017}
{Zhu}, D., {Sun}, M., \& {Wang}, T. 2017,
  \href{http://dx.doi.org/10.3847/1538-4357/aa76e7}{\color{magenta}\apj},
  \href{http://adsabs.harvard.edu/abs/2017ApJ...843...30Z}{843, 30}

\end{thebibliography}

\bsp	
\label{lastpage}
\end{document}